\documentclass[11pt,a4paper]{article}
\usepackage{jheppub}

\usepackage{amssymb}
\usepackage{graphicx,psfrag,color}
\usepackage{dcolumn}
\usepackage{bm}
\usepackage{mathbbol} 
\usepackage{mathabx}
\usepackage{mathtools,cuted}

\title{On Lorentz invariant complex scalar fields}

\author{Gustavo Rigolin}
\affiliation{Departamento de F\'isica, Universidade Federal de
S\~ao Carlos,\\ S\~ao Carlos, SP 13565-905, Brazil}
\emailAdd{rigolin@ufscar.br}

\abstract{ 
We obtain a Lorentz covariant wave equation who\-se complex wave function transforms 
under a Lorentz boost according to the following rule,
$\Psi(x)\rightarrow e^{\frac{i}{\hbar}f(x)}\Psi(x)$. We show that 
the spacetime dependent phase $f(x)$ is the most natural relativistic 
extension of the phase associated with
the transformation rule for the non-relativistic Schr\"odinger wave function when
it is subjected to a Galilean transformation. We then generalize the previous 
analysis by postulating that $\Psi(x)$ transforms according to the above rule
under proper Lorentz transformations (boosts or spatial rotations). 
This is the most general 
transformation rule compatible with a Lorentz invariant physical theory 
whose observables are bilinear
functions of the field $\Psi(x)$. 
We use the previous wave equations to describe several physical systems. 
In particular, we solve the bound state and scattering problems of two particles
which interact both electromagnetically and gravitationally 
(static electromagnetic and gravitational fields). 
The former interaction is modeled via the minimal coupling prescription while the 
latter enters via an external potential.
We also formulate logically consistent 
classical and quantum field theories associated with these Lorentz covariant 
wave equations. We show that it is possible to make those theories 
equivalent to the Klein-Gordon theory 
whenever we have self-interacting terms that 
do not break their Lorentz invariance or if we introduce electromagnetic interactions via the minimal coupling prescription.
For interactions that break Lorentz invariance, we show that the present 
theories imply that particles and antiparticles
behave differently at decaying processes, 
with the latter being more unstable. This suggests a possible 
connection between Lorentz invariance-breaking interactions and 
the matter-antimatter asymmetry problem.
}


\arxivnumber{2010.13767 [hep-ph]}
\notoc

\begin{document}
\maketitle
\flushbottom


\section{Introduction}

A complex scalar field $\Phi(x)$ is usually defined as a 
function of the spacetime coordinates $x=(ct,\mathbf{r})$ such that it
remains invariant under a symmetry operation, i.e., $\Phi(x)=\Phi'(x')$, where 
$\Phi'(x')$ is the field after we apply the symmetry operation. In 
its more general standard definition, one may also multiply the
transformed field by a \textit{constant} complex number of modulus 
one \cite{bal98,gre00,gri95,man86,gre95}. 
A paradigmatic example of a physical system described by 
a complex scalar field is a charged 
Klein-Gordon particle, which transforms according to the above rule 
under proper Lorentz transformations (boosts or space rotations). 

And what about the non-relativistic wave function, which is a solution to the 
Schr\"odin\-ger
equation \cite{sch26}? 
Is it a complex scalar field in the above sense? Strictly speaking, it is not. 
This comes about because under a Galilean transformation it changes 
according to the following prescription \cite{bal98},
$\Psi(x) = e^{\frac{i}{\hbar}\theta(x')} \Psi'(x')$,
where the phase $\theta(x')$ is \textit{not a constant}, being a function of
the spacetime coordinates. We require $\Psi(x)$ to transform in this way
in order to have the Schr\"odinger equation invariant 
under a Galilean transformation \cite{bal98}. 

The Schr\"odinger field $\Psi(x)$ illustrates 
that it is perfectly possible to build a logically consistent field theory assuming a more general transformation law for a complex scalar field under a symmetry operation,
where the phase $\theta$ depends on the spacetime coordinates. 
The extension of the latter observation 
to the relativistic domain is the leitmotif of the present work. 
We want to determine the wave equation whose wave function transforms 
similarly to Schr\"odinger's wave function and which is invariant (Lorentz covariant)
under proper Lorentz transformations. 

We can also understand the main idea of this work, as described above, as the search for
the answer to the following simple question:
What is the Lorentz covariant wave equation associated with a complex scalar field $\Psi(x)$ 
whose bilinear $\Psi(x) \Psi^*(x)=|\Psi(x)|^2$, rather than $\Psi(x)$ itself, is assumed to be a Lorentz invariant (scalar)?
Note that here we are demanding not $\Psi(x)$ but rather $\Psi(x) \Psi^*(x)$ to be a Lorentz scalar. 
The above question can be rephrased as follows:
What is the Lorentz covariant wave equation associated with a complex scalar field $\Psi(x)$ 
that changes after a Lorentz transformation according to the following rule: 
$\Psi(x) \rightarrow e^{\frac{i}{\hbar}f(x)}\Psi(x)$?
Here $f(x)$ is an arbitrary real spacetime dependent phase that is chosen 
to guarantee the Lorentz covariance of the wave equation.

The motivation underlying the above question is related to the fact that almost all observables in quantum field theory
are bilinear functions of the fields. Thus, a scalar quantum field theory built from the start by
assuming that the bilinear functions rather than the fields themselves are invariant should, in principle, be consistent with all known experimental facts.

The first step towards answering the above questions is a critical examination of the physical
meaning of the non-relativistic phase $\theta(x)$ above, which allows us
to naturally infer its relativistic extension. With the relativistic 
transformation rule for
$\Psi(x)$ at hand, we can search for the 
wave equation whose wave function transforms according to it and which is 
covariant under a Lorentz boost.
By adding a few extra assumptions, such that we should recover the Schr\"odinger 
wave equation at the non-relativistic limit, we show that 
the relativistic wave equation we obtain is unique. We also show 
its connection to the
Klein-Gordon equation and we apply it to the description of a variety 
of physical problems at the first quantization level.

We then move to the construction of the classical and quantum field theories related to
this relativistic wave equation. As we show in the following pages, 
it is possible to build consistent relativistic classical and quantum field theories 
if we employ the more general definition of a complex scalar field as outlined above.  
We also determine the scenarios under which the  
complex scalar field theories here presented are equivalent to the  
Klein-Gordon theory. For Lorentz  
invariant self-interactions and for electromagnetic interactions, 
the predictions of the present theory and those derived from the Klein-Gordon theory
are the same.

A few surprises appeared during our journey to the classical and,
in particular, to the quantum field theories here presented. First, 
as a logical consequence of those theories we noticed that we can
ascribe ``negative'' masses to antiparticles, while still
keeping their energies positive.
We then proved that this intriguing result, naturally emerging from the present theory, 
does not lead to any conflict with present day experimental facts. 
We also observed that particles and antiparticles with the same wave number 
no longer had the same energies and that we could make them have different momenta as well 
by properly tuning the free parameters of the field theories here developed. 
Despite all these peculiarities, and as we already noted above, 
at the level of electromagnetic interactions the 
present theories were shown to be equivalent to the Klein-Gordon one.
Moreover, we  showed that  the ``negative'' masses for antiparticles lead to interesting experimental 
predictions if we add Lorentz invariance-breaking interactions to the theory.
In this scenario, particles and antiparticles do not behave symmetrically and the process of baryogenesis emerges
naturally (antiparticles are more unstable). This surprising and interesting result could not have been 
anticipated without fully developing the quantum field theory associated with the Lorentz covariant 
Schr\"odinger equation given in the first part of this work. 
We also noted that for gravitational interactions,
the present theory can be tuned such that 
particles and antiparticles either attract or repel each other.

A few other subjects are addressed in this work, all connected in one way 
or another to the wave equations here derived. 
These results, together with the major ones highlighted 
above, are for ease of access systematically put together below before we start the more
technical part of this work. 

\section{Outline of this work}

This work can be divided into three main parts. The first one, comprising 
sections \ref{secLCSE} to \ref{PredictionsLCSE}, 
develops the main idea of this manuscript in the 
framework of first quantization. Being mo\-re specific, the following results
are achieved in the first part:

\begin{enumerate}
\item In section \ref{secLCSE} we derive a 
wave equation, which we dub Lo\-ren\-tz covariant Schr\"odinger equation,
that is covariant under proper Lorentz transformations
and whose wave function transforms under a Lorentz boost according to eq. (\ref{transformationL}),
\begin{displaymath}
\Psi(\mathbf{r},t) \rightarrow e^{\frac{i}{\hbar}\theta(\mathbf{r},t)} \Psi(\mathbf{r},t). 
\end{displaymath}
This is the relativistic extension of the transformation law (\ref{transformation})
associated with the non-relativistic
Schr\"odinger equation.
\item In section \ref{basicLCSE} we obtain by elementary methods the 
four-current density associated with the Lorentz covariant 
Sc\-hr\"odinger equation. We also show how to write it in 
a manifestly covariant way and we show its relation to the 
Klein-Gordon equation.
\item In section \ref{PredictionsLCSE} we apply the Lorentz covariant 
Schr\"o\-dinger equation to describe a multitude of physical systems. 
Possible applications to condensed matter systems are also briefly discussed.
The main points of this section are:
(a) We investigate the free particle solutions (external potential $V=0$)
of the Lorentz covariant Schr\"o\-dinger equation, 
which already hint 
towards the possibility of the existence of antiparticles possessing negative masses 
and positive energies.
(b) We then study the case of a non null constant potential ($V\neq 0$), 
where the dependence of the particle's mass on its value is highlighted.  
(c) For a constant potential, we also show that the Lorentz covariant 
Schr\"odinger equation can be seen as formally equivalent to the complex variable extension
of the telegraph equation. This implies that for certain values of $V$ we can have a  distortionless and dispersionless wave packet evolution.
(d) The bound state problem associated with a Coulomb-like potential $V$ is presented and solved.  
(e) We apply the minimal coupling prescription to the Lorentz covariant 
Schr\"odinger equation in order to include electromagnetic interactions in a gauge
invariant way. Of particular interest is the exact solution to the bound state
problem where we have the simultaneous action of static electromagnetic and gravitational fields between two particles.  
The electromagnetic interaction is modeled via the minimal coupling 
prescription while the gravitational interaction enters via the external potential
$V$. 
(f) We also perturbatively study the scattering between two scalar particles of different masses and charges when both the electromagnetic and gravitational interactions are simultaneously taken into account.
\end{enumerate}

The second part of this work, contained in sections \ref{cft} and \ref{qft}, and in 
the appendixes \ref{apA} and \ref{apC}, 
develops the classical and quantum field theories associated with the
Lorentz covariant Schr\"odinger equation. Specifically:

\begin{enumerate}
\item In section \ref{cft} we present the Lagrangian formulation of the 
Lorentz covariant Schr\"odin\-ger equation and we also investigate 
the continuous and discrete symmetries related to it. 
Of particular notice is the need to
modify the charge conjugation operation such that the mass changes 
its sign when we implement this symmetry operation. 
Only in this way can we properly exchange the roles
of particles with antiparticles and, at the same time, guarantee the validity of the CPT theorem. 
\item In section \ref{qft} we formulate the quantum field theory associated with 
the Lorentz covariant Schr\"odinger Lagrangian. We
show that it is possible to build a canonically second quantized theory out of the 
Lorentz covariant Schr\"odinger fields that
is logically consistent. In the framework of second quantization, the necessity 
to identify particles and antiparticles
with masses having different signs is further clarified. It is shown 
that particles and antiparticles have different but positive-definite 
expressions describing their relativistic energies. 
We then prove the e\-qui\-valence between the Lorentz covariant Schr\"odinger theory and
the Klein-Gordon theory when we add relativistic invariant self-inte\-racting terms to the Lagrangian density.
We also show the equivalence between both theories when electromagnetic interactions
are present. The details of those proofs are given in the appendix \ref{apC}. 
However, if we include interactions that do not respect Lorentz invariance, 
we prove by giving explicit examples that both theories lead to different predictions. 
We discuss a possible connection between those Lorentz invariance-breaking interactions and the asymmetry of matter and antimatter seen in the present day universe. 
\item In the appendix \ref{apA} we explore the main features associated with a non-canonical quantization of the Lorentz covariant Schr\"odinger fields,
whose origin is tra\-ced back to requiring the relativistic energies associated with particles and antiparticles to be the same. 
This leads to the violation of the microcausality condition and, interestingly, to the emergence of an instantaneous gra\-vi\-ta\-ti\-o\-nal-like interaction between the scalar 
particles of the theory. 
We also speculate about the possibility of arriving at
a consistent quantum field theory of gravitation 
if we work with quantum field theories 
whose main assumption is the
existence of particles and antiparticles having both positive energies but positive and negative masses.
\end{enumerate}

In section \ref{glcse}, the third and last part of this work, we 
generalize the Lorentz covariant Schr\"odinger equation, 
presenting a new Lorentz covariant wave equation where the
four spacetime coordinates and its derivatives appear on an equal footing.
For the free particle case we have
\begin{displaymath}
\partial_\mu\partial^\mu\Psi - 2i\kappa^\mu\partial_\mu\Psi 
= 0.
\end{displaymath}
Note that $\kappa^\mu$ is not a four-vector\footnote{The four constants
$\kappa^0,\kappa^1,\kappa^2,\kappa^3$
are invariants under proper Lorentz transformations, being the equivalent in the 
present theory of
the mass ``m'' in the Klein-Gordon theory. In other words, 
it is a postulate of the present theory that $\kappa^0,\kappa^1,\kappa^2,\kappa^3$ 
are invariants under
proper Lorentz transformations.}, it is a shorthand notation for the four
constants $\kappa^0,\ldots, \kappa^3$. 
The above equation is Lorentz covariant
if we use the more general transformation law for $\Psi$ as given below.
This should
be contrasted with the original Lorentz covariant Schr\"odinger equation
developed in the first part of this work,
\begin{displaymath}
\partial_\mu\partial^\mu \Psi 
- i\frac{2mc}{\hbar}\partial_0 \Psi 
=0,
\end{displaymath}
where we have a first order time derivative of the wave function 
and no first order space derivatives. To arrive at the generalized 
Lorentz covariant Schr\"odinger equation we assume that under a proper Lorentz transformation (boosts or spatial rotations) the wave 
function changes 
according to the following prescription,
\begin{displaymath}
\Psi(\mathbf{r},t) \rightarrow e^{\frac{i}{\hbar}f(\mathbf{r},t)} \Psi(\mathbf{r},t),
\end{displaymath}
with $f(\mathbf{r},t)$ being uniquely determined by the requirement of the covariance
of the wave equation under spatial rotations and boosts. In other words, 
in addition to assuming that $f(\mathbf{r},t)\neq 0$ for Lorentz boosts, as we did
in the first part of this manuscript, we now 
assume that $f(\mathbf{r},t)\neq 0$ for spatial rotations as well.

We also show in section \ref{glcse} the conditions under which  
the generalized Lorentz covariant Schr\"odinger Lagrangian 
leads to the same predictions of the Klein-Gordon theory.
It turns out that for Lorentz invariant self-interactions and electromagnetic 
interactions, we can make the two theories agree by properly
adjusting the values of the constant coefficients ($\kappa^\mu$)
appearing in the generalized Lagrangian density. 
Interestingly, and contrary to the original Lorentz covariant Schr\"odinger Lagrangian,
there still remains a free parameter at our disposal after enforcing the equivalence of the generalized theory and the Klein-Gordon one. 
Finally, we show that  
the inclusion of interaction terms in the Lagrangian 
that are not Lorentz invariant 
destroys the equivalence
between both theories.

\section{Galilean invariance and Lorentz covariance}
\label{secLCSE}

Our main goal in this section is to derive a 
wave equation that is Lorentz covariant if its solution, the wave function, 
transforms similarly to the way the solution to the 
non-rela\-ti\-vis\-tic Schr\"odinger equation transforms 
under a Ga\-li\-le\-an boost. To achieve that, we first review how and why 
the solution to the Schr\"odinger equation transforms according to eq.~(\ref{transformation}). We then propose a natural 
relativistic generalization of this transformation rule and  
derive the most general linear wave equation that is
covariant under proper Lorentz transformations if its wave function transforms
according to the relativistic extension of eq.~(\ref{transformation}).

\subsection{Galilean invariance of the Schr\"odinger equation}

First off we need to clearly explain what one usually means by stating that
the Schr\"odinger equation is invariant under a Galilean transformation \cite{bal98}. 

Let $S$ and $S'$ be two inertial reference frames whose Cartesian coordinate systems are 
connected by the following Ga\-li\-le\-an boost,  
\begin{eqnarray}
\mathbf{r} &=& \mathbf{r'} + \mathbf{v}t', \label{g1} \\
t&=& t' \label{g2}.
\end{eqnarray}
Here $\mathbf{r}=(x,y,z)$ and $\mathbf{r'}=(x',y',z')$ are the Cartesian coordinates of systems $S$ and $S'$, respectively,
while $t$ and $t'$ are the respective time coordinates measured in those inertial frames. 
The constant vector $\mathbf{v}$ gives
the relative velocity of $S'$ with respect to $S$.

In reference frame $S$ the Schr\"odinger wave equation for a particle of mass $m$ subjected to the  
potential $V(\mathbf{r},t)$ is
\begin{equation}
i\hbar\frac{\partial \Psi(\mathbf{r},t)}{\partial t} = -\frac{\hbar^2}{2m}\nabla^2\Psi(\mathbf{r},t)+
V(\mathbf{r},t)\Psi(\mathbf{r},t) \label{se}, 
\end{equation}
where $\Psi(\mathbf{r},t)$ is the particle wave function and 
$\nabla^2 = \frac{\partial^2}{\partial x^2}+\frac{\partial^2}{\partial y^2}+\frac{\partial^2}{\partial z^2}$ 
is the Laplacian written in Cartesian coordinates.  

Now, using eqs.~(\ref{g1}) and (\ref{g2}) together with the chain rule we get 
\begin{eqnarray}
\nabla^2 &=& \nabla'^2, \label{nab} \\
\frac{\partial }{\partial t} &=& \frac{\partial }{\partial t'} - \mathbf{v} \cdot \nabla', \label{delt}
\end{eqnarray}
where the dot means the scalar product and $\nabla'$ is the gradient operator in system $S'$,
namely, $\nabla' = (\frac{\partial}{\partial x'},\frac{\partial}{\partial y'},\frac{\partial}{\partial z'})$.
Substituting eqs.~(\ref{nab}) and (\ref{delt}) into (\ref{se}) and assuming that the wave function in $S$
is connected to the wave function in $S'$ according to the following prescription,
\begin{equation}
\Psi(\mathbf{r},t) = e^{\frac{i}{\hbar}\theta(\mathbf{r'},t')} \Psi'(\mathbf{r'},t'),
\label{transformation}
\end{equation}
we obtain the invariance of the Schr\"odinger equation,
\begin{equation}
i\hbar\frac{\partial \Psi'(\mathbf{r'},t')}{\partial t'} = -\frac{\hbar^2}{2m}\nabla'^2\Psi'(\mathbf{r'},t')+
V'(\mathbf{r'},t')\Psi'(\mathbf{r'},t') \label{se2}, 
\end{equation}
if, and only if, 
\begin{equation}
\theta(\mathbf{r'},t') = \frac{mv^2}{2}t' + m \mathbf{v\cdot\mathbf r'} + cte,
\label{thetaG}
\end{equation}
where $cte$ is a real constant that is usually assumed to be zero and $v^2=\mathbf{v\cdot v}$. 
We also assume that
the potential transforms as $V(\mathbf{r},t)=V'(\mathbf{r'},t')$. The Coulomb potential,
for instance, is the paradigmatic example of a potential that transforms according
to the above prescription under a Galilean boost.

Note that strictly speaking we can only make the Schr\"o\-din\-ger equation invariant under a Galilean boost
by imposing that the wave function ``fails'' 
to be invariant thereunder. In other words, the 
prescription 
$\Psi(\mathbf{r},t) = \Psi'(\mathbf{r'},t')$ connecting the wave functions
in coordinate systems $S$ and $S'$ does not suffice to make the Schr\"odinger equation 
invariant under a Ga\-li\-le\-an transformation. 
Nevertheless, the probability density  $|\Psi(\mathbf{r},t)|^2$ is invariant under the transformation given in
eq.~(\ref{transformation}) and thus invariant under a Galilean boost.

We can also understand eq.~(\ref{transformation}) as a way to obtain the solution to the 
Schr\"odinger equation in reference frame $S'$ if the solution in $S$ is already known,
\begin{eqnarray}
\Psi'(\mathbf{r'},t') 
=e^{-\frac{i}{\hbar}\theta(\mathbf{r'},t')}\Psi(\mathbf{r},t) 
= e^{-\frac{i}{\hbar}\theta(\mathbf{r'},t')}\Psi(\mathbf{r'+v} t',t').
\label{transformation2}
\end{eqnarray}
This prescription relating the solution in $S'$ to the one in $S$ is crucial for the consistency 
of the description of the same physical system in different Galilean 
inertial frames. The following 
simple example 
illustrates this point and further clarifies how to apply eq.~(\ref{transformation2}) in order to get
$\Psi'(\mathbf{r'},t')$ from $\Psi(\mathbf{r},t)$.

\subsubsection{Example: The plane wave}

Let us assume that in $S$ we have a free particle with momentum $p$. For simplicity we 
set $V(\mathbf{r},t)=0$ and assume, without loss of generality, that we are dealing with a one dimensional 
problem.  Up to an overall global phase,
the solution to the Schr\"odinger equation (\ref{se})
describing such a particle is the plane wave
\begin{equation}
\Psi(x,t) = e^{i(kx-\omega t)}=e^{\frac{i}{\hbar}\left(px-\frac{p^2}{2m} t\right)},
\label{pwave}
\end{equation}
where $p^2/2m$ is the particle's energy. 

Let us now solve the same problem in the
reference frame $S'$, which we assume is moving away from $S$ with constant velocity 
$\mathbf{v}=(v,0,0)$, with $v>0$. This means that $S'$ is moving along the positive direction of
the x-axis. In this frame, the momentum of the particle is no longer $p$ but $p-mv$ and the solution to
the Schr\"odinger equation (\ref{se2}) is 
\begin{equation}
\Psi'(x',t') = e^{i(k'x'-\omega' t')}=e^{\frac{i}{\hbar}\left((p-mv)x'-\frac{(p-mv)^2}{2m} t'\right)},
\label{pwave2}
\end{equation}
where now $(p-mv)^2/2m$ is the particle's energy. 

It is not difficult to see that the naive prescription 
$\Psi(x,t)$ $\rightarrow$ $\Psi'(x',t')$ connecting the wave functions from
these two reference frames will not transform the right hand side 
of eq.~(\ref{pwave}) into the
right hand side of (\ref{pwave2}). 
However, noting that eq.~(\ref{thetaG}) for this specific
problem is
\begin{equation}
\theta(x',t') = \frac{mv^2}{2}t' + m v x',
\end{equation}
where we have set $cte=0$, we get after eq.~(\ref{transformation2})
\begin{eqnarray}
\Psi'(x',t') &=&  e^{-\frac{i}{\hbar}\theta(x',t')}\Psi(x,t) 
= e^{-\frac{i}{\hbar}\theta(x',t')}\Psi(x'+ v t',t') \nonumber \\
&=& 
e^{-\frac{i}{\hbar}(\frac{mv^2}{2}t' + m v x')}
e^{\frac{i}{\hbar}\left(p(x'+vt')-\frac{p^2}{2m} t'\right)} 
=e^{\frac{i}{\hbar}\left((p-mv)x'-\frac{(p-mv)^2}{2m} t'\right)},
\end{eqnarray}
which is exactly eq.~(\ref{pwave2}). This simple example clearly illustrates why 
prescription (\ref{transformation}) is mandatory to properly connect the solutions
to the Schr\"odinger equation obtained in two different inertial reference 
frames.\footnote{We employed the term invariance to denote that the Schr\"odinger equation 
does not change its form under a Galilean boost because this is 
the usual practice in non-relativistic treatises on quantum mechanics (see ref. \cite{bal98},
for example).
From now on we reserve the term invariant to quantities such as the rest mass
of a particle, the speed of light, the electric charge, or any scalar quantity that
has the same numerical value in any inertial reference frame. Equations that look
the same in all inertial frames after we apply a given transformation will now be called covariant.
}

\subsection{Lorentz covariant Schr\"odinger equation}

Looking carefully at eq.~(\ref{thetaG}), we can understand the terms multiplying 
$t'$ and $\mathbf{r'}$, namely, $mv^2/2$ and $m\mathbf{v}$, as the contribution to the energy and momentum 
of the particle of mass $m$ that is related to the fact that reference frame $S'$ is moving
from $S$ with constant velocity $\mathbf{v}$. For example, assume the particle is at rest in $S'$.
From the point of view of $S$ it is moving away with constant velocity $\mathbf{v}$ and thus with 
kinetic energy $mv^2/2$ and momentum $m\mathbf{v}$. 

The above interpretation is the key that opens the door to the relativistic version of 
eqs.~(\ref{transformation}) and (\ref{thetaG}), which will ultimately allow us
to obtain a natural modification to the Schr\"odinger equation that makes it Lorentz covariant.

Calling $\gamma = 1/\sqrt{1-v^2/c^2}$ the Lorentz factor, $m$ the particle's rest mass, and
$c$ the speed of light in vacuum, the relativistic kinetic energy and momentum for a particle of mass $m$
are, respectively, $(\gamma -1)mc^2$ and $\gamma m\mathbf{v}$, where $\mathbf{v}$ is the particle's velocity.
With that in mind, we postulate that under a Lorentz transformation connecting
two different inertial frames $S$ and $S'$ with relative velocity $\mathbf{v}$, 
the wave functions describing a particle of mass $m$ in $S$ and $S'$
are connected by the following relation,
\begin{equation}
\Psi(\mathbf{r},t) = e^{\frac{i}{\hbar}\theta(\mathbf{r'},t')} \Psi'(\mathbf{r'},t'),
\label{transformationL}
\end{equation}
with
\begin{equation}
\theta(\mathbf{r'},t') = (\gamma - 1)mc^2 t' + \gamma m \mathbf{v\cdot\mathbf r'} + cte.
\label{thetaL}
\end{equation}
Note that we can set the real constant $cte$ to zero without losing in generality. 
Equation (\ref{transformationL}) is formally identical to eq.~(\ref{transformation})
and when $c\rightarrow \infty$ eq.~(\ref{thetaL}) tends to (\ref{thetaG}) since
$\lim_{c\rightarrow \infty}$ $\gamma$ $= 1$ and 
$\lim_{c\rightarrow \infty} [(\gamma-1)mc^2] = mv^2/2$.

With the wave function's transformation law given by 
eqs.~(\ref{transformationL}) and (\ref{thetaL}), we are naturally led to ask the
following question: 
\begin{quote}
What is the form of the wave equation that is covariant under a Lorentz boost  
and whose wave function transforms under such a boost according to
eqs.~(\ref{transformationL}) and (\ref{thetaL})?
\end{quote}

We can answer that question unambiguously, i.e., 
we can get one and only one wave equation, if we add the following
extra very natural assumptions in our search for the Lorentz
covariant Schr\"odinger equation:
\begin{itemize}
\item It tends to the non-relativistic Schr\"odinger equation
when $c \rightarrow \infty$.
\item It is isotropic, namely, covariant under three-dimensional spatial rotations in the same 
sense as the non-relativistic Schr\"odinger equation is.
\item It is a homogeneous linear partial differential equation of order not greater than two and 
with constant coefficients multiplying the derivatives.  
\end{itemize}

The first assumption above guarantees that we recover the non-relativistic quantum mechanics
when the physical system studied moves with small velocities. The second assumption assumes that 
there is no privileged orientation in space, a symmetry that is also respected by the non-rela\-ti\-vis\-tic 
Schr\"odinger equation.
The third and last extra assumption keeps the superposition principle valid in the relativistic
domain and restricts our search to the simplest wave equations, namely, those that are homogeneous and 
that have at most second order derivatives and constant coefficients multiplying those derivatives.  
Note that we allow the coefficient multiplying the wave function $\Psi(\mathbf{r},t)$ 
to be non-constant since we want to recover the term $V(\mathbf{r},t)\Psi(\mathbf{r},t)$
of the non-relativistic Schr\"odinger equation.  
Using this terminology, the non-relativistic Schr\"odinger equation
is a homogeneous linear partial differential equation of order two (in the spatial variables) with constant
coefficients multiplying the derivatives.

Before we start our search for the Lorentz covariant Sch\-r\"odinger equation, we need first to set up some
conventions and notation. The contravariant four-vector $x^\mu$ is defined 
such that $(x^0,x^1,x^2,x^3)=(ct,x,y,z)$. Working with the metric $g_{\mu\nu}=\mbox{diag}\{1,-1,-1,-1\}$, where 
$g_{\mu\nu}$ is a diagonal $4\times 4$ matrix, the covariant four-vector is $x_\mu=g_{\mu\nu}x^\nu$ $=$ $(ct,-x,$ $-y,-z)$.
We assume the Einstein summation convention with Greek indexes running from $0$ to $3$ 
and Latin ones running from $1$ to $3$. 
The scalar product between two four-vectors is defined as 
$x^\mu y_\mu$, which makes it invariant under Lorentz boosts and spatial rotations. We also define 
the covariant four-gradient as 
$
\partial_\mu = 
\left(\frac{\partial}{\partial x^0},\frac{\partial}{\partial x^1},\frac{\partial}{\partial x^2},\frac{\partial}{\partial x^3}
\right)
$
and the contravariant four-gradient by $\partial^\mu=g^{\mu\nu}\partial_\nu$, where $g^{\mu\nu}=g_{\mu\nu}$
since we are dealing with Minkowski spacetime. 

In the four-vector notation just defined, 
the most general
homogeneous linear partial differential equation in the variables $x^\mu$, of order less or equal to two, and
with constant coefficients multiplying the derivatives is
\begin{equation}
a^{\mu\nu}\partial_\mu\partial_\nu\Psi(x) + b^{\mu}\partial_\mu\Psi(x) + f(x)\Psi(x)=0.
\label{GeneralEq}
\end{equation}
Here $\Psi(x)=\Psi(\mathbf{r},t)$, $a^{\mu\nu}$
and $b^{\mu}$ are constants (relativistic invariants), 
and $f(x)=f(\mathbf{r},t)$, i.e., 
it may depend on time and on the spatial coordinates. 
The function $f(x)$ is proportional to 
the relativistic version of the potential that affects a particle described by
the non-relativistic
Schr\"odinger equation. 
Later we will fix its value using the first extra assumption above.

By invoking the isotropic condition for the free particle case, i.e., the wave equation should be 
covariant under any spatial orthogonal 
transformation belonging to the group $SO(3)$ when $f(x)=0$, 
we can show that several of the $a^{\mu\nu}$ and $b^{\mu}$ constants are zero.
To see this note that the free particle Schr\"odinger equation is covariant under spatial rotations if 
\begin{equation}
\Psi(\mathbf{r},t) = e^{i\alpha} \Psi'(\mathbf{r'},t),
\label{3Drotations}
\end{equation}
where $\alpha$ is a constant (usually set to zero) and $\mathbf{r'}$ is connected to $\mathbf{r}$
by an orthogonal transformation belonging to the $SO(3)$ group, namely, $\mathbf{r'}=M\mathbf{r}$, with $M \in SO(3)$.

The covariance of the free particle Schr\"odinger equation 
under transformations belonging to the group $SO(3)$ can be easily proved using eq.~(\ref{3Drotations}) and 
by noting that the Laplacian does not change under such transformations.  
We are thus led to assume that the wave functions that are solutions to the Lorentz covariant Schr\"odinger equation 
also obey eq.~(\ref{3Drotations}) when we
spatially rotate the system of coordinates. 
This requirement is our second extra assumption pointed out above.

Let us start analyzing how the term $b^{\mu}\partial_\mu\Psi(x)$ of eq.~(\ref{GeneralEq}) changes under a spatial
rotation. Writing this term explicitly we have
\begin{equation}
b^0\partial_0\Psi(\mathbf{r},t) + b^1\partial_1\Psi(\mathbf{r},t) +
b^2\partial_2\Psi(\mathbf{r},t) + b^3\partial_3\Psi(\mathbf{r},t).
\label{bterm}
\end{equation}
If we rotate the system of coordinates counterclockwise by $\pi$ radians about the $x^3$ axis, it is not difficult
to see that new components of the vector $x^\mu$ becomes
$x^{\mu'}=(x^{0'},x^{1'},x^{2'},x^{3'})$ $=$ $(x^0,-x^1,-x^2,x^3)$. Applying the chain rule this implies that
\begin{equation}
\partial_{\mu'}=(\partial_{0'},\partial_{1'},\partial_{2'},\partial_{3'})=(\partial_{0},-\partial_{1},-\partial_{2},\partial_{3}).
\label{delprime}
\end{equation}
Substituting eqs.~(\ref{3Drotations}) and (\ref{delprime}) into (\ref{bterm}),
and neglecting the irrelevant constant phase, we obtain 
\begin{equation}
b^0\partial_{0'}\Psi'(\mathbf{r'},t) - b^1\partial_{1'}\Psi'(\mathbf{r'},t) -
b^2\partial_{2'}\Psi'(\mathbf{r'},t) + b^3\partial_{3'}\Psi'(\mathbf{r'},t).
\label{bterm2}
\end{equation}
We can only make eqs.~(\ref{bterm2}) and (\ref{bterm}) look the same, i.e., guarantee
covariance under this particular rotation, if $b^1=-b^1$ and $b^2=-b^2$. This implies that
$b^1=b^2=0$. A similar analysis, where we fix, for instance, the $x^1$ axis and rotate 
the other two spatial axis leads to $b^3=0$. This reduces the term given by
eq.~(\ref{bterm}) to
\begin{equation}
b^0\partial_0\Psi(\mathbf{r},t).
\label{bterm3}
\end{equation}

Turning our attention to the term $a^{\mu\nu}\partial_\mu\partial_\nu\Psi(x)$ 
of eq. (\ref{GeneralEq}), we note that the same reasoning that led to $b^{j}=0$
leads to $a^{0j}+a^{j0}=0$, for $j=1,2,3$. This must be the case since the derivatives 
accompanying the constants $a^{0j}$ or $a^{j0}$ are of the form $\partial_0\partial_j$
and $\partial_j\partial_0$, with only one spatial coordinate. 
Assuming that we can change the order of the derivatives, and this can 
always be done for well behaved wave functions $\Psi(\mathbf{r},t)$, we get 
$(a^{0j}+a^{j0})\partial_0\partial_j\Psi(\mathbf{r},t)$ for that piece of the wave equation
where one and only one of the superscript index is zero. Using the same arguments that led
to eq.~(\ref{delprime}), we can always rotate the axis to get $\partial_j'=-\partial_j$ for a given
$j$. For example, if we rotate the system of coordinates counterclockwise by $\pi$ radians about the $x^3$ axis,
we get that $(a^{01}+a^{10})\partial_0\partial_1\Psi(\mathbf{r},t)$ changes to
$-(a^{01}+a^{10})\partial_0'\partial_1'\Psi'(\mathbf{r'},t)$ (and similarly for $j=2$).
The covariance can only be guaranteed if $a^{01}+a^{10}=0$ and $a^{02}+a^{20}=0$. By rotating the 
system about the $x^1$ axis by $\pi$ radians, we see that $a^{03}+a^{30}=0$ in order to preserve
the covariance of the wave equation under this particular rotation.

To deal with the remaining purely spatial terms $a^{ij}\partial_i\partial_j$ $\Psi(x)$,  
we rotate the system of coordinates by $\pi/2$ instead of $\pi$ radians. For instance, 
if we rotate the system of coordinates counterclockwise by $\pi/2$ radians about the $x^3$ axis, 
the components of the vector $x^\mu$ in the rotated frame is
$x^{\mu'}=(x^{0'},x^{1'},$ $x^{2'},x^{3'})=(x^0,x^2,-x^1,x^3)$. And if we now apply the 
chain rule we get
\begin{equation}
\partial_{\mu'}=(\partial_{0'},\partial_{1'},\partial_{2'},\partial_{3'})=(\partial_{0},\partial_{2},-\partial_{1},\partial_{3}).
\label{partialprime}
\end{equation}
Substituting eq.~(\ref{partialprime}) into 
\begin{equation}
(a^{12}+a^{21})\partial_1\partial_2\Psi(\mathbf{r},t)  
\label{a12}
\end{equation}
and using eq.~(\ref{3Drotations}), we obtain, up to an irrelevant global phase,
eq.~(\ref{a12}) in the rotated frame,
\begin{equation}
-(a^{12}+a^{21})\partial_2'\partial_1'\Psi'(\mathbf{r'},t).  
\label{a12r}
\end{equation}
Since the order of the derivatives can be exchanged freely for physical wave functions,
we can only have covariance under this rotation if $a^{12}+a^{21}=0$. Repeating the previous
analysis where we rotate the system either about $x^2$ or $x^1$ by $\pi/2$ radians, we
get that we should have $a^{13}+a^{31}=0$ and  $a^{23}+a^{32}=0$ to keep the rotational
covariance of the wave equation (\ref{GeneralEq}).

By a similar argument we can show that $a^{11}=a^{22}=a^{33}$. For example, the 
counterclockwise $\pi/2$ rotation about the $x^3$ axis leads to eq.~(\ref{partialprime}).
This implies, together with eq.~(\ref{3Drotations}), that the quantity
\begin{equation}
a^{11}\partial_1\partial_1\Psi(\mathbf{r},t)+a^{22}\partial_2\partial_2\Psi(\mathbf{r},t)
\end{equation}
changes to 
\begin{equation}
a^{11}\partial_2'\partial_2'\Psi'(\mathbf{r'},t)+a^{22}\partial_1'\partial_1'\Psi(\mathbf{r'},t)
\end{equation}
after this rotation.
By demanding the covariance under this rotation we get that $a^{11}=a^{22}$.
Repeating the previous argument for a rotation of $\pi/2$ about the $x^1$ axis gives
$a^{22}=a^{33}$.

Putting all those results together, eq.~(\ref{GeneralEq}) becomes
\begin{equation}
a^{00}\partial_0\partial_0\Psi(x) - a^{11}\partial_j\partial^j\Psi(x) + b^{0}\partial_0\Psi(x) + f(x)\Psi(x)=0,
\label{GeneralEq2}
\end{equation}
where we have used that $\partial_j=-\partial^j$. Note that this equation is covariant under all
rotations since the only remaining spatial derivatives, namely $-\partial_j\partial^j = \nabla^2$, is  the 
Laplacian, which does not change its form under rotations.

So far we have not investigated what restrictions a Lo\-ren\-tz boost impose 
on the remaining constants shown in eq. (\ref{GeneralEq2}). In order to do that,
it is convenient to rewrite eq.~(\ref{GeneralEq2}) as follows,
\begin{equation}
A\partial^2_0\Psi(x) - B\partial_j\partial^j\Psi(x) + C\partial_0\Psi(x) + f(x)\Psi(x)=0.
\label{GeneralEq3}
\end{equation}
In this notation, our task is to determine what are the values and relations among $A$, $B$, and $C$ that arise after we impose the covariance of eq.~(\ref{GeneralEq3}) and
assume that $\Psi(x)$ transforms
according to the prescription (\ref{transformationL}) under a Lorentz boost. Note that 
we are assuming 
that $f(x)$ is a relativistic scalar, namely, it transforms to $f'(x')$ under a Lorentz boost.

Since we already proved the covariance of eq.~(\ref{GeneralEq3}) under three-dimensional spatial
rotations, we do not lose in generality by assuming that the velocity $v$ of the inertial reference frame
$S'$ with respect to $S$ is directed along the $x^1$ axis. With such a choice for $\mathbf{v}$ the 
variables $x^\mu$ are connected to $x^{\mu'}$ by the following Lorentz transformation,
\begin{eqnarray}
x^0 &=& \gamma(x^{0'} + \beta x^{1'}), \label{l0} \\
x^1 &=& \gamma(x^{1'} + \beta x^{0'}), \label{l1} \\
x^2 &=& x^{2'}, \label{l2} \\
x^3 &=& x^{3'}, \label{l3}
\end{eqnarray}
where 
\begin{eqnarray*}
\beta = \frac{v}{c} & \mbox{and} & \gamma =\frac{1}{\sqrt{1-\beta^2}}. 
\end{eqnarray*}
By applying the chain rule we get
\begin{eqnarray}
\partial_0 &=& \gamma(\partial_{0'} - \beta \partial_{1'}), \label{d0} \\
\partial_1 &=& \gamma(\partial_{1'} - \beta \partial_{0'}), \label{d1} \\
\partial_2 &=& \partial_{2'}, \label{d2} \\
\partial_3 &=& \partial_{3'}. \label{d3}
\end{eqnarray}

Inserting eqs.~(\ref{d0})-(\ref{d3}) into (\ref{GeneralEq3}), using
eqs.~(\ref{transformationL}) and (\ref{thetaL}), and carrying out the derivatives,
the wave equation 
(\ref{GeneralEq3}) can be written up to an overall phase as follows,
\begin{eqnarray}
A'&\hspace{-.1cm}\partial^2_{0'}\Psi'(x') + B'\partial^2_{1'}\Psi'(x') 
+ B \partial^2_{2'}\Psi'(x') + B \partial^2_{3'}\Psi'(x')   \nonumber \\ 
+ &\hspace{-1cm} C'\partial_{0'}\Psi'(x') + D'\partial_{0'}\partial_{1'}\Psi'(x') + E' \partial_{1'}\Psi'(x')
\nonumber \\  
+ &\hspace{-3.0cm}F'\Psi'(x') + f'(x')\Psi'(x') = 0,  
\label{GeneralEq4}
\end{eqnarray}
where
\begin{eqnarray}
A' &=&  \gamma^2(A+\beta^2 B), \label{al} \\
B' &=&  \gamma^2(\beta^2 A + B), \label{bl} \\
C' &=&  \gamma \left\{ \frac{i2mc}{\hbar}  \left[ (1-\gamma) A -\gamma \beta^2 B \right] + C \right\}, \label{cl} \\
D' &=&  - 2\gamma^2\beta (A + B), \label{dl} \\
E' &=&  - \gamma \beta \left\{ \frac{i2mc}{\hbar}  \left[ (1-\gamma) A -\gamma B \right] + C \right\}, \label{el} \\
F' &\!\!=\!\!&\!\!  \frac{imc}{\hbar}\!\!\left\{\! \frac{imc}{\hbar}\! \left[ (1\!-\!\gamma)^2 A +\gamma^2\beta^2 B \right] 
\!+\! (1\!-\!\gamma) C\! \right\}\!. \label{fl}
\end{eqnarray}

Equation (\ref{GeneralEq4}) is what wave equation (\ref{GeneralEq3}) becomes 
after a Lorentz boost for arbitrary values of the constants $A$, $B$, and $C$.
In order to make eqs.~(\ref{GeneralEq4}) and (\ref{GeneralEq3}) look the same, i.e.,
in order to have Lorentz covariance, we must have $D'=0$, $E'=0$, and $F'=0$.
The first condition, $D'=0$, implies $B=-A$, as can be seen by looking at eq.~(\ref{dl}).
Inserting this last relation into eq.~(\ref{el}) we get that $E'=0$ if 
$C = -i2mcA/\hbar$. With this value for $C$ and using that $B=-A$, we automatically
get $F'=0$. Also, using that $A=-B$ we get $C'=C=-i2mcA/\hbar$, $A'=A$, and $B'=-A=B$.
By gathering all these results together we can write eq.~(\ref{GeneralEq4}) as follows,
\begin{equation}
A \partial^2_{0'}\!\Psi'\!(x') - B \partial_{j'}\partial^{j'}\!\Psi'\!(x') 
+ C\partial_{0'}\!\Psi'\!(x') 
+ f'\!(x')\Psi'\!(x') = 0,  
\label{GeneralEq5}
\end{equation}
which explicitly shows the Lorentz covariance
of the wave equation (\ref{GeneralEq3}). If we now use that 
\begin{eqnarray}
B=-A  & \mbox{and} & C = -i\frac{2mc}{\hbar}A,
\end{eqnarray}
and that we must have $A\neq 0$, we can cast eq.~(\ref{GeneralEq3}) as follows,
\begin{equation}
 \partial_{\mu}\partial^{\mu}\Psi(x) 
-i\frac{2mc}{\hbar} \partial_{0}\Psi(x)  
+ \frac{f(x)}{A}\Psi(x) = 0.  
\label{GeneralEq5b}
\end{equation}

To fix the value of $A$, we need the first extra assumption, namely, we need to
impose that eq.~(\ref{GeneralEq3}), or equivalently (\ref{GeneralEq5b}), 
tends to the non-relativistic Schr\"odinger equation
when $c\rightarrow \infty$. This can be accomplished more easily by first rewriting 
eq.~(\ref{GeneralEq5b}) in the non-relativistic notation.  
Noting that the d'Alembertian 
$\partial_{\mu}\partial^{\mu} = \frac{1}{c^2}\frac{\partial^2}{\partial t^2} - \nabla^2$,
$\partial_0 = \frac{1}{c}\frac{\partial}{\partial t}$,
and $\Psi(x)=\Psi(\mathbf{r},t)$,
we can rewrite eq.~(\ref{GeneralEq5b}) as
\begin{equation}
\frac{1}{c^2}\frac{\partial^2\Psi(\mathbf{r},t)}{\partial t^2} \!-\! \nabla^2 \Psi(\mathbf{r},t) 
\!-\! i\frac{2m}{\hbar}\frac{\partial \Psi(\mathbf{r},t)}{\partial t} \!+\! \frac{f(\mathbf{r},\!t)}{A}\Psi(\mathbf{r},t)\!=\!0.
\label{GeneralEq6}
\end{equation}

If we take the non-relativistic limit of eq.~(\ref{GeneralEq6}), assuming that $\Psi(\mathbf{r},t)$
and its first and second order derivatives does not diverge, the 
limit $c\rightarrow \infty$ gives
\begin{equation}
- \nabla^2 \Psi(\mathbf{r},t) 
- i\frac{2m}{\hbar}\frac{\partial \Psi(\mathbf{r},t)}{\partial t} 
+ \lim_{c\rightarrow \infty}\left[\frac{f(\mathbf{r},t)}{A}\right]\Psi(\mathbf{r},t)=0.
\label{GeneralEq6b}
\end{equation}
On the other hand, the non-relativistic Schr\"odinger equation (\ref{se}) can be put in the
following form,
\begin{equation}
- \nabla^2\Psi(\mathbf{r},t) - i\frac{2m}{\hbar}\frac{\partial \Psi(\mathbf{r},t)}{\partial t} 
+\frac{2m}{\hbar^2}V(\mathbf{r},t)\Psi(\mathbf{r},t) = 0 \label{se3}. 
\end{equation}

Comparing eqs.~(\ref{GeneralEq6b}) and (\ref{se3}), we see that they are equal if
\begin{equation}
\lim_{c\rightarrow \infty}\left[\frac{f(\mathbf{r},t)}{A}\right] = \frac{2m}{\hbar^2}V(\mathbf{r},t).
\label{limit1}
\end{equation}
Since $A$ does not depend on $\mathbf{r}$ and $t$, we can without loss of generality 
set 
\begin{eqnarray}
\lim_{c\rightarrow \infty} f(\mathbf{r},t)&=&\frac{2m}{\hbar^2}V(\mathbf{r},t), \label{fc} 
\\
\lim_{c\rightarrow \infty} A &=& 1, \label{ac}
\end{eqnarray}
as the conditions upon $f(\mathbf{r},t)$ and $A$ that allow us to
get the non-relativistic Schr\"odinger equation as the exact non-relativistic 
limit of the Lorentz covariant Schr\"odinger equation.

We can go even further and prove that $A=1$, for whatever value of $c$, if we
note that eq.~(\ref{ac}) implies that
\begin{equation}
A = 1 + \sum_{j=1}^{N} \frac{A_j}{c^j},
\label{ag}
\end{equation}
where $N\geq 1$, $A_j$ is a constant, and $c^j$ is a positive power of the speed of light $c$.
In addition to that, eq.~(\ref{fc}) implies that $A$ must be dimensionless, since otherwise
$\frac{f(\mathbf{r},t)}{A}\Psi(\mathbf{r},t)$ will have a different dimension when compared to
the other three terms of eq.~(\ref{GeneralEq6}). Now, the only dimensional constants appearing
so far, and in particular in the free particle case, are the mass of the particle $m$, the speed of 
light $c$, and Planck's constant $\hbar$. This means that $A_j=\alpha_jm^{x_j}\hbar^{y_j}$,
where $\alpha_j$ is a pure number and $x_j, y_j$ are exponents to be determined such that $A_j/c^j$ 
becomes dimensionless. 

The dimensions of $m,c$, and $\hbar$ are  
$[m]=M$, $[c]=LT^{-1}$, and $[\hbar]=ML^2T^{-1}$, where $M$ means mass, $L$ length, and
$T$ time. 
Thus, $[A_j/c^j]$ $=$ $\alpha_j[m^{x_j}][\hbar^{y_j}][c^{-j}] = M^{x_j+y_j}L^{2y_j-j}$ 
$T^{-y_j+j}$.
$A_j/c^j$ is dimensionless if $x_j+y_j=0$, $2y_j-j=0$, and $-y_j+j=0$. The last two equations
give $j=2y_j$ and $j=y_j$, which is only possible if $j=y_j=0$. This result when inserted into the first
relation gives $x_j=0$. We see then that $A_j/c^j=\alpha_j$, a pure number, and that eq.~(\ref{ag}) 
reads
\begin{equation}
A = 1 + \sum_{j=1}^{N} \alpha_j.
\label{ag2}
\end{equation}
But since $\lim_{c\rightarrow \infty} A = 1$, we must have 
\begin{equation}
\lim_{c\rightarrow \infty} \sum_{j=1}^{N} \alpha_j = 0. 
\end{equation}
This implies that
\begin{equation}
\sum_{j=1}^{N} \alpha_j = 0 
\label{ag3}
\end{equation}
for any value of $c$ because $\alpha_j$, being a pure number, does not depend on $c$.
Equation (\ref{ag3}) together with (\ref{ag2}) lead to the desired result, namely,
$A=1$ for any value of $c$.

Using that $A=1$ we can write the Lorentz covariant
Schr\"odinger equation (\ref{GeneralEq6}) as
\begin{equation}
\frac{1}{c^2}\frac{\partial^2\Psi(\mathbf{r},t)}{\partial t^2} - \nabla^2 \Psi(\mathbf{r},t) 
- i\frac{2m}{\hbar}\frac{\partial \Psi(\mathbf{r},t)}{\partial t} + f(\mathbf{r},\!t)\Psi(\mathbf{r},t)\!=\!0,
\label{extra1}
\end{equation}
where the non-relativistic limit of $f(\mathbf{r},t)$ is given by eq.~(\ref{fc}).
Finally, if we abuse notation and 
make the convention that $V(\mathbf{r},t)$ represents both the non-relativistic
 potential as well as its relativistic extension, we can write the Lorentz covariant
Schr\"odinger equation (\ref{GeneralEq6}) as follows,
\begin{equation}
\boxed{
\frac{1}{c^2}\frac{\partial^2\Psi}{\partial t^2} - \nabla^2 \Psi 
- i\frac{2m}{\hbar}\frac{\partial \Psi}{\partial t} + \frac{2mV}{\hbar^2}\Psi=0,
}
\label{GeneralEq7}
\end{equation}
where we have written $\Psi$ and $V$ instead of $\Psi(\mathbf{r},t)$ and 
$V(\mathbf{r},t)$ to simplify notation. Equation
(\ref{GeneralEq7}) is the Lorentz covariant wave equation that we were
searching for and that satisfies all the assumptions laid out in the beginning of this
section.\footnote{We should remark that the results here reported can be readily adapted
to two independent real fields. This is true because we can map a single complex field 
to two real and independent scalar fields. See, for instance,
sections 2.2 and 3.2 of ref. \cite{man86} and section 4.2 of ref. \cite{gre95}.}

Note that all time derivatives above are associated with the ``geometrical time''
of the reference frame $S$, where all measurements and observables are made and defined
for a given experiment in $S$. 
Putting it simply, $t$ is the time an observer at the inertial
frame $S$ records looking at his or her
watch. This should be compared with the following free particle 
wave equation, derived in refs. \cite{stu42,nam50,hor73,fan78} using a whole set of different assumptions than those employed here,
\begin{equation}
\frac{1}{c^2}\frac{\partial^2\Psi}{\partial t^2} - \nabla^2 \Psi 
- i\frac{2m}{\hbar}\frac{\partial \Psi}{\partial \tau}=0.
\label{properT}
\end{equation}

In eq.~(\ref{properT}) $\tau$ is an invariant free parameter usually 
associated with the proper time of the system being studied. As such, it does not
change under a Lorentz boost and eq.~(\ref{properT}) is Lorentz covariant if
its wave function transforms according to the standard definition of a complex scalar function, i.e., 
$\Psi'(\mathbf{r'},t')=\Psi(\mathbf{r},t)$. In the context of 
the Lorentz covariant Schr\"odinger equation, however, $\tau$ is no longer a free parameter 
and we must
set $\tau=t$ in order to have a consistent Lorentz covariant wave equation whose
wave function transforms according to eq.~(\ref{transformationL}) 
under a Lorentz boost.

It is important to stress at this point that we have made no attempt to uniquely determine
the relativistic invariant $f(\mathbf{r},t)$, or equivalently, $V(\mathbf{r},t)$.
In this sense, wave equation (\ref{GeneralEq7}) is not strictly unique. 
However, for the free particle case, 
this point is not relevant and we do have a unique
wave equation stemming from the three assumptions given at the beginning of this
section. On the other hand, 
when we have an external field, we just need eq.~(\ref{fc}) to be satisfied 
in order to recover the non-relativistic Schr\"odinger equation from the wave equation
(\ref{extra1}). Equivalently, we must have $V(\mathbf{r},t)$ tending to the potential
energy associated to this external field when $c\rightarrow \infty$ to recover the 
non-relativistic Schr\"odinger equation from (\ref{GeneralEq7}).

Most of the time in this work we will be dealing with the free particle case,
in particular when we second quantize the Lorentz covariant Schr\"odinger equation,
and with the minimal coupling prescription, when modeling how a charged particle 
interacts with electromagnetic fields.
Therefore, in those instances 
the issue of the non-uniqueness of the relativistic invariant $f(\mathbf{r},t)$
is not a problem since we will be assuming $f(\mathbf{r},t)=0$.

On the other hand, we will also solve several bound state or scattering problems 
where we will assume that $V(\mathbf{r},t)$ 
is given by its non-relativistic version. For 
particles with small velocities, this is a very
good approximation. Also, 
this is the approach we will follow
when dealing with an external gravitational field.\footnote{Although we do not 
need in this work the exact relativistic expression for $V(\mathbf{r},t)$, we make 
the following conjecture aiming at an effective invariant modeling of it. 
What we show below is just a sketch and should not be taken as 
the final and definitive prescription to the relativistic version of $V(\mathbf{r},t)$. 
The argument goes as follows. 
When we are presented with the non-relativistic Schr\"odinger 
equation (\ref{se}), it is implicitly assumed that the external source associated with
the potential energy $V$ is at rest. In this rest frame we can define  
a unitary and dimensionless 
four-vector $\mathrm{n}^\mu=(1,0,0,0)$ that is proportional
to the four-momentum of the source (the source momentum is zero since it is at rest).
Since $V$ has dimension of energy, 
we can also postulate the existence of a four-vector $\mathrm{v}^\mu$ such that 
in this frame it is given by $\mathrm{v}^\mu=(V,0,0,0)$. 
Now, it is clear that $\mathrm{n}^\mu\mathrm{v}_\mu=V$ and it is tempting to replace $V$ 
in eq.~(\ref{GeneralEq7}) by $\mathrm{n}^\mu\mathrm{v}_\mu$ as the relativistic scalar
whose non-relativistic limit tends to the potential $V$ 
appearing in the Schr\"odinger equation.
}

It is worth mentioning that we can also write eq.~(\ref{GeneralEq7}) in two other useful ways, 
one that shows its resemblance to the non-relativistic Schr\"odinger equation,
\begin{equation}
-\frac{\hbar^2}{2mc^2}\frac{\partial^2\Psi}{\partial t^2} + i\hbar\frac{\partial \Psi}{\partial t} =  
-\frac{\hbar^2}{2m}\nabla^2 \Psi  + V\Psi,
\label{GeneralEq8}
\end{equation}
and the other using the four-vector notation,
\begin{equation}
\partial_\mu\partial^\mu \Psi 
- i\frac{2mc}{\hbar}\partial_0 \Psi + \frac{2mV}{\hbar^2}\Psi=0.
\label{GeneralEq9}
\end{equation}

\textit{Remark 1:} If we were dealing strictly with the free particle case, we would not be able to fix the
value of $A$. In this scenario $f(\mathbf{r},t)=0$ and any value of $A\neq 0$ would give 
eq.~(\ref{GeneralEq7}), with $V(\mathbf{r},t)=0$, from eq.~(\ref{GeneralEq6}).

\textit{Remark 2:} Again, 
if we restrict ourselves to the free particle case, the non-relativistic  
Schr\"odinger equation follows directly from eq.~(\ref{GeneralEq7}) when $c\rightarrow \infty$. 
There is no need to go through the discussion contained between eqs.~(\ref{GeneralEq6}) and (\ref{GeneralEq7}).

\textit{Remark 3:} We can write eq.~(\ref{GeneralEq7}) more compactly if we define
$\tilde{\partial}_\mu = \partial_\mu - \frac{mc}{\hbar}\delta_\mu$ and
$\tilde{\partial}^\mu = \partial^\mu - \frac{mc}{\hbar}\epsilon^\mu$
where  
$\delta_\mu = \epsilon^\mu = (i,1/\sqrt{3},1/\sqrt{3},1/\sqrt{3})$. Note that 
$\tilde{\partial}^\mu$ is not $g^{\mu\nu}\tilde{\partial}_\nu$. With this convention 
eq.~(\ref{GeneralEq9}) is $\tilde{\partial}_\mu\tilde{\partial}^\mu\Psi+\frac{2mV}{\hbar^2}\Psi=0$.

\textit{Remark 4:} It is possible, though, to redefine $\partial_{\mu}$ such that 
$\partial^{\mu} = g^{\mu\nu}\partial_\nu$ is still valid. This is accomplished if 
$\bar{\partial}_\mu = \partial_\mu + \frac{imc}{\hbar}\zeta_\mu$, with 
$\zeta_\mu  = (-1,0,0,0)$, and if we measure the potential energy from a different origin,
starting from $mc^2/2$ instead of zero. 
In other words, using $\bar{\partial}_\mu$ and $\bar{V}=V+mc^2/2$,
eq.~(\ref{GeneralEq7}) becomes
$\bar{\partial}_\mu\bar{\partial}^\mu\Psi+\frac{2m\bar{V}}{\hbar^2}\Psi=0$,
where 
$\bar{\partial}^{\mu} = g^{\mu\nu}\bar{\partial}_\nu$.

\subsubsection{Example: the plane wave}

It is instructive at this point to study the free particle case in the light of eq.~(\ref{GeneralEq7}),
the Lorentz covariant Schr\"odinger equation. For simplicity, we deal with the 
one dimensional case, assume that in the reference frame $S$ the particle of mass $m$
has momentum $\mathbf{p}=(p,0,0)$, and that $S'$ moves away from $S$ with velocity $\mathbf{v}=(v,0,0)$,
where $v>0$.

Setting $V(x,t)=0$ and writing
\begin{equation}
\Psi(x,t) = e^{\frac{i}{\hbar}(px-Kt)},
\label{psir}
\end{equation}
it is not difficult to see that eq.~(\ref{psir}) is a solution to
eq.~(\ref{GeneralEq7}) if
\begin{equation}
K = -mc^2 +\sqrt{m^2c^4 + p^2c^2}.
\label{kinetic}
\end{equation}
Note that we also have another possible relation between $K$ and $p$
giving a solution to eq.~(\ref{GeneralEq7}), i.e.,
$K = -mc^2 -\sqrt{m^2c^4 + p^2c^2}$. We will deal with this other possible solution later,
where we will attempt to give a physical interpretation to it that has intriguing consequences. 

Looking at eq.~(\ref{kinetic}) and remembering that a relativistic particle satisfies
\begin{displaymath}
E^2=m^2c^4+p^2c^2,
\end{displaymath}
$E$ being the total energy of the particle, we immediately see
that $K$ is the particle's kinetic energy, where we have subtracted from $E$ the particle's
rest energy $mc^2$. We can thus rewrite eq.~(\ref{psir}) as
\begin{equation}
\Psi(x,t) = e^{\frac{i}{\hbar}(mc^2 t + px - Et)}.
\label{psir2}
\end{equation}

Using eqs.~(\ref{transformationL}) and (\ref{thetaL}),
the wave function in the reference frame $S'$ is, up to a constant phase,
\begin{eqnarray}
\Psi'(x',t') &=& e^{-\frac{i}{\hbar}[(\gamma -1)mc^2t'+\gamma m v x']}\Psi(x,t) \nonumber \\
& = & 
e^{-\frac{i}{\hbar}[(\gamma -1)mc^2t'+\gamma m v x']} 
\Psi\left(\gamma(x'+vt'),\gamma(t'+v x'/c^2)\right),
\label{psi'}
\end{eqnarray}
where we have used eqs.~(\ref{l0}) and (\ref{l1}) to obtain the last line.
Employing eq.~(\ref{psir2}) to evaluate eq.~(\ref{psi'}) we arrive at
\begin{equation}
\Psi'(x',t') = e^{\frac{i}{\hbar}[mc^2t'+\gamma (p^1 - \beta p^0)x'
-c \gamma (p^0-\beta p^1)t']},
\label{psi'2}
\end{equation}
where $p^0=E/c$ and $p^1=p$ are the first two components of the 
four-momentum $p^\mu=(p^0,p^1,p^2,p^3)$ $=$ $(E/c,p_x,p_y,p_z)$.

Under a Lorentz boost in the $x$ direction the four-momentum $p^\mu$ transforms as
\begin{eqnarray}
p^{0'} &=& \gamma(p^{0} - \beta p^{1}), \label{p0} \\
p^{1'} &=& \gamma(p^{1} - \beta p^{0}), \label{p1} \\
p^{2'} &=& p^{2}, \label{p2} \\
p^{3'} &=& p^{3}. \label{p3}
\end{eqnarray}
Using eqs.~(\ref{p0}) and (\ref{p1}), eq.~(\ref{psi'2}) becomes
\begin{eqnarray}
\Psi'(x',t') &=& e^{\frac{i}{\hbar}[mc^2t'+p^{1'}x' -c p^{0'}t']}, \nonumber \\
& = & e^{\frac{i}{\hbar}[mc^2t'+p'x' -E't']},
\label{psi'3}
\end{eqnarray}
where $p' = p^{1'}$ and $E'=p^{0'}c$ are, respectively, the particle's momentum and total energy 
in reference frame $S'$. 

As expected, eq.~(\ref{psi'3}), which is a solution to the 
Lorentz covariant Schr\"odinger equation in the reference frame $S'$, can be obtained from 
eq.~(\ref{psir2}) by simply changing the unprimed quantities $x,t,p$, and $E$ to the
respective primed ones.

\section{Basic properties of the Lorentz covariant Schr\"odinger equation}
\label{basicLCSE}

\subsection{Probability four-current density}

If we write the wave function $\Psi(\mathbf{r},t)$ in its polar form,
\begin{equation}
\Psi(\mathbf{r},t) = R(\mathbf{r},t)e^{iS(\mathbf{r},t)/\hbar} = \sqrt{\rho(\mathbf{r},t)}e^{iS(\mathbf{r},t)/\hbar},
\label{psiPolar}
\end{equation}
the transformation
law (\ref{transformationL}) for the wave function under a Lorentz boost becomes
\begin{eqnarray}
R(\mathbf{r},t) &=& R'(\mathbf{r'},t') \hspace{.2cm} \mbox{or} \hspace{.2cm} \rho(\mathbf{r},t) = \rho'(\mathbf{r'},t'), \label{transformationL2} \\
S(\mathbf{r},t) &=& S'(\mathbf{r'},t') + (\gamma -1)mc^2 t' + \gamma m \mathbf{v\cdot r'},
\label{thetaL2}
\end{eqnarray}
where we have set $cte=0$ in eq.~(\ref{thetaL}).

Using eqs.~(\ref{transformationL2}) and (\ref{thetaL2}), we want to prove that the following object, 
\begin{eqnarray}
J^\mu &=& \left(  \rho c - \frac{\rho}{m}\frac{\partial S}{c\partial t}, \frac{\rho}{m}\frac{\partial S}{\partial x}, 
\frac{\rho}{m}\frac{\partial S}{\partial y}, \frac{\rho}{m}\frac{\partial S}{\partial z}\right) 
=\left(  \rho c - \frac{\rho}{m}\partial_0 S, \frac{\rho}{m}\partial_1 S, 
\frac{\rho}{m}\partial_2 S, \frac{\rho}{m}\partial_3 S\right), 
\label{Jmu}
\end{eqnarray}
is a contravariant four-vector and that it satisfies the continuity equation, $\partial_\mu J^\mu = 0$, 
whenever the potential $V(\mathbf{r},t)$ is real.
These two properties allow us to identify 
$J^\mu$ as the probability four-current density of the Lorentz covariant Schr\"odinger equation.

Note that the three spatial terms of $J^\mu$, namely $\mathbf{J}$ $=$ $(J^1,$ $J^2, J^3)$, are formally identical to the three 
vector components of the probability current density of the non-relativistic 
Schr\"o\-din\-ger equation \cite{bal98},
\begin{equation}
\mathbf{J} = -\frac{i\hbar}{2m}\left( \Psi^*\nabla\Psi - \Psi\nabla\Psi^*\right).
\label{Jj}
\end{equation}
Equivalently, $\mathbf{J}$ can also be seen as  
formally equal to the three spatial terms of the Klein-Gordon equation 
four-current density \cite{gre00}.

The time component of $J^\mu$, on the other hand, can be written as
\begin{equation}
J^0 = \rho c +  \frac{i\hbar}{2m}\left( \Psi^*\partial_0\Psi - \Psi\partial_0\Psi^*\right).
\label{j0}
\end{equation}
The time component of
the Klein-Gordon four-current density \cite{gre00} is formally equal to the second term  
of the right hand side of eq.~(\ref{j0}). The first term, $\rho c$, is the extra ingredient we
need to guarantee that $J^\mu$ is a four-vector if $\Psi(\mathbf{r},t)$ transforms according to
eqs.~(\ref{transformationL2}) and (\ref{thetaL2}).

Let us start the proof that $J^\mu$ is a four-vector. For simplicity, and without loss of generality,
we assume a Lorentz boost along the $x^1$ direction. This implies that $\mathbf{v}=(v,0,0)$.
Thus, using eqs.~(\ref{d0}), (\ref{transformationL2}), (\ref{thetaL2}), and carrying out the derivatives,
the time component of $J^\mu$ can be written as follows,
\begin{equation}
J^0 =  \rho' c -\frac{\rho'}{m}\gamma\left[ \partial_{0'}S' + (\gamma - 1)mc -\beta\partial_{1'}S' -\beta\gamma mv \right].
\end{equation}
Noting that $v=\beta c$ and that $\gamma^2(1-\beta^2)=1$, we get after a couple of simplifications,
\begin{eqnarray}
J^0 &=&  \rho' c -\frac{\rho'}{m}\gamma\left[ \frac{mc}{\gamma} (1-\gamma) + \partial_{0'}S' -\beta\partial_{1'}S' \right] \nonumber \\
    &=&  \gamma\left[ \left( \rho' c -\frac{\rho'}{m} \partial_{0'}S'\right) + \beta\left(\frac{\rho'}{m} \partial_{1'}S'\right) \right] \nonumber \\
    &=& \gamma [ J^{0'} + \beta J^{1'}]. \label{j0l}
\end{eqnarray}
In a similar way we get
\begin{eqnarray}
J^1 &=&  \gamma ( J^{1'} + \beta J^{0'} ), \label{j1l} \\
J^2 &=&  J^{2'}, \label{j2l} \\
J^3 &=&  J^{3'}. \label{j3l}
\end{eqnarray}
Equations (\ref{j0l})-(\ref{j3l}) are the laws a contravariant vector must obey after
a Lorentz boost in the $x^1$ 
direction [cf. eqs.~(\ref{l0})-(\ref{l3})], proving that $J^\mu$ is indeed a four vector.

Let us move to the proof that $J^\mu$ is a conserved current. Computing the four-divergence of eq.~(\ref{Jmu}) we get
\begin{equation}
\partial_\mu J^\mu = c\partial_0\rho -\frac{1}{m}\partial_\mu \rho \partial^\mu S - \frac{\rho}{m}\partial_\mu\partial^\mu S.
\label{divergence}
\end{equation}
To make progress, we need to insert the polar form of the wave function, eq.~(\ref{psiPolar}), into the 
Lorentz covariant Schr\"o\-din\-ger equation, eq.~(\ref{GeneralEq7}). Carrying out the derivatives and noting that
$R = \sqrt{\rho}$, we can show that eq.~(\ref{GeneralEq7}) is equivalent to the following two coupled equations, 
%
\begin{eqnarray}
\rho\partial_\mu\partial^\mu\rho - \frac{1}{2}\partial_\mu\rho\partial^\mu\rho  
- \frac{2\rho^2}{\hbar^2} 
(\partial_\mu S \partial^\mu S - 2mc\partial_0 S -2m V_R ) & = & 0, \label{real} \\  
\rho\partial_\mu\partial^\mu S + \partial_\mu\rho\partial^\mu S - mc\partial_0 \rho + \frac{2m\rho V_I}{\hbar}  & = & 0.
\label{imaginary}
\end{eqnarray}
%
Equations (\ref{real}) and (\ref{imaginary}) are, respectively, what we get by equating the real and imaginary parts of eq.~(\ref{GeneralEq7})
to zero. We have also written 
\begin{displaymath}
V = V_R + i V_I,
\end{displaymath}
since a complex potential can be used to phenomenologically model, for example,
dissipative processes. 

If we divide eq.~(\ref{imaginary}) by $m$ we can write it as
\begin{equation}
c\partial_0 \rho  - \frac{1}{m} \partial_\mu\rho\partial^\mu S- \frac{\rho}{m}\partial_\mu\partial^\mu S = \frac{2\rho V_I}{\hbar}.
\label{imaginary2}
\end{equation}
Comparing eqs.~(\ref{imaginary2}) and (\ref{divergence}) we see that 
\begin{equation}
\partial_\mu J^\mu = \frac{2\rho}{\hbar}V_I.
\label{divergence2}
\end{equation}
Equation (\ref{divergence2}) is the continuity equation with a source or sink modeled by the complex part of the  
potential $V$. If $V$ is real, as one would expect for closed systems described by Hermitian operators, we arrive 
at
\begin{equation}
\partial_\mu J^\mu = 0,
\label{divergence3}
\end{equation}
proving that $J^\mu$ is the conserved current of the Lorentz covariant Schr\"odinger equation.

It is worth mentioning that $\partial_\mu J^\mu$ tends exactly to the continuity equation
of the non-relativistic Schr\"odinger equation. This can be seen writing eq.~(\ref{divergence3})
explicitly in terms of its derivatives and using the definition of $J^0$ as given in eq.~(\ref{Jmu}),
\begin{equation}
\partial_\mu J^\mu = \frac{1}{c}\frac{\partial J^0}{\partial t}+ \nabla \cdot \mathbf{J} = 
\frac{\partial \rho}{\partial t} -\frac{1}{mc^2}\frac{\partial }{\partial t}
\left( \rho\frac{\partial S}{\partial t}\right) + \nabla \cdot \mathbf{J}  = 0.
\label{continuity}
\end{equation}
Taking the limit where $c\rightarrow \infty$ we get
\begin{equation}
\frac{\partial \rho}{\partial t} + \nabla \cdot \mathbf{J}= 0,
\label{continuity2}
\end{equation}
which is the continuity equation associated to the 
non-relati\-vis\-tic Schr\"odinger equation.

\subsection{Manifest covariance}

Our goal here is to recast the Lorentz covariant
Schr\"odinger equation in a manifestly covariant way. 
This can be accomplished by noting that the Lorentz covariant Schr\"odinger equation, eq.~(\ref{GeneralEq7}), 
is equivalent to eqs.~(\ref{real}) and (\ref{imaginary}). Moreover, eq.~(\ref{imaginary}) is equivalent to
the continuity equation (\ref{divergence2}), which is already written in a manifestly covariant form,
\begin{equation}
\partial_\mu J^\mu = \frac{2\rho}{\hbar}V_I.
\label{divergence4}
\end{equation}

What remains to be done is to rewrite eq.~(\ref{real}) in a manifestly covariant way.
Using eq.~(\ref{Jmu}), it is not difficult to see that
\begin{equation}
J_\mu J^\mu = \rho^2c^2 -\frac{2\rho^2c}{m}\partial_0 S + \frac{\rho^2}{m^2}\partial_\mu S \partial^\mu S.
\label{jmujmu}
\end{equation}
After multiplying eq.~(\ref{jmujmu}) by $m^2/\rho^2$ we arrive at
\begin{equation}
\frac{m^2}{\rho^2} ( J_\mu J^\mu - \rho^2c^2 ) =  \partial_\mu S \partial^\mu S - 2mc\partial_0 S.
\label{jmujmu2}
\end{equation}
Using eq.~(\ref{jmujmu2}), we can rewrite eq.~(\ref{real}) as
\begin{equation}
 \rho\partial_\mu\partial^\mu\rho - \frac{1}{2}\partial_\mu\rho\partial^\mu\rho - \frac{2m^2}{\hbar^2} 
(  J_\mu J^\mu - \rho^2 c^2 )  + \frac{4m\rho^2}{\hbar^2} V_R  =  0. \label{real2}
\end{equation}

Equations (\ref{divergence4}) and (\ref{real2}) are the manifestly covariant e\-qua\-tions we were looking for. 
They are, respectively, the imaginary and real parts of the Lorentz covariant 
Schr\"o\-din\-ger equation, 
written in a manifestly covariant way, that one obtains after inserting the wave function in its 
polar form (\ref{psiPolar}) into eq.~(\ref{GeneralEq7}). 

\subsection{Connection to the Klein-Gordon equation}

Looking at eq.~(\ref{GeneralEq7}), it is not difficult 
to realize that the non-relativistic limit of the 
Lorentz covariant Schr\"odinger equation  
gives exactly the usual Schr\"odinger equation. Indeed,
taking the limit where the speed of light $c\rightarrow \infty$, the first term of eq.~(\ref{GeneralEq7})
tends to zero and what remains is the non-relativistic Schr\"odinger equation. This is most clearly seen  
using eq.~(\ref{GeneralEq8}), where the remaining terms after the limit in which $c\rightarrow \infty$ is
the non-relativistic Schr\"odinger equation written in its usual form. Note that to obtain the non-relativistic
Schr\"odinger equation from the Klein-Gordon e\-qua\-tion, a much more involved limiting process 
is needed \cite{gre00}. 

On the other hand, if the mass $m \rightarrow 0$, eq.~(\ref{GeneralEq7}) is the standard 
relativistic invariant wave equation for a massless scalar particle or, equivalent, 
the massless Klein-Gordon e\-qua\-tion. It is worth mentioning that when $m=0$, 
$\Psi$ is a relativistic scalar in the conventional way, i.e., $\Psi\rightarrow \Psi$ 
after a Lorentz transformation 
[cf. eqs.~(\ref{transformationL}) and (\ref{thetaL}) for $m=0$].

Our goal now is to search for the phase transformation that we must implement on $\Psi$  to go
from the Lorentz covariant Schr\"odinger equation, eq.~(\ref{GeneralEq7}), to the Klein-Gordon equation
with $m\neq 0$ \cite{gre00}. In order to determine the correct phase transformation, we make the following
ansatz,
\begin{equation}
\Psi(\mathbf{r},t) = e^{-\frac{i}{\hbar}f(t)}\Phi(\mathbf{r},t),
\label{transformationKG}
\end{equation}
where $f(t)$ depends only on the time. The reason to work with 
a function $f$ that does not depend on the position vector $\mathbf{r}$ is
related to the fact that the spatial derivatives of eq.~(\ref{GeneralEq7}) are
already the ones appearing in the Klein-Gordon equation. To arrive at the Klein-Gordon
equation starting with eq.~(\ref{GeneralEq7}), we just need to be able to find an $f$
that allows us to get rid of the first order time derivative after the transformation above.

Inserting eq.~(\ref{transformationKG}) into eq.~(\ref{GeneralEq7}) and carrying out the de\-ri\-va\-ti\-ves
we get, up to a global phase,
\begin{eqnarray}
\frac{1}{c^2}\frac{\partial^2\Phi}{\partial t^2} &-& \nabla^2 \Phi 
-\frac{2i}{\hbar} \left( m + \frac{\dot{f}}{c^2}\right)\frac{\partial \Phi}{\partial t} 
- \left(\frac{i\ddot{f}}{c^2\hbar} + \frac{\dot{f}^2}{c^2\hbar^2} 
+\frac{2m\dot{f}}{\hbar^2}  - \frac{2mV}{\hbar^2} \right) \Phi=0,
\label{KG1}
\end{eqnarray}
where the dot and double dots over $f$ denote the first and second order derivatives
with respect to time $t$. To obtain the Klein-Gordon equation, we must first impose that
\begin{equation}
m + \frac{\dot{f}}{c^2} = 0,
\end{equation}
which leads to 
\begin{equation}
f(t) = -mc^2t + cte.
\label{ft}
\end{equation}
This condition guarantees that eq.~(\ref{KG1}) has no first order derivatives of $\Phi$
with respect to time.  

Inserting eq.~(\ref{ft}) into eq.~(\ref{KG1}) we get
\begin{equation}
\frac{1}{c^2}\frac{\partial^2\Phi}{\partial t^2} - \nabla^2 \Phi 
+ \left(\frac{m^2c^2}{\hbar^2}  +  \frac{2mV}{\hbar^2} \right) \Phi=0.
\label{KG2}
\end{equation}
Equation (\ref{KG2}) can be seen as a generalization of the Klein-Gordon equation, effectively describing
a scalar field $\Phi$ in the presence of an external potential $V$. If we set $V=0$, we recover the
Klein-Gordon equation for massive scalar fields \cite{gre00},
\begin{equation}
\frac{1}{c^2}\frac{\partial^2\Phi}{\partial t^2} - \nabla^2 \Phi 
+ \frac{m^2c^2}{\hbar^2}  \Phi=0.
\label{KG3}
\end{equation}

\begin{figure}[!ht]
\centering
\includegraphics[width=12cm]{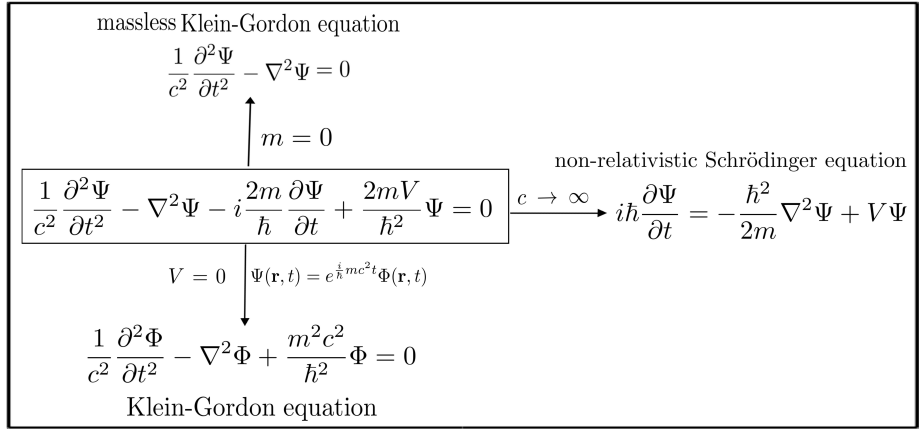}
\caption{\label{fig1} Relationship among the Lorentz covariant Schr\"odinger equation and the 
other standard scalar wave equations of quantum mechanics.}
\end{figure}

We have thus proved that if $\Psi$ satisfies the Lorentz covariant Schr\"odinger equation (\ref{GeneralEq7}), 
the wave function $\Phi$, connected to $\Psi$ by the phase transformation 
\begin{equation}
\Psi(\mathbf{r},t) = e^{\frac{i}{\hbar}mc^2t}\Phi(\mathbf{r},t),
\label{transformationKG2}
\end{equation}
satisfies the generalized Klein-Gordon equation (\ref{KG2}). In figure \ref{fig1}
we show pictorially how the Lorentz covariant Schr\"o\-din\-ger equation are connected to 
the non-relativistic Schr\"o\-din\-ger equation and to the Klein-Gordon equation.

\section{Predictions of the Lorentz covariant Schr\"odinger equation}
\label{PredictionsLCSE}

\subsection{The free particle}
\label{vzero}

Here, and in sections \ref{vconst}, \ref{telegraph}, and \ref{secD}, we assume, without losing generality, that 
we deal with a 
one-dimensional problem. 

The free particle solution (plane wave solution) 
to the Lorentz covariant Schr\"odinger equation is obtained assuming $V(x,t)=0$. Inserting
the ansatz 
\begin{equation}
\Psi(x,t) = e^{\frac{i}{\hbar}(px - Kt)}
\label{free}
\end{equation}
into eq.~(\ref{GeneralEq7}), we see that eq.~(\ref{free}) is a solution to eq.~(\ref{GeneralEq7}) if
\begin{equation}
K = K_{\pm} = -mc^2 \pm \sqrt{m^2c^4 + p^2c^2}.
\label{kpm}
\end{equation}

The physical meaning of $K_+$ is straightforward. It is the relativistic kinetic energy of 
a free particle of mass $m>0$, where $E=K_+ + mc^2=\sqrt{m^2c^4 + p^2c^2}$ is its
total relativistic energy. 
The physical interpretation of $K_-$ is subtler than that of $K_+$, at least at the stage of first quantization. 
A rigorous justification of what will be said about $K_-$  here and in section \ref{vconst} 
will be given in section \ref{qft}, where we
develop the quantum field theory of the Lorentz covariant Schr\"odinger equation. 
Our goal here is to motivate and to qualitatively explore the main features of eq.~(\ref{GeneralEq7}) that
lead to the results outlined below and rigorously established when we ``second quantize''
the Lorentz covariant Schr\"odinger equation.

If we could redefine the zero of the potential $V$ in such a way that 
$K_\pm$ changed to $\tilde{K}_\pm = \pm \sqrt{m^2c^4 + p^2c^2}$, the same interpretation associated
to the negative and positive energy solutions of the Klein-Gordon equation would be valid here.
We would associate $\tilde{K}_+>0$, the positive energy solution of the wave equation,   
to a particle of mass $m$, energy $\tilde{K}_+$, and 
traveling forward in time ($t>0$). The negative solution would be reinterpreted as
an antiparticle with the same mass $m$ and positive energy 
$-\tilde{K}_-=\tilde{K}_+$ traveling backwards 
in time: 
\begin{equation}
e^{\frac{i}{\hbar}(px - \tilde{K}_-t)} = e^{\frac{i}{\hbar}[px + \tilde{K}_-(-t)]} 
= e^{\frac{i}{\hbar}[px - \tilde{K}_+(-t)]}.
\label{anti1}
\end{equation}

However, as we show in section \ref{vconst}, we cannot redefine the zero of the potential independently of $\mathbf{p}$ to get
$\tilde{K}_\pm = \pm \sqrt{m^2c^4 + p^2c^2}$. This means that for each value of the magnitude of $\mathbf{p}$ 
we would need to work with a different
zero of the potential for this interpretation to make sense. 
We believe this solution is unsatisfactory and in order to have a $\mathbf{p}$-independent interpretation of $K_{\pm}$
we are obliged to take the following route.

If we stick with the zero of the potential at $V=0$, we readily see that $K_-\neq - K_+$ and 
the previous interpretation attributed to the negative energy $K_-$ does not follow. 
This can be remedied by the following 
procedure.
Looking at eq.~(\ref{anti1}), and comparing the far left term with the far right one, we note that
$\tilde{K}_-t = \tilde{K}_+(-t)$. This last equality is clearly not satisfied by $K_-$ and $K_+$. 
We can only satisfy it if the antiparticle that travels backwards in time also has
a negative mass, namely, $K_{-}$ changes to $\bar{K}_- = mc^2 - \sqrt{m^2c^4 + p^2c^2}$.
In this scenario we have $\bar{K}_-t = K_+(-t)$ and the particle of 
mass $m$, positive energy $K_+$, and traveling forward in time, must have an 
antiparticle with a negative mass traveling backwards in time with energy
$- \bar{K}_- = K_+$.

This interpretation is not too wide of the mark if we study the solutions to 
the Lorentz covariant Schr\"odinger e\-qua\-tion when we set $V=0$ and change $m \rightarrow -m$. 
In this case it is not difficult to see that eq.~(\ref{GeneralEq7}) 
has the following solutions,
\begin{equation}
\Psi(x,t) = e^{\frac{i}{\hbar}(px - \bar{K}t)},
\label{free2}
\end{equation}
with
\begin{equation}
\bar{K} = \bar{K}_\pm = mc^2 \pm \sqrt{m^2c^4 + p^2c^2}.
\label{kbar}
\end{equation}

We now have that $\bar{K}_-=-K_+$, which allows us to write
\begin{equation}
e^{\frac{i}{\hbar}(px - \bar{K}_- t)} = e^{\frac{i}{\hbar}[px + \bar{K}_-(-t)]} 
= e^{\frac{i}{\hbar}[px - K_+ (-t)]}.
\label{anti2}
\end{equation}
Therefore, to the particle with mass $m$, energy $K_+$, traveling forward in time, 
we must have an antiparticle with mass $-m$, energy $-\bar{K}_-=K_+$, traveling backwards in time.  
The same interpretation can be attached to the pair of solutions with energies given by
$\bar{K}_+$ and $K_-$ since here we also have $-K_-=\bar{K}_+$. 
The solution with $\bar{K}_+=mc^2 + \sqrt{m^2c^4 + p^2c^2}$
describes a particle of
mass $-m$, traveling forward in time, and with positive energy $\bar{K}_+$, while the solution with
$K_-$ is its antiparticle with mass $m$ traveling backwards in time.

We can intuitively understand why this interpretation will rigorously appear in 
the second quantization of the Lorentz covariant Schr\"odinger equation 
by the following argument. 
Looking at eq.~(\ref{GeneralEq7}) when $V=0$,
we note that if we change $m$ to $-m$ we get the following
equation,
\begin{equation}
\frac{1}{c^2}\frac{\partial^2\Psi}{\partial t^2} - \nabla^2 \Psi 
+ i\frac{2m}{\hbar}\frac{\partial \Psi}{\partial t} =0.
\label{-m}
\end{equation}
On the other hand, by taking the complex conjugate of eq. (\ref{GeneralEq7}) we arrive at
\begin{equation}
\frac{1}{c^2}\frac{\partial^2\Psi^*}{\partial t^2} - \nabla^2 \Psi^* 
+ i\frac{2m}{\hbar}\frac{\partial \Psi^*}{\partial t} =0. 
\label{*}
\end{equation}

Comparing eqs.~(\ref{-m}) and (\ref{*}), we see that they are formally the same,
having the same mathematical structure. In other words, the solution
$\Psi^*$ of the complex conjugate equation is equivalent to the solution 
$\Psi$ of the negative mass equation. In this way, when second quantizing the
complex Lagrangian leading to eq.~(\ref{GeneralEq7}) and its complex conjugate version, 
the conserved charge obtained via Noether's theorem will 
be formed by particles with positive and negative masses.

It is instructive to remark that the Klein-Gordon equation does not possess this feature,
even when we deal with its complex version. Looking at eq.~(\ref{KG3}), we see that the Klein-Gordon
equation is unchanged if $m\rightarrow -m$ and that its complex conjugate
version is formally equivalent to eq. (\ref{KG3}). The key ingredient that makes the 
Lorentz covariant Schr\"o\-di\-nger equation change its structure when $m\rightarrow -m$,
or when we take its complex conjugate, is the presence of the term 
\begin{displaymath}
i\frac{2m}{\hbar}\frac{\partial \Psi}{\partial t},
\end{displaymath} 
which has the imaginary number $i$
and which is linear in the mass $m$. The Klein-Gordon equation has no term linear in $m$ or
explicitly depending on $i$.\footnote{In the context of 
Foldy-Wouthuysen transformations, the interpretation of 
the positive and negative energy solutions of the Dirac equation as describing particles 
with positive and ``negative'' masses is also possible 
for weak external fields \cite{fol50}. Furthermore, for optical and quantum
hydrodynamic systems, effective negative masses associated with particle-like excitations
in those systems were experimentally reported \cite{tuc99,pol01,cav12,con15,kha17}.}

\subsection{Particle in a constant potential}
\label{vconst}

We now assume that the potential $V(x,t)$ is a non-null 
constant $V$. Inserting
the ansatz given by eq.~(\ref{free})
into (\ref{GeneralEq7}), we have that it is a solution to eq.~(\ref{GeneralEq7}) if
\begin{equation}
K = K_{\pm} = -mc^2 \pm \sqrt{m^2c^4 + p^2c^2 + 2mc^2V}.
\label{kpmV}
\end{equation}

Adding $mc^2$ to both sides of eq.~(\ref{kpmV}) and squaring it leads to
\begin{equation}
E^2 - p^2c^2 =  m^2c^4   + 2mc^2V,
\label{invE}
\end{equation}
where we have defined
\begin{displaymath}
E = K  + mc^2.
\end{displaymath}
The right hand side of eq.~(\ref{invE}) is a Lorentz invariant and can be compactly written
as
\begin{equation}
p^\mu p_\mu = m^2c^2 + 2mV = m^2c^2 \left( 1 + \frac{2V}{mc^2} \right),  
\label{pmupmu}
\end{equation}
where $p^\mu=(E/c,\mathbf{p})$ is the four-momentum.

In the particle's rest frame $\mathbf{p} = 0$ and evaluating eq.~(\ref{pmupmu})
in this frame we get for the particle's rest energy,
\begin{equation}
E_0 = mc^2 \sqrt{1+ \frac{2V}{mc^2}} = m'c^2.
\label{eo}
\end{equation}
We see that the presence of the potential $V$ changes the rest mass of the particle to
\begin{equation}
m' = m \sqrt{1+ \frac{2V}{mc^2}}.
\label{restmass}
\end{equation}

For small values of $V$, such that 
\begin{equation}
\frac{V}{mc^2} \ll 1,
\label{condition1}
\end{equation}
we have
\begin{equation}
m' \approx m + \frac{V}{c^2}.
\label{smallV}
\end{equation}
Equation (\ref{smallV}) means that the mass of any particle in the presence
of a small potential changes equally by the same quantity, namely, $V/c^2$. 
For high values of $V$, the changes in the mass are different for each type of particle
(it depends on the value of $m$) and
is given by eq.~(\ref{restmass}). 
These results tell us that the mass of a particle described
by the Lorentz covariant Schr\"odinger equation depends on the background external field 
to which it is subjected and which is phenomenologically modeled by the potential $V$.
This feature might be useful in the effective description of condensed matter systems, where
the ``spacetime'' structure is given by an effective Minkowski spacetime with limiting velocity $c_{eff} < c$  
and effective masses differing from the ``bare'' masses defined when $V=0$. 
Note that this feature is also seen in the generalized Klein-Gordon equation (\ref{KG2})
since it also depends on the potential $V$.\footnote{The possibility of an external
potential that is not related to electromagnetic or gravitational fields and 
that may change the rest mass of a particle is introduced by Stueckelberg in ref. \cite{stu41}.}

We can also obtain a stronger condition than (\ref{condition1}) to the validity of eq.~(\ref{smallV}) and 
in a way that connects the magnitude of the potential $V$ with the magnitude of the particle's momentum
$|\mathbf{p}|=|p|$. This is achieved by rewriting eq.~(\ref{kpmV}) as
\begin{equation}
\left(\frac{E}{pc}\right)^2 = \left( \frac{mc^2}{pc} + \frac{V}{pc} \right)^2 + 1 - \left(\frac{V}{pc}\right)^2
\end{equation}
and noting that if 
\begin{equation}
\left(\frac{V}{pc}\right)^2 \ll 1, 
\end{equation}
we get that $1 - \left(\frac{V}{pc}\right)^2 \approx 1$ and thus 
\begin{equation}
\left(\frac{E}{pc}\right)^2 \approx \left( \frac{mc^2}{pc} + \frac{V}{pc} \right)^2 + 1.
\label{Eapprox}
\end{equation}
Rewriting eq.~(\ref{Eapprox}) as
\begin{equation}
E^2  - p^2c^2 \approx \left(m + \frac{V}{c^2} \right)^2c^4 = m'^2c^4,
\label{Eapprox2}
\end{equation}
we immediately see that it is the energy-momentum relation for a particle of mass
$m'$, where $m'$ is given by eq.~(\ref{smallV}). Therefore, if
\begin{equation}
|V| \ll |\mathbf{p}| c,
\end{equation}
a particle with mass $m$ when subjected to the constant field $V$ will behave like a free particle of
mass $m + V/c^2$.

In order to better appreciate the meaning of $K_+$, as given by eq.~(\ref{kpmV}), we expand it
in powers of $1/c^2$
and compare the resulting expansion with what one would expect in the non-relativistic limit. A simple calculation leads to
\begin{equation}
K_{+} = \frac{p^2}{2m} + V + \mathcal{O}\left(\frac{1}{c^2}\right),
\label{kplus}
\end{equation}
where $\mathcal{O}(1/c^2)$ denotes terms of order higher or equal to $1/c^2$.
Looking at eq.~(\ref{kplus}), we see that in the limit of $c\rightarrow \infty$ we obtain that $K_+$
is the non-relativistic kinetic energy plus the potential energy of the particle. 
This is consistent with assigning to 
$K_+$ the interpretation that it is the total relativistic energy of the particle minus its ``bare'' rest energy
$mc^2$ when we get back to the relativistic domain. A similar interpretation is attached to 
$K_-$ if we remember that we should consider it the solution to the Lorentz covariant
Schr\"odinger equation with a negative mass. In other words, $K$ can only be interpreted as 
the particle's relativistic kinetic energy when $V=0$.

We end this section proving, as claimed in section \ref{vzero}, that  
it is not possible to redefine the zero of the potential independently of $\mathbf{p}$ to get
$K_\pm = \pm \sqrt{m^2c^4 + p^2c^2}$. 
If we try to obtain a scenario where $K_\pm =\pm \sqrt{m^2c^4 + p^2c^2}$, we have to solve
for $V$ the following equation (cf. eq.~(\ref{kpmV})),
\begin{equation}
-mc^2 \pm \sqrt{m^2c^4 + p^2c^2 + 2mc^2V} = \pm \sqrt{m^2c^4 + p^2c^2}. 
\end{equation}
Its solutions are 
\begin{equation}
V = \frac{mc^2}{2} \pm \sqrt{m^2c^4 + p^2c^2}. 
\label{solveV}
\end{equation}
Looking at eq.~(\ref{solveV}), we see that it is not possible to have $V$ independent of 
the magnitude of $|\mathbf{p}|$. Moreover, even allowing for a p-dependence, we  cannot get
the job done since we need different values of $V$ to set either $K_+$ or $K_-$ to be 
$\sqrt{m^2c^4 + p^2c^2}$ or $-\sqrt{m^2c^4 + p^2c^2}$, respectively (note the
$\pm$ in the expression for $V$ above). This completes the 
proof of the claim given in section \ref{vzero}.

\subsection{Complex variable telegraph equation}
\label{telegraph}

If we look at eq.~(\ref{pmupmu}), we note that as we decrease the value of 
the external potential $V$, we decrease the value of the Lorentz invariant $p^\mu p_\mu$.
Eventually, for a sufficiently low potential we will get $p^\mu p_\mu = 0$.
In this scenario, when $V=-mc^2/2$,  it is expected that a particle of ``bare'' mass $m$ 
behaves like a massless
particle. Moreover, a wave packet in this case is expected to propagate without distortion
and without spreading. This is indeed the case as we prove below.

The one dimensional Lorentz covariant Schr\"odinger e\-qua\-tion, eq.~(\ref{GeneralEq7}), 
can be written as
\begin{equation}
\frac{1}{c^2}\frac{\partial^2\Psi}{\partial t^2} - \frac{\partial^2\Psi}{\partial x^2} 
- i\frac{2m}{\hbar}\frac{\partial \Psi}{\partial t} + \frac{2mV}{\hbar^2}\Psi=0.
\label{CTele0}
\end{equation}
This equation has the same structure as the Heaviside telegraph 
equation \cite{hea87a,hea87b}, 
a real partial differential equation which under certain conditions has a solution in which
a distortionless propagation of a wave packet is possible. 
Equation (\ref{CTele0}) is a complex version of the telegraph equation and it  
inherits the main features of the real Heaviside telegraph equation. As we show below, under 
a particular condition for the external potential $V$, eq.~(\ref{CTele0})
has a solution in which a wave packet propagates distortionless and
without dispersion. 

We should also mention that when $V=0$, eq.~(\ref{CTele0}) is similar to 
the complex telegraph equation postulated by Sancho in his search
for a non-relativistic wave equation leading to a non-instantaneous spread of a localized 
wave packet \cite{pan97},
\begin{equation}
\frac{2m\tau}{\hbar}\frac{\partial^2\Psi}{\partial t^2} - \frac{\partial^2\Psi}{\partial x^2} 
- i\frac{2m}{\hbar}\frac{\partial \Psi}{\partial t} =0, 
\label{pan1}
\end{equation}
with $\tau$ being a free parameter with dimension of time. Equation (\ref{pan1})
was obtained by extending to the complex plane the real telegraph equation 
\cite{hea87a,hea87b}
and adjusting the coefficients multiplying the derivatives such that the 
Schr\"odinger equation is obtained for an ``instantaneous'' relaxation time ($\tau = 0$).
In ref. \cite{pan97} it was shown that we indeed have a non-instantaneous spread of the wave function
for a localized initial condition whenever $\tau \neq 0$. Comparing eqs.~(\ref{CTele0})
and (\ref{pan1}), we see that they become equal if $\tau = \hbar/(2mc^2)$.

If we insert the ansatz [cf. eq.~(\ref{transformationKG2})]
\begin{equation}
\Psi(x,t) = e^{\frac{i}{\hbar}{mc^2t}}\Phi(x,t)
\end{equation}
into eq.~(\ref{CTele0}) we get
\begin{equation}
\frac{1}{c^2}\frac{\partial^2\Phi}{\partial t^2} - \frac{\partial^2\Phi}{\partial x^2} 
+ \left(\frac{m^2c^2}{\hbar^2}  +  \frac{2mV}{\hbar^2} \right) \Phi=0.
\label{CTele}
\end{equation}
Choosing
\begin{displaymath}
V = -\frac{mc^2}{2}, 
\end{displaymath}
eq.~(\ref{CTele}) becomes
\begin{equation}
\frac{1}{c^2}\frac{\partial^2\Phi}{\partial t^2} - \frac{\partial^2\Phi}{\partial x^2} 
=0.
\label{CTele2}
\end{equation}
This is the standard wave equation for a massless scalar particle in one dimension.
Its general solution is
\begin{displaymath}
\Phi(x,t) = F(x-ct) + G(x+ct), 
\end{displaymath}
where $F(x-ct)$ and $G(x+ct)$ are non-dispersive traveling waves to the right and to the left,
respectively. 

The general solution to eq.~(\ref{CTele0}) is thus
\begin{equation}
\Psi(x,t) = e^{\frac{i}{\hbar}{mc^2t}}\left[F(x-ct) + G(x+ct)\right].
\end{equation}
For a wave packet traveling to the right we have $G(x+ct)=0$ and thus
\begin{equation}
\Psi(x,t) = e^{\frac{i}{\hbar}{mc^2t}}F(x-ct).
\end{equation}
This leads to a probability density that propagates without
dispersion and distortion to the right and with the speed of light $c$,
\begin{equation}
\rho(x,t) = \Psi^*(x,t)\Psi(x,t) = |F(x-ct)|^2.
\end{equation}

We can also test the previous result by studying the propagation of a wave packet
for decreasing values of the potential $V$. As we decrease $V$, less dispersive is the evolution of the wave
packet, until we reach the critical value $V=-mc^2/2$ where a dispersionless 
evolution of the wave packet is achieved. 

A possible way to achieve such a huge negative potential 
would be to work with a scalar charged particle subjected to a constant electric potential $\varphi$.
If $q$ is its charge, $V=q\varphi$ and equating it to $-mc^2/2$ gives $\varphi=-mc^2/(2q)$. 
We can estimate the order of magnitude of the electric potential using the rest mass of the electron and 
its charge. This gives $\varphi \approx - 2.54 \times 10^5 V$. In other words, one needs
an electric potential of magnitude of the order of 100 thousands volts to start to see a dispersionless propagation. 

However, if we work with greater charges or lower mas\-ses, the lower the voltage needed. For some 
condensed matter systems such as a semiconductor, the effective mass of the electron can be as low as $1\%$
of its rest mass. In these materials, an electric potential of the order of $-10^3 V$ would be enough to see
a dispersionless evolution.

\subsection{Free particle wave packets}
\label{secD}

We now investigate how wave packets evolve according to the Lorentz covariant 
Schr\"odinger equation when we set $V=0$, comparing its evolution to the
ones predicted by the Klein-Gordon equation and by the non-relativistic Schr\"odinger
e\-qua\-tion when we set the Hamiltonian equal to $\!\sqrt{\!m^2c^4 \!+\! \hat{p}^2c^2}$, with 
$\hat{p} = -i\hbar \partial/\partial x$. In all cases we will be dealing with the positive 
energy solutions.

If we write the wave function as
\begin{equation}
\Psi(x,t) = \frac{1}{\sqrt{2\pi\hbar}}\int f_{\!_\Psi}(p)e^{\frac{i}{\hbar}[px - K(p)t]}dp,
\label{wp}
\end{equation}
$f_{\!_\Psi}(p)$ is the momentum distribution of the wave packet and $K(p)$, a function of 
the momentum $p$, is determined by inserting eq.~(\ref{wp}) into the corresponding wave equation 
and imposing that $\Psi(x,t)$ should be a solution to the wave equation. The integration above 
includes the entire real line, i.e., $p$ runs from $-\infty$ to $\infty$.

If we insert eq.~(\ref{wp}) into the one dimensional Lorentz covariant Schr\"odinger equation for a free particle, 
namely, eq.~(\ref{CTele0}) with $V=0$, we guarantee that eq.~(\ref{wp}) is a solution to it if
\begin{equation}
K(p) = -mc^2+\sqrt{m^2c^4+p^2c^2},
\label{dispersionLCSE}
\end{equation}
where $K(p)$ is the positive energy solution.
Similarly, inserting eq.~(\ref{wp}) into eq.~(\ref{CTele}) with $V=0$, where the latter 
equation is the one dimensional Klein-Gordon equation for a particle with mass $m$, 
we get 
\begin{equation}
K(p) = \sqrt{m^2c^4+p^2c^2}.
\label{dispersionKG}
\end{equation}

Comparing the two dispersion relations above, eqs.~(\ref{dispersionLCSE}) and (\ref{dispersionKG}),
we note that they differ by $-mc^2$, which does not depend on $p$. This means that 
the probability density  $|\Psi(x,t)|^2$ computed from eq.~(\ref{wp}) is the same whether we
work with the Lorentz covariant Schr\"odinger equation or with the Klein-Gordon equation.

It is also instructive if we compute the dispersion relation for the following wave equation,
\begin{equation}
i\hbar \frac{d}{dt}|\Psi\rangle = \sqrt{m^2c^4 + \hat{p}^2c^2} |\Psi\rangle,
\label{SE}
\end{equation}
which is the standard Schr\"odinger equation written in the bra-ket notation with  
$\hat{H} = \sqrt{m^2c^4 + \hat{p}^2c^2}$. 
Here $\hat{H}$ is the relativistic Hamiltonian for a free particle where
the momentum $p$ is replaced by the operator $\hat{p}=-i\hbar \partial/\partial x$.
If we project it onto the momentum space we can easily operate with  
$\sqrt{m^2c^4 + \hat{p}^2c^2}$ on the momentum eigenstates and solve the corresponding
differential equation. Fourier transforming back to the position representation we realize that
$\Psi(x,t)$ evolves according to eq.~(\ref{wp}) and with a dispersion relation given by eq.~(\ref{dispersionKG}).
In other words, the Lorentz covariant Schr\"odinger equation, the Klein-Gordon equation, 
and the wave equation (\ref{SE}) predict the same wave-packet dynamics in the 
free particle regime.

\subsection{Time independent potentials}

We proceed with the investigation of the main predictions of the Lorentz 
covariant Schr\"odin\-ger equation, studying now its general features when the particle 
of mass $m$ is subjected to a time independent, and not 
necessarily constant, potential. Specifically, we want to address the bound state
solutions to eq.~(\ref{GeneralEq7}) when $V(\mathbf{r},t)=V(\mathbf{r})$. 

Inserting the ansatz
\begin{equation}
\Psi(\mathbf{r},t) = e^{-iKt/\hbar}\psi(\mathbf{r})
\label{stationary}
\end{equation}
into eq.~(\ref{GeneralEq7}), where $K$ is a constant, we get
the following time-independent equation,
\begin{equation}
\frac{-\hbar^2}{2m}\nabla^2\psi(\mathbf{r}) + V(\mathbf{r})\psi(\mathbf{r}) = 
\lambda\psi(\mathbf{r}),
\label{t-ind}
\end{equation}
where
\begin{equation}
\lambda =\left( \frac{K^2}{2mc^2} + K \right).
\label{eigenvalue}
\end{equation}

Equation (\ref{t-ind}) is formally identical to the time-in\-de\-pen\-dent 
non-relativistic Schr\"odin\-ger 
equation, with $\lambda$ being its eigenvalue. 
We thus see that any solution to the latter 
is also a solution to eq.~(\ref{t-ind}), 
the time-independent Lorentz covariant Schr\"odinger equation. Due to the linearity of 
eq.~(\ref{GeneralEq7}), its general solution for time independent potentials is 
the superposition of the stationary solutions (\ref{stationary}),
\begin{equation}
\Psi(\mathbf{r},t) = \sum_{n}e^{-iK_n t/\hbar}\psi_n(\mathbf{r}),
\end{equation}
where $\psi_n(\mathbf{r})$ is an eigenvector of eq.~(\ref{t-ind}) with eigen\-va\-lue
$\lambda_n$ and, for bound state problems, $K_n$ is the following solution to eq.~(\ref{eigenvalue}),
\begin{equation}
K_n = -mc^2 + \sqrt{m^2c^4 + 2mc^2\lambda_n}.
\label{Kn}
\end{equation}

To better appreciate the physical meaning of $K$, we expand it in powers of $1/c^2$,
\begin{equation}
K = \lambda -\frac{1}{2}\frac{\lambda^2}{mc^2} + \mathcal{O}(1/c^4).
\label{K}
\end{equation}
In the non-relativistic limit, when $c\rightarrow \infty$, $K=\lambda$ and we recognize 
it as the energy of the non-relativistic system, with the rest energy already absent.
The standard notation in the non-relativistic regime is to employ the letter $E$ instead of $\lambda$
to represent the eigenvalues of eq.~(\ref{t-ind})
since in this regime they are the system's total energy (no rest energy included). 
Similarly, we can understand $K$ as the relativistic bound energy of the system, with
the rest energy already excluded.

\subsubsection{Example: the Hydrogen atom}

Let us study the solution to eq.~(\ref{t-ind}) when a particle of mass $m$
and charge $-e$ (an electron, for instance) 
is subjected to the electrostatic Coulomb potential generated by a 
particle of mass $M$ and charge $e>0$ (a proton, for example). 
In the SI units the potential is 
\begin{equation}
V(\mathbf{r}) = - \frac{e^2}{4\pi\epsilon_{\!\,_0}}\frac{1}{r},
\label{vr}
\end{equation}
where $r$ is the distance between the two particles and $\epsilon_0$ is the vacuum permittivity.
If $M\gg m$, we can consider the mass $M$ at rest and $m$ in eq.~(\ref{t-ind}) is, for all practical purposes, 
equal to the mass of the lightest particle. Otherwise we should understand 
$m$ in eq.~(\ref{t-ind}) as the reduced mass $mM/(m+M)$. We are also leaving out of our analysis
any spinorial properties of the electron and proton. We are effectively dealing with
charged scalar particles.

Inserting eq.~(\ref{vr}) into (\ref{t-ind}) we see that we formally have the 
non-relativistic Schr\"o\-din\-ger equation for the Hydrogen atom, whose eigenvalues are \cite{gre00}
\begin{equation}
\lambda_n = -\frac{mc^2}{2}\frac{\alpha^2}{n^2}.
\label{lambda}
\end{equation}
Here $n\geq 1$ is a positive integer and 
$\alpha$, the fine structure constant, is 
\begin{displaymath}
\alpha = \frac{1}{4\pi\epsilon_0}\frac{e^2}{\hbar c}.
\end{displaymath}
Inserting eq.~(\ref{lambda}) into (\ref{Kn}) we get
\begin{eqnarray}
K_n &=& -mc^2 \left(  1 -\sqrt{1 - \frac{\alpha^2}{n^2}}  \right) \label{Klcse}\\
    &=& -\frac{mc^2}{2} \frac{\alpha^2}{n^2} \left(  1 + \frac{\alpha^2}{4n^2}  \right) + \mathcal{O}(\alpha^6) \label{Klcse2},
\end{eqnarray}
where the last line is the expansion of the exact value of $K_n$ up to fourth order in 
the fine structure constant.

It is interesting to compare the previous result with the solution to the same problem using the Dirac equation 
minimally coupled to the electromagnetic field. In this case we get after subtracting the particle's rest energy \cite{gre00}
%
\begin{eqnarray}
K_{n,j+1/2}^{D} &=& -mc^2 \left\{  1 - \left[ 1 + \left( \frac{\alpha}{n-(j+\frac{1}{2})+\sqrt{(j+\frac{1}{2})^2-\alpha^2}} \right)^2 \right]^{-1/2}  
\right\} \nonumber \\
&=& -\frac{mc^2}{2} \frac{\alpha^2}{n^2} \left[  1 + \frac{\alpha^2}{n^2} \left( \frac{n}{j+\frac{1}{2}} - \frac{3}{4} \right) \right] 
    + \mathcal{O}(\alpha^6), 
\label{KD}
\end{eqnarray}
where $n\geq 1$ is a positive integer and $j+1/2 = 1,2,\ldots,n$.
The last expression above is $K_{n,j+1/2}^{D}$ expanded up to fourth order in $\alpha$.

Computing the energy of the ground state predicted by the Lorentz covariant Schr\"odin\-ger equation, eq.~(\ref{Klcse}) with $n=1$,
and by the Dirac equation, eq.~(\ref{KD}) with $n=1$ and $j+1/2=1$, we get the same result to all orders of $\alpha$,
\begin{equation}
K_1 = K_{1,1}^D = -mc^2(1-\sqrt{1-\alpha^2}).
\label{D0}
\end{equation}
Moreover, if we compute the energies predicted by the Dirac equation for $n\geq 1$ using the highest possible value for the
total angular momentum, i.e., if we set $j+1/2=n$ in eq.~(\ref{KD}), we get 
\begin{equation}
K_{n,n}^D = -mc^2\left(1-\sqrt{1-\frac{\alpha^2}{n^2}}\right) = K_n.
\end{equation}
These are exactly the energies predicted by the Lorentz covariant Schr\"odinger equation [cf. eq.~(\ref{Klcse})],
a remarkable match with the most energetic bound energies predicted by the Dirac equation for a given $n$. 

It is worth noting that we have neglected the spin properties of the electron and of the proton and 
we have not minimally coupled the Lorentz covariant Schr\"odinger equation with the electromagnetic field
to arrive at those results. We just have inserted the Coulomb potential as given above directly into the 
Lorentz covariant Schr\"odinger equation. It is really impressive that such a simple approach leads exactly 
to a subset of
the bound energies predicted by the Dirac equation. And of course, 
to break the degeneracy and get bound energies depending on the total angular momentum
of the electron,
we either have to add the spin-orbit coupling by hand in the Lorentz covariant Schr\"odinger equation 
or deal directly with a spin-$1/2$ relativistic wave equation as Dirac did. 

Before we move on to the next section, we want to compare the bound energies given by the
Lorentz covariant Sch\-r\"o\-din\-ger equation with the ones given by solving the Klein-Gordon
equation under the same conditions and minimally coupled with the electromagnetic field. 
Subtracting the rest energy
we have \cite{gre00},

\begin{eqnarray}
K_{n,l}^{KG} &=& -mc^2 \left\{  1 - \left[ 1 + \left( \frac{\alpha}{n-(l+\frac{1}{2})+\sqrt{(l+\frac{1}{2})^2-\alpha^2}} \right)^2 \right]^{-1/2}  
\right\} \nonumber \\
&=& -\frac{mc^2}{2} \frac{\alpha^2}{n^2} \left[  1 + \frac{\alpha^2}{n^2} \left( \frac{n}{l+\frac{1}{2}} - \frac{3}{4} \right) \right] 
    + \mathcal{O}(\alpha^6), 
    \label{KG-1}
\end{eqnarray}
where $n\geq 1$ is a positive integer and $l = 0,1,2,\ldots,n-1$. 
The last line is $K_{n,l}^{KG}$ expanded up to fourth order in $\alpha$.
Equation (\ref{KG-1}) is similar to (\ref{KD}), with $j+1/2$ changed to $l+1/2$, where $l$ labels the orbital
angular momentum of the mass $m$.  

Calculating the ground state energy predicted by the Klein-Gordon equation, 
namely, eq.~(\ref{KG-1}) with $n=1$ and $l=0$,
we obtain
\begin{eqnarray}
K^{KG}_{1,0} & = & -mc^2 \left[ 1 - \sqrt{\frac{1}{2} \left( 1 + \sqrt{1-4\alpha^2}  \right)  } \right]  \\  
& = & -\frac{mc^2}{2} \alpha^2 \left(  1  +  \frac{5}{4}\alpha^2\right) + \mathcal{O}(\alpha^6). 
\label{KG-0}
\end{eqnarray}
Looking at eq.~(\ref{KG-0}), we see that the ground state energy predicted by the Klein-Gordon equation minimally
coupled to the electromagnetic field is not equal to eq.~(\ref{D0}), the ground state energy given by the Dirac equation 
and by the Lorentz covariant Schr\"odinger equation. Even to order of $\alpha^4$ the values of the ground state energies 
are already different. In other words, at the level of the ground state energy, the Lorentz covariant Schr\"odinger equation
gives a better description of the Hydrogen atom than the Klein-Gordon equation minimally coupled to the electromagnetic
field. It is worth mentioning, nevertheless,
that if we insert $V$ given
by eq.~(\ref{vr}) into the generalized Klein-Gordon equation, eq.~(\ref{KG2}), we will get 
the same bound energies reported here for the Lorentz covariant Schr\"odinger equation [eq.~(\ref{Klcse})].

\subsection{Minimal coupling with the electromagnetic field }
\label{mincoup}

Our goal now is to study the Lorentz covariant Schr\"odinger equation,
as given by eq.~(\ref{GeneralEq7}), minimally coupled to an electromagnetic 
field. We also want to study 
the simultaneous action of the electromagnetic and gravitational fields 
on a particle of mass $m$ and charge $q$, where the 
static Newtonian gravitational potential enters 
in eq.~(\ref{GeneralEq7}) via the potential energy $V(\mathbf{r})$.\footnote{See 
refs. \cite{wil74,tou76,par80,fis81} for general-relativistic and 
post-Newtonian corrections in Hydrogen-like systems when the spin of the interacting particles are also included in the analysis.}

The electromagnetic minimal coupling, 
in SI units and in the metric signature we have been using, 
is obtained by replacing in the wave equation all derivatives $\partial_\mu$
according to the following prescription \cite{gre00},
\begin{equation}
\partial_\mu \rightarrow D_\mu = \partial_\mu + \frac{iq}{\hbar}A_\mu,
\label{minimalcoupling}
\end{equation}
where the covariant four-vector potential is
\begin{equation}
A_\mu = \left( \frac{\varphi}{c}, -\mathbf{A} \right). 
\label{prescription}
\end{equation}
Here $\varphi$ and $\mathbf{A}=(A^1,A^2,A^3)$ are, respectively,
the electric and vector potentials characterizing an electromagnetic field.

Applying prescription (\ref{prescription}) to eq.~(\ref{GeneralEq9}) we have 
\begin{equation}
D_\mu D^\mu \Psi -i\frac{2mc}{\hbar}D_0\Psi +\frac{2mV}{\hbar^2}\Psi = 0, 
\label{GeneralEq10}
\end{equation}
where $V$ should be thought as the particle's potential energy associated to an
external field whose origin is not electromagnetic.
All electromagnetic interactions are embedded in the minimal coupling assumption.

Noting that after a Lorentz boost $D_\mu$ transforms as a covariant four-vector and that 
$D_\mu D^\mu$ is a Lorentz
invariant, we can show that eq.~(\ref{GeneralEq10}) is Lorentz covariant  
if $\Psi$ transforms according to  eq.~(\ref{transformationL}). In addition to that,
it is not difficult to see that if we implement the gauge transformation
\begin{equation}
\tilde{A}_\mu(x) = A_\mu(x) + \partial_\mu \chi(x)
\label{gauge1}
\end{equation}
we get
\begin{equation}
\tilde{D}_\mu \tilde{D}^\mu \tilde{\Psi} -i\frac{2mc}{\hbar}\tilde{D}_0\tilde{\Psi} +\frac{2mV}{\hbar^2}\tilde{\Psi} = 0, 
\label{GeneralEq11}
\end{equation}
where 
\begin{displaymath}
\tilde{D}_\mu = \partial_\mu + \frac{iq}{\hbar}\tilde{A}_\mu
\end{displaymath}
and
\begin{equation}
\tilde{\Psi}(x) = e^{-\frac{iq}{\hbar}\chi(x)}\Psi(x).
\label{gauge2}
\end{equation}
In other words, the Lorentz covariant Schr\"odinger equation minimally coupled 
to the electromagnetic field is invariant with respect to a local gauge
transformation given by eq. (\ref{gauge1}) if the wave function transforms according to
eq.~(\ref{gauge2}).

\subsubsection{Bound state solutions in a ``pure'' static electric field}
\label{pureE}

We now set $V=0$, i.e., we have a ``pure'' electromagnetic problem, 
and assume that the particle of mass $m$ and charge $q=-e<0$ is subjected to
an attractive Coulomb potential generated by a ``fixed'' charge $e$. This is essentially 
the Hydrogen atom where we disregard the spinorial aspects of the electron and the proton.
In this scenario 
\begin{equation}
A^\mu=(\varphi/c,0,0,0), 
\label{c1}
\end{equation}
with
\begin{equation}
\varphi = \frac{1}{4\pi\epsilon_{\!\,_0}}\frac{e}{r}.
\label{c2}
\end{equation}

Inserting eqs.~(\ref{c1}) and (\ref{c2}) into eq.~(\ref{GeneralEq10}), and
separating the time variable from the spatial ones using the ansatz given
by eq.~(\ref{stationary}), we get 
\begin{equation}
-\hbar^2c^2\nabla^2\psi(\mathbf{r}) = \left[  (K-\mathcal{V}(\mathbf{r})+mc^2)^2 - m^2c^4  \right]\psi(\mathbf{r}),
\label{eqCoulomb}
\end{equation}
where 
\begin{equation}
\mathcal{V}(\mathbf{r}) = - \frac{e^2}{4\pi\epsilon_{\!\,_0}}\frac{1}{r} = -\hbar c \frac{\alpha}{r}
\label{vr2}
\end{equation}
is the particle's electrostatic potential energy.

If we make the substitution $K+mc^2 \rightarrow E$, 
eq.~(\ref{eqCoulomb}) becomes the stationary Klein-Gordon equation
minimally coupled to the electromagnetic field via the Coulomb potential \cite{gre00}. 
Therefore, the bound energies for the present problem is simply $K_{n,l}=E_{n,l}^{KG}-mc^2$, where 
$E_{n,l}^{KG}$ are the bound energies for the Klein-Gordon equation. The explicit form
of $K_{n,l}$ is given by eq.~(\ref{KG-1}).

\subsubsection{Bound state solutions in a ``pure'' gravitational field}
\label{pureG}

We now set $A^\mu = 0$ and 
\begin{equation}
V(\mathbf{r}) = -\frac{GmM}{r} = -\hbar c \frac{\tilde{\alpha}}{r},
\label{Vgravitation}
\end{equation}
where $G$ is the gravitational constant, $M$ is the mass of the 
``fixed" particle generating the gravitational field that acts on $m$,
and $\tilde{\alpha}=GmM/\hbar c$ is  the ``gravitational fine structure
constant''.  

Inserting Eq~(\ref{Vgravitation}) into eq.~(\ref{GeneralEq10}), and using that
$A^\mu =0$, the stationary equation we get is given by eq.~(\ref{t-ind}), whose
bound state energies are given by eq.~(\ref{Klcse}) with $\alpha$ changed to
$\tilde{\alpha}$. Note that for the Hydrogen atom the leading term in the energy is of the order 
$\tilde{\alpha}^2mc^2 \sim 10^{-83}mc^2$, irrelevant if compared to $\alpha^2mc^2 \sim 10^{-5}mc^2$,
the leading contribution coming	from the Coulomb potential.

\subsubsection{Bound state solutions in static electric and gravitational fields}

It is very instructive to study the case in which we have a charged scalar particle described by
the Lorentz covariant Schr\"odinger equation minimally coupled to the electromagnetic field and 
also subjected to a gravitational field. The leading contribution coming from a static gravitational field
is obtained setting $V(\mathbf{r})$ as given by eq.~(\ref{Vgravitation}). The Coulomb potential is 
modeled via eqs.~(\ref{c1}) and (\ref{c2}). 

As we will show below, the bound energies to this problem are not simply
independent contributions coming from the electromagnetic and gravitational fields.
We will see that the way the system ``feels'' the gravitational field \textit{depends} on the electromagnetic
field acting on it. In other words, the bound energies depend not only on 
isolated functions of $\alpha$ and $\tilde{\alpha}$ but also on functions of the
product $\alpha\tilde{\alpha}$. 
The simultaneous presence of both fields leads to bound energies that couples the electromagnetic and gravitational fields
in a non-trivial way.

Inserting eqs.~(\ref{c1}), (\ref{c2}), and (\ref{Vgravitation}) into (\ref{GeneralEq10}), and employing the
ansatz (\ref{stationary}), the stationary Lorentz covariant Schr\"odinger equation can be written as
\begin{eqnarray}
\nabla^2\psi(\mathbf{r}) &=& -\frac{1}{\hbar^2c^2}\left\{  \left[ K-\mathcal{V}(\mathbf{r})\right]^2 \right. \nonumber \\
&& \left. +2 mc^2 \left[K - \mathcal{V}(\mathbf{r}) - V(\mathbf{r})\right] \right\}\psi(\mathbf{r}).
\label{eqCG}
\end{eqnarray}

If we express the Laplacian operator in spherical coordinates and 
write $\psi(\mathbf{r})= \psi_{n,l,m}(\mathbf{r})$ $= u_{n,l}(r)Y^m_l(\theta,\phi)$, 
where $r=|\mathbf{r}|$ and $\theta$ and $\phi$ are, respectively, 
the polar and azimuthal angles of the spherical-polar coordinates,  
eq.~(\ref{eqCG}) decouples into the following two equations:
\begin{eqnarray}
\frac{1}{u_{n,l}}\frac{d}{dr}\left(  r^2 \frac{du_{n,l}}{dr} \right) +\frac{r^2f(r)}{\hbar^2c^2} &=& l(l+1), 
\label{radial} \\
\frac{1}{Y^m_l}\left[  \frac{1}{\sin\theta}\frac{\partial}{\partial \theta}\left( \sin\theta \frac{\partial Y^m_l}{\partial \theta}\right) 
 + \frac{1}{\sin^2\theta}\frac{\partial^2Y^m_l}{\partial\phi^2} \right] \!&\!=\!& \!-l(l+1). 
 \label{angular} 
\end{eqnarray}
The solutions to eq.~(\ref{angular}) are the spherical harmonics \cite{gri95}, where $l=0,1,2,\ldots$ and $m=0, \pm 1, \pm 2,\ldots, \pm l$.

The radial equation (\ref{radial}), where
\begin{equation}
f(r) = \left[ K-\mathcal{V}(\mathbf{r})\right]^2 +2 mc^2 \left[K - \mathcal{V}(\mathbf{r}) - V(\mathbf{r})\right],
\end{equation}
becomes 
\begin{equation}
\left[ \frac{d^2}{dr^2} - \frac{l(l+1)}{r^2} + \frac{f(r)}{\hbar^2c^2}\right]R_{n,l} = 0
\label{radial2}
\end{equation}
after using that 
\begin{displaymath}
u_{n,l}(r) = \frac{R_{n,l}(r)}{r}.
\end{displaymath}
Computing explicitly $f(r)$ we can rewrite eq.~(\ref{radial2}) as
\begin{eqnarray}
\left[ \frac{d^2}{dr^2} - \frac{l(l+1)-\alpha^2}{r^2} \right. &+& \left. \frac{2(E\alpha + mc^2\tilde{\alpha})}{\hbar c r}  
- \frac{m^2c^4-E^2}{\hbar^2c^2}\right]R_{n,l} = 0, \nonumber \\
& & \, \label{radial3}
\end{eqnarray}
where
\begin{equation}
E = K + mc^2.
\label{kinetic2}
\end{equation}

We will restrict ourselves to bound energies such that $-2mc^2 < K < 0$. This is equivalent to
working with $|E|<mc^2$. With such values for $E$, the constant 
\begin{equation}
B = \sqrt{\frac{4(m^2c^4-E^2)}{\hbar^2c^2}} 
\end{equation}
is a real number. Now, defining
\begin{eqnarray}
\varrho &=& B r, \\
\lambda &=& \frac{2(E\alpha + mc^2\tilde{\alpha})}{B \hbar c}, 
\label{lambda2}\\
\mu &=& \sqrt{(l+1/2)^2-\alpha^2},
\end{eqnarray}
eq.~(\ref{radial3}) becomes 
\begin{equation}
\left[ \frac{d^2}{d\varrho^2} - \frac{\mu^2-1/4}{\varrho^2} + \frac{\lambda}{\varrho} -\frac{1}{4} \right]R_{n,l}(\varrho) = 0.
\label{radial4}
\end{equation}

Equation (\ref{radial4}) is formally the same one obtains when solving the Klein-Gordon equation 
minimally coupled to the electromagnetic field \cite{gre00}. The only difference between the present problem
and the one given in ref. \cite{gre00} is the absence of the term proportional to $\tilde{\alpha}$
in eq.~(\ref{lambda2}).  Therefore, we can follow the same steps given in ref. \cite{gre00} to arrive at the
following solution to eq.~(\ref{radial4}),
\begin{equation}
R_{n,l}(\varrho)=Ne^{-\varrho/2}\varrho^{\mu+1/2}\,_1F_1(a,b;\varrho),
\end{equation}
where $\,_1F_1(a,b;\varrho)$ is the confluent hypergeometric function, $N$ a normalization constant, 
and 
\begin{eqnarray}
b &=& 2\mu + 1, \\
a &=& \mu +\frac{1}{2} - \lambda.
\label{lambda3}
\end{eqnarray}

The function $\,_1F_1(a,b;\varrho) \rightarrow \infty$ when $\varrho \rightarrow \infty$
and thus in order to
get a finite and normalizable solution, we have to truncate the series that defines 
the confluent hypergeometric function. This is achieved if \cite{gre00}
\begin{equation}
a = -n' = 0,1,2,3,\ldots
\label{nprime}
\end{equation}

Equation (\ref{nprime}) leads to the quantization of the bound energies, whose values are given by solving for $E$
eq.~(\ref{lambda3}),
\begin{equation}
\mu +\frac{1}{2} - \lambda = -n',
\end{equation}
where $\lambda$ is given by eq.~(\ref{lambda2}). Picking the solution in which
$|E|<mc^2$ and using eq.~(\ref{kinetic2}) we finally get
%
\begin{equation}
K_{n,l} = -mc^2\left\{1+ \frac{\alpha\tilde{\alpha}}{w^2+\alpha^2} - 
\left[ \left(1 + \frac{\alpha^2}{w^2}\right) \left(1 + \frac{\tilde{\alpha}^2}{w^2+\alpha^2-\tilde{\alpha}^2}\right) \right]^{-1/2} \right\},
\label{knl}
\end{equation}
%
where
\begin{equation}
w = n -(l+1/2) + \sqrt{(l+1/2)^2 - \alpha^2}
\label{w}
\end{equation}
and
\begin{equation}
n = n'+l+1.
\label{n}
\end{equation}
Equation (\ref{n}) together with the fact that $l=0,1,2,\ldots$
lead to $n=1,2,3, \ldots$ and $l=0,1,2, \ldots, n-1$.

It is not difficult to see that if $\tilde{\alpha}=0$ we recover the pure electromagnetic solution
(\ref{KG-1}) and that if $\alpha=0$ a simple calculation shows that we get back to the pure gravitational solution, 
namely, eq.~(\ref{Klcse}) 
with $\alpha$ changed to $\tilde{\alpha}$. The more interesting scenario occurs when both the 
electric and gravitational fields are turned on. In this case the effect of the gravitational field acting on the
particle $m$ is affected by the presence of the electric field. This is clear looking at eq.~(\ref{knl}), where
we see that the bound energies depend also on products of $\alpha\tilde{\alpha}$. 

We can better appreciate this feature if we expand eq. (\ref{knl}) in terms of $\alpha$ and
$\tilde{\alpha}$, 
\begin{eqnarray}
K_{n,l} &=& -\frac{mc^2\alpha^2}{2n^2} - \frac{mc^2\alpha^4}{2n^4}\left( \frac{n}{l+1/2}-\frac{3}{4}\right) + \mathcal{O}(\alpha^6) \nonumber \\
&& -\frac{mc^2\alpha\tilde{\alpha}}{n^2} + \mathcal{O}(\alpha^3\tilde{\alpha})
-\frac{mc^2\tilde{\alpha}^2}{2n^2} +  \mathcal{O}(\tilde{\alpha}^4).
\label{expansion}
\end{eqnarray}
The first two terms in the right hand side of eq.~(\ref{expansion}) are, respectively, the non-relativistic energy and the dominant 
relativistic correction due to the presence of a Coulomb potential alone, as discussed in section \ref{pureE}. 
The third term in the second line of eq.~(\ref{expansion}) is the non-relativistic
energy due to the presence of a gravitational field alone (cf. section \ref{pureG}).
The first term in the second line of eq.~(\ref{expansion}),
\begin{equation}
-\frac{mc^2\alpha\tilde{\alpha}}{n^2},
\label{mixedTerm}
\end{equation}
is the dominant contribution to the bound energy coming from the simultaneous
presence of an electric field and a gravitational field. For the Hydrogen atom
we have 
\begin{displaymath}
\alpha\tilde{\alpha}mc^2 \sim 10^{-44}mc^2 \sim \alpha^{20}mc^2,
\end{displaymath}
which is too small to be detected using today's technology. On the other hand,
the simultaneous existence of electric and gravitational fields lead to a contribution to the
bound energies about 40 orders of magnitude greater than the expected one 
due to the presence of a gravitational field alone, which 
is of order $\tilde{\alpha}^2mc^2 \sim 10^{-83}mc^2$. 

In order to have a gravitational contribution to the bound energies 
of order $\alpha^4$, which is the order of magnitude of the
dominant relativistic correction to the bound energies due to an electric field, 
we would need $\tilde{\alpha}\sim \alpha^3$. This is achieved with a ``heavy proton'' of
mass of the order of $10^8kg$.

From a fundamental point of view, however, the fact that this simple model leads
to a non-trivial influence of the gravitational field on how an electromagnetic field
acts upon a particle of mass $m$ deserves further investigation. It is not unlikely, as we show in
section \ref{scattering}, that 
the scattering between two charged particles, one of which is very massive, might lead
to detectable predictions within the present model.\footnote{Note that by solving the standard
Schr\"odinger equation for a particle subjected simultaneously to static electric and gravitational fields
we get for the bound energies $K_n^{Sch} = -mc^2\alpha^2/(2n^2) -mc^2\tilde{\alpha}^2/(2n^2)
-mc^2\alpha\tilde{\alpha}/n^2$. In other words, the mixed term given by eq.~(\ref{mixedTerm}) is already
present at the non-relativistic level. Equation (\ref{knl}), on the other hand, 
is the relativistic bound energies $K_{n,l}$ we obtain by solving the Lorentz covariant Schr\"odinger equation
for the same physical system. 
As can be seen by looking at eq.~(\ref{expansion}), $K_{n,l}$ tends to $K_n^{Sch}$
for small values of $\alpha$ and $\tilde{\alpha}$.}

\subsubsection{Bound state solutions in a static electric field and a constant gravitational field}

If we have a Coulomb field minimally coupled to a charged particle of mass $m$, as given by eqs.~(\ref{c1}) and (\ref{c2}), 
together with a constant gravitational field acting on it, which is achieved by setting $V$
a constant in eq.~(\ref{Vgravitation}), we can repeat the steps detailed above and get the following expression
for the bound energies,
\begin{equation}
K_{n,l} = -mc^2\left[1 - 
\left(1 + \frac{\alpha^2}{w^2}\right)^{-1/2} 
\left(1 + \frac{2V}{mc^2}\right)^{1/2} \right],
\label{Gconst}
\end{equation}
where $w$ was defined in eq.~(\ref{w}).

Looking at eq.~(\ref{Gconst}), we see that it is similar to eq.~(\ref{KG-1}), 
the pure electrostatic solution to the Lorentz covariant Schr\"o\-din\-ger equation. 
However, the term $(1 + \alpha^2/w^2)^{-1/2}$ does not appear alone now. 
It comes multiplied by 
$[1 + 2V/(mc^2)$ $]^{1/2}$, 
whose origin stems from the presence of a constant gravitational field.
Also, if $V\sim mc^2\alpha^2$ 
we would get corrections to the bound energies of order
$\alpha^4$, which could in principle be detected.

\subsubsection{Bound state solutions in a gravitational field and a constant electric field}

For completeness we present the bound energies when we have a constant electric field,
which is achieved by setting eq.~(\ref{vr2}) to a constant, and using eq.~(\ref{Vgravitation})
to model the gravitational field. The bound energies in this case are
\begin{equation}
K_{n,l} = \mathcal{V} -mc^2 \left(1 - 
\sqrt{1 - \frac{\tilde{\alpha}^2}{n^2}}\right),
\end{equation}
where $n=1,2,3,\ldots$ This case is less interesting since it is simply the 
pure gravitational case with bound energies displaced by $\mathcal{V}$, the
electric potential energy due to a constant electric field.

\subsubsection{Solutions in constant electric and gravitational fields}

We are now interested in the plane wave solutions arising from solving
eq.~(\ref{GeneralEq10}) when both the electric and gravitational fields 
acting on the particle are constant. The time-independent equation we need to solve 
is given by eq.~(\ref{eqCG}), with $\mathcal{V}$ and $V$ treated as two constants.
Inserting the ansatz
\begin{displaymath}
\psi(\mathbf{r})= e^{\frac{i}{\hbar}\mathbf{p}\cdot \mathbf{r}} 
\end{displaymath}
into eq.~(\ref{eqCG}) we get
\begin{equation}
K_{\pm} = \mathcal{V} -mc^2 \pm \sqrt{p^2c^2 + m^2c^4 + 2mc^2 V}, 
\label{kpmCG}
\end{equation}
where $p=|\mathbf{p}|$.
Defining $m'=m\left(1+\frac{2V}{mc^2}\right)^{1/2}$, $E = K_\pm + mc^2$,
and squaring eq.~(\ref{kpmCG}), we get 
\begin{equation}
p^\mu p_\mu = m'^2c^2,
\label{pmuCG}
\end{equation}
where 
\begin{displaymath}
p^\mu = \left(\frac{E-\mathcal{V}}{c},\mathbf{p}\right).
\end{displaymath}

Looking at eq.~(\ref{kpmCG}) we see that the constant electric field
affects $K_+$ and $K_-$ equally, displacing both energies by the same 
quantity $\mathcal{V}$. On the other hand, the constant gravitational field
affects $K_+$ and $K_-$ differently, leading to the same results already
discussed in section \ref{vconst}. 

It is worth mentioning the following two points. First,  we can make $K_+=-K_-$
working with the Lorentz covariant Schr\"odinger equation minimally coupled to
a constant electric field. Specifically,
if we set $\mathcal{V}=mc^2$ we accomplish this task. This feature cannot
be achieved without working in the minimal coupling scenario or by applying an external
gravitational field, as we can see by looking at eq.~(\ref{kpmCG}). See also
sections \ref{vzero} and \ref{vconst} for more details.

Second, eq.~(\ref{kpmCG}) leads to a dispersion relation whose dependence on the value of 
the external constant gravitational field affects a particle's wave packet
dynamics in a non-trivial way. 
This point becomes clearer inserting $K(p)$ as given by eq.~(\ref{kpmCG})
into eq.~(\ref{wp}), which gives the time evolution of a one-dimensional wave packet
in the presence of constant electric and gravitational fields. 
The wave packet dynamics will differ from that of a free particle ($\mathcal{V}=V=0$) 
due to the presence of the term $2mc^2V$ inside the square root of the dispersion relation (\ref{kpmCG}).
This factor introduces non-trivial changes when we integrate eq.~(\ref{wp}) to obtain the 
wave packet dynamics. Note that the effect of a constant electric field is trivial. It does not 
change the wave packet dynamics since it only adds a global phase to $\Psi(x,t)$.

\subsection{Coulomb and gravitational scattering}
\label{scattering}

Our goal here is to compute perturbatively the differential cross-section for a
beam of charged particles of mass $m$ hitting charged particles of mass $M$ at rest. 
The charges of the incident and target particles are respectively $q$ and $Q$, 
the kinetic energy of the incident particles are $K<mc^2$,
and we model the interaction of those particles
via static electric and gravitational fields. 

The stationary Lorentz covariant Schr\"odinger equation describing the simultaneous action of the Coulomb
and gravitational fields is given by eq.~(\ref{eqCG}), which can be rewritten as
\begin{equation}
\nabla^2\psi(\mathbf{r}) + k^2\psi(\mathbf{r}) = J(\mathbf{r})\psi(\mathbf{r}),
\label{scat1}
\end{equation}
with
\begin{eqnarray}
k^2 &=& (E^2-m^2c^4)/(\hbar^2c^2), \label{scat2} \\
J(\mathbf{r}) & = & \frac{-[\mathcal{V}(\mathbf{r})]^2+2E\mathcal{V}(\mathbf{r})+2mc^2V(\mathbf{r})}
{\hbar^2c^2}. \label{scat3}
\end{eqnarray}
For the present problem we have
\begin{eqnarray}
\mathcal{V}(\mathbf{r}) & = & \frac{qQ}{4\pi\epsilon_{\!\,_0} r} = -\hbar c \frac{\alpha}{r} = \mathcal{V}(r), \label{v1}\\
V(\mathbf{r}) & = & -\frac{GmM}{r} = -\hbar c \frac{\tilde{\alpha}}{r} = V(r) \label{v2},
\end{eqnarray}
where $|\mathbf{r}|=r$. Note that now we have that 
$\alpha=-qQ/(4\pi\epsilon_{\!\,_0} \hbar c)$. 
For $\alpha>0$ the charges attract each other and for $\alpha<0$ we have a repulsive
electrostatic interaction. The same interpretation for the sign of $\tilde{\alpha}$
applies. For ordinary matter, we always have $\tilde{\alpha}>0$.

Using elementary techniques we can transform eq.~(\ref{scat1}) to its integral form \cite{gri95}, 
\begin{equation}
\psi(\mathbf{r}) = \psi_0(\mathbf{r}) + \int\mathcal{G}(\mathbf{r} - \mathbf{r'})J(\mathbf{r'})\psi(\mathbf{r'})d^3r',
\label{scat4}
\end{equation}
where  $\psi_0(\mathbf{r})$ is the solution to eq.~(\ref{scat1}) with $J(\mathbf{r})=0$ and  
\begin{equation}
\mathcal{G}(\mathbf{r}) = -\frac{e^{ik|\mathbf{r}|}}{4\pi|\mathbf{r}|}
\end{equation}
is the Green's function of the Helmholtz equation, i.e.,
solution to $\nabla^2\mathcal{G}(\mathbf{r}) + k^2\mathcal{G}(\mathbf{r})=\delta^{(3)}(\mathbf{r})$,
with $\delta^{(3)}(\mathbf{r})$ being the three dimensional Dirac delta function.

Assuming we are dealing with localized potentials situated at the origin ($\mathbf{r'}=0$) and noting
that in a scattering experiment we want the wave function far away from the target, i.e., 
$|\mathbf{r}| \gg |\mathbf{r'}|$, we can write the Green's function as \cite{gri95}
\begin{equation}
\mathcal{G}(\mathbf{r} - \mathbf{r'}) = -\frac{e^{ikr}}{4\pi r}e^{-i\mathbf{k}\cdot\mathbf{r'}}
+ \mathcal{O}\left(\frac{1}{|\mathbf{r}|^2}\right),
\label{scat5}
\end{equation}
where $\mathbf{k} = k\mathbf{\hat{r}}$.
Moreover, considering that the incident particles move along the $z$ axis with a well defined momentum and energy, 
the stationary wave function describing them can be written as 
\begin{equation}
 \psi_0(\mathbf{r}) = e^{i\mathbf{k'}\cdot \mathbf{r}},
\end{equation}
where $\mathbf{k'} = k\mathbf{\hat{z}}$. 

Now, employing the first Born approximation (weak potential approximation), 
which implies that
\begin{equation}
\psi(\mathbf{r}) \approx \psi_0(\mathbf{r})
\end{equation}
inside the integral of eq.~(\ref{scat4}),
we get after eqs.~(\ref{scat3}) and (\ref{scat5}),
\begin{equation}
\psi(\mathbf{r}) = e^{ikz} + f(\theta)\frac{e^{ikr}}{r}. 
\end{equation}
The scattering amplitude is given by
\begin{equation}
f(\theta) = 
\frac{1}{\hbar^2c^2\kappa}\int_0^\infty\!\! r\sin(\kappa r)[\mathcal{V}(r)]^2dr 
- \frac{2E}{\hbar^2c^2\kappa}\int_0^\infty\!\! r\sin(\kappa r)\mathcal{V}(r)dr 
- \frac{2m}{\hbar^2\kappa}\int_0^\infty\!\! r\sin(\kappa r)V(r)dr,
\label{scat6}
\end{equation}
where
\begin{equation}
\kappa = 2k\sin(\theta/2)
\label{kappa}
\end{equation}
and $\theta$ is the polar angle with respect to the incident $z$ direction,
defining the polar angular position of the detector measuring the scattered particles. 
To arrive at eq.~(\ref{scat6}) we relied on the fact that the potentials are spherically
symmetric. 
This allowed us to straightforwardly integrate over the solid angle $d\Omega'=\sin(\theta')d\theta' d\phi'$, where 
$d^3r'=r'^2dr'd\Omega'$, $\boldsymbol{\kappa}\cdot \mathbf{r}=\kappa r \cos\theta$, and 
$\boldsymbol{\kappa}=\mathbf{k'}-\mathbf{k}$ in eq.~(\ref{scat4}).

It is worth mentioning that the origin of the first and second terms of eq.~(\ref{scat6})
can be traced back to applying the minimal coupling prescription 
to the Lorentz covariant Schr\"odinger equation, while the third one is related to the existence of 
an external potential whose origin is not electromagnetic. Moreover, for small energies ($E\ll mc^2$) 
the second and third terms of eq.~(\ref{scat6}) tend, at the first Born approximation level,
to the scattering amplitude obtained by solving 
the non-relativistic Schr\"odinger equation for a particle 
subjected to the external potential $\mathcal{V}(r)+V(r)$. The first term, 
on the other hand, is absent from any non-relativistic treatment of this problem, being
a purely relativistic contribution to the scattering amplitude. 

By inserting eqs.~(\ref{v1}) and (\ref{v2}) into (\ref{scat6}) we can explicitly compute the three
remaining integrals. This leads to 
\begin{equation}
f(\theta) = \frac{\pi\alpha^2}{2\kappa} + \frac{2E\alpha}{\hbar c \kappa^2} + \frac{2mc^2\tilde{\alpha}}{\hbar c \kappa^2} 
=\frac{\pi\hbar c\alpha^2}{4\sqrt{E^2-m^2c^4}\sin(\theta/2)} \!+\! 
\frac{\hbar c (E\alpha+mc^2\tilde{\alpha})}{2(E^2-m^2c^4)\sin^2(\theta/2)}, 
\label{ftheta}
\end{equation}
where we used eqs.~(\ref{scat2}) and (\ref{kappa}) to arrive at the last line. 
Note that we have a term that is second order in $\alpha$ in eq.~(\ref{ftheta}). This means that
we must also compute the second Born approximation to obtain all the other second order
terms contributing to the scattering amplitude.

With the aid of eq.~(\ref{ftheta}), the differential cross-section can be written as \cite{gri95}
\begin{equation}
\frac{d\sigma}{d\Omega} = |f(\theta)|^2.
\end{equation}

In order to compare eq.~(\ref{ftheta}) with its non-relativistic version, the one coming from
solving the standard Schr\"o\-din\-ger equation at the same level of approximation, and to properly
identify its dominant relativistic corrections, it is convenient to write eq.~(\ref{ftheta})
in terms of the kinetic energy of the particle,
\begin{equation}
K = E - mc^2. \label{kinetic3}
\end{equation}

Using eq.~(\ref{kinetic3}), expanding up to first order in $K/(mc^2)$, and retaining terms of lowest order 
in $\alpha$ and $\tilde{\alpha}$, eq.~(\ref{ftheta}) becomes,
\begin{equation}
f(\theta) = \frac{\hbar c}{4K\sin^2(\theta/2)}\left[ \alpha + \tilde{\alpha} +\frac{K}{2mc^2}(\alpha - \tilde{\alpha}) \right]
+ \mathcal{O}\left[\alpha^o\tilde{\alpha}^s\left(\frac{K}{mc^2}\right)^2\right] 
+ \mathcal{O}(\alpha^n\tilde{\alpha}^q), 
\label{scatLS}
\end{equation}
where $o,s,n,q$ are integers such that $o+s = 1$ and $n+q\geq 2$. If we solve the non-relativistic Schr\"odinger
equation for a particle subjected to the potential $\mathcal{V}(r)+V(r)$, we get for the first Born approximation
\begin{eqnarray}
f^{Sch}(\theta) = \frac{\hbar c}{4K\sin^2(\theta/2)}( \alpha + \tilde{\alpha} ) 
= \frac{1}{4K\sin^2(\theta/2)}\left( -\frac{qQ}{4\pi\epsilon_{\!\,_0}} + GmM \right).
\label{scatSch}
\end{eqnarray}
Comparing eqs.~(\ref{scatLS}) and (\ref{scatSch}) we see that the dominant relativistic correction to the
scattering amplitude is 
\begin{eqnarray}
f^{Rel}(\theta) = \frac{\hbar c}{8mc^2\sin^2(\theta/2)}(\alpha - \tilde{\alpha}) 
= \frac{1}{8mc^2\sin^2(\theta/2)}\!\left(\! -\frac{qQ}{4\pi\epsilon_{\!\,_0}} - GmM \right). 
\end{eqnarray}
It is interesting to observe that the leading relativistic correction to the scattering amplitude due to the 
gravitational field has a different sign when compared to the non-relativistic term  
($\tilde{\alpha} \rightarrow -\tilde{\alpha}$). Loosely speaking, the relativistic correction 
looks like a ``negative'' mass interacting with a positive one, i.e., we have an effective 
gravitational repulsive force.

It is important to notice that the pieces of eq.~(\ref{scatLS}) proportional to $\tilde{\alpha}$ are 
relevant only if $|\alpha| \approx |\tilde{\alpha}|$. This implies that
\begin{equation}
M \approx \frac{1}{4\pi\epsilon_{\!\,_0}G}\frac{qQ}{m}.
\label{massM}
\end{equation}

For a beam of electrons incident on a target composed of particles of mass $M$ and with the same charge of the electron, 
eq.~(\ref{massM}) leads to $M\approx 3.79 \times 10^{12}$ kg $\approx 6.35 \times 10^{-13}M_\oplus$, where
$M_\oplus$ is the mass of the Earth. Looking at eq.~(\ref{massM}) we see that we can decrease the 
mass $M$ by increasing the mass of the incident particles. For instance, for incident particles with
$m = 1$ $\mu$g (one micro-gram) we get $M\approx 3.45$ $\mu$g if both particles have the same charge of the electron. Moreover,
by tuning the values of the charges $q$ and $Q$, we can also obtain manageable values for the masses $m$ and $M$ that might lead to
an experimental test of eq.~(\ref{scatLS}) if we build on 
state-of-the-art experimental techniques that
can detect the gravitational attraction between millimeter-sized particles with 
masses of the 
order of $100$ mg \cite{asp20}.

Finally, if we look at eq.~(\ref{ftheta}), we see that it is possible to completely suppress the first order contribution to 
the scattering amplitude if
\begin{equation}
E\alpha+mc^2\tilde{\alpha} = 0.
\end{equation}
Solving the previous equation by noting that $E=\gamma mc^2$, with $\gamma = 1/\sqrt{1-v^2/c^2}$ and $v$ being the speed of the incident particle,
we get
\begin{equation}
\frac{mM}{qQ} = \frac{\gamma}{4\pi\epsilon_{\!\,_0}G},
\label{massM2}
\end{equation}
after inserting the definitions of $\alpha$ and $\tilde{\alpha}$.

This means that by properly setting $m$, $M$, $q$, and $Q$ such that eq.~(\ref{massM2})
is satisfied, only second order effects will be present in the scattering amplitude. This 
feature can be employed to indirectly test the influence of gravitation at the quantum level.
If after the proper tuning of the masses and charges no first order scattering effect is seen,
this can only be attributed to the concomitantly action of the static gravitational and electric fields
between the incident and target particles. And any observed second order scattering can be compared to
the predictions coming from the Lorentz covariant Schr\"odinger equation: the first term
of eq.~(\ref{ftheta}) and the ones coming from the second Born approximation, two of which are 
proportional to $\tilde{\alpha}^2$ and $\alpha\tilde{\alpha}$, genuine  
quantum contributions to the scattering amplitude 
related to the presence of a gravitational field.

\subsection{Justifying the way we modeled the gravitational interaction}

We start assuming 
that we are dealing only with stationary electromagnetic fields and in the minimal coupling
scenario. Specifically,
we restrict ourselves to the Coulomb potential and set $V=0$ in eq.~(\ref{GeneralEq10}). 
In this case the four-vector potential is
given by eq.~(\ref{c1}) and eq.~(\ref{GeneralEq10}) can be written as 
\begin{eqnarray}
-\frac{\hbar^2}{2m}\nabla^2\Psi + \mathcal{V}\Psi &=& i\hbar\frac{\partial\Psi}{\partial t} 
- \frac{\hbar^2}{2mc^2}\frac{\partial^2\Psi}{\partial t^2} + \frac{i\hbar \mathcal{V}}{mc^2}\frac{\partial\Psi}{\partial t} 
+ \frac{\mathcal{V}^2}{2mc^2}\Psi.
\label{Srel}
\end{eqnarray}

Using the ansatz (\ref{stationary}) it is not difficult to see that 
$\partial \Psi/\partial t = -iK \Psi/\hbar$ and that 
$\partial^2 \Psi/\partial t^2 = -K^2 \Psi/\hbar^2$, where $K$ is the bound 
energy of the system. Using experimental data, as Schr\"odinger probably did, or relying, for example, on the Bohr model,
we know that 
\begin{displaymath}
|K| \sim \frac{mc^2\alpha^2}{2}.
\end{displaymath}
This leads to
\begin{eqnarray}
\left| \frac{\partial \Psi}{\partial t} \right| \sim \frac{mc^2\alpha^2}{2\hbar}|\Psi| & \mbox{and}& 
\left| \frac{\partial^2 \Psi}{\partial t^2} \right| \sim \frac{m^2c^4\alpha^4}{4\hbar^2}|\Psi|.
\label{derivadast}
\end{eqnarray}

Moreover, using the Bohr radius $r_0$, where
\begin{displaymath}
\frac{1}{r_0} = \frac{mc\alpha}{\hbar},
\end{displaymath}
we have
\begin{equation}
|\mathcal{V}| \sim mc^2\alpha^2.
\label{energiaPotV}
\end{equation}

Thus, using eqs.~(\ref{derivadast}) and (\ref{energiaPotV}), the dominant order of magnitude for the 
four terms on the right hand side of eq.~(\ref{Srel}) are
\begin{eqnarray}
\left| i\hbar \frac{\partial \Psi}{\partial t}\right| &\sim& \frac{mc^2\alpha^2}{2}|\Psi|, \label{alpha2} \\
\left| \frac{\hbar^2}{2mc^2} \frac{\partial^2 \Psi}{\partial t^2}\right| &\sim& \frac{mc^2\alpha^4}{8}|\Psi|, \\
\left| \frac{i\hbar\mathcal{V}}{mc^2} \frac{\partial \Psi}{\partial t}\right| &\sim& \frac{mc^2\alpha^4}{2}|\Psi|, \\
\left| \frac{\mathcal{V}^2}{2mc^2} \frac{\partial \Psi}{\partial t}\right| &\sim& \frac{mc^2\alpha^4}{2}|\Psi|.
\end{eqnarray}
Of the four terms above, only the first one is of order $\alpha^2$ while the remaining three are of order $\alpha^4$.
This fact together with eq.~(\ref{energiaPotV}) fully justify 
why we can neglect as a first approximation all but the first term in the right hand side of eq.~(\ref{Srel}). 
Proceeding in such a way, we obtain the non-relativistic
Schr\"odinger equation,
\begin{eqnarray}
-\frac{\hbar^2}{2m}\nabla^2\Psi + \mathcal{V}\Psi &=& i\hbar\frac{\partial\Psi}{\partial t} + \mathcal{O}(mc^2\alpha^4|\Psi|).
\end{eqnarray}

Furthermore, the term $\mathcal{V}\Psi$ is the dominant term when it comes to the presence of a static electric field acting on
a charged particle. Thus, it is not unreasonable to assume that $V\Psi$ should be the dominant term when we have other 
static fields acting on the particle, the Newtonian gravitational field being an example (see 
sections \ref{mincoup} and \ref{scattering}). And in this case, since the ``gravitational fine structure constant'' $\tilde{\alpha}$ is much smaller
than $\alpha^2$, we must work with all terms appearing in eq.~(\ref{Srel}) when investigating the simultaneous action of
electromagnetic and gravitational fields on a scalar charged particle, similarly to what we have done in sections \ref{mincoup} and \ref{scattering}.\footnote{As we show in this
work, antiparticles apparently possess negative masses and we can adjust the present theory such that particles and
antiparticles repel or attract  each other gravitationally. Therefore, all the results
above where we considered two particles interacting gravitationally 
can be readily extended to the case where we have a particle and an antiparticle repelling each other
by properly choosing the sign of $\tilde{\alpha}$.}

\section{Lagrangian formulation}
\label{cft}

We now start the second part of this paper. In this section we will 
show the Lagrangian formalism associated with the Lorentz covariant Schr\"odinger equation.
The wave function $\Psi$ and its complex conjugate $\Psi^*$ will be considered two 
independent ``classical'' fields. Our goal will be to develop the main features of the classical 
field theory associated with the Lagrangian that leads to the Lorentz covariant Schr\"odinger equation,
in particular those features needed to prepare the ground for 
section \ref{qft}. In that section we will implement the ``second''
quantization of the classical fields here studied and develop the 
quantum field theory of the Lorentz covariant Schr\"odinger equation.  
We also show in section \ref{cft} how smoothly one goes from the relativistic 
conserved quantities to the ones derived from the non-relativistic Schr\"odinger equation.
By simply taking the $c\rightarrow \infty$ limit we promptly recover the non-relativistic
quantities from the relativistic ones.

Treating $\Psi$ and $\Psi^*$ as two independent fields, we define the
Lagrangian describing these two fields as
\begin{equation}
L = \int d^3x \mathcal{L}(\Psi,\Psi^*,\partial_\mu \Psi,\partial_\mu \Psi^*).
\end{equation}
Here $\mathcal{L}$ is the Lagrangian density, assumed to depend on the fields and at most on
their first derivatives, $d^3x = dx^1dx^2dx^3$ is the
infinitesimal spatial volume, and, unless stated otherwise, the integration is taken over all space.
We assume that the fields and their derivatives vanish at the boundaries of integration. Note that
$\mathcal{L}$ is such that $\int d^3x \mathcal{L}$ has the dimension of energy.

We define the action as
\begin{equation}
S = \int dt L = \frac{1}{c} \int d^4x \mathcal{L}(\Psi,\Psi^*,\partial_\mu \Psi,\partial_\mu \Psi^*),
\label{action}
\end{equation}
where $d^4x = dx^0d^3x$ is the infinitesimal four-volume. Applying the variational principle to
the action, i.e., demanding that the infinitesimal variation of the action vanishes,
\begin{equation}
\delta S = 0,
\end{equation}
we get the following Euler-Lagrange equations,
\begin{eqnarray}
\frac{\partial \mathcal{L}}{\partial \Psi} &=& \partial_\mu\left( \frac{\partial \mathcal{L}}{\partial(\partial_\mu \Psi)}\right), \label{euler1} \\
\frac{\partial \mathcal{L}}{\partial \Psi^*} &=& \partial_\mu\left( \frac{\partial \mathcal{L}}{\partial(\partial_\mu \Psi^*)}\right), 
\label{euler2}
\end{eqnarray}
where the Einstein summation convention is implied for repeated indexes. In obtaining eqs.~(\ref{euler1}) and (\ref{euler2})
we varied the fields $\Psi$ and $\Psi^*$ as two independent variables and assumed 
that they vanished at the boundaries.

When inserted into eqs.~(\ref{euler1}) and (\ref{euler2}), 
the simplest Lagrangian density leading to the Lorentz covariant Schr\"o\-di\-nger equation and its complex conjugate is
\begin{equation}
\mathcal{L}_{asym} =  \frac{\hbar^2}{2m}\partial_\mu\Psi\partial^\mu\Psi^* + i\hbar c \Psi^* \partial_0 \Psi - V |\Psi|^2, 
\label{asym}
\end{equation}
where $|\Psi|^2=\Psi\Psi^*$ and for simplicity we write $(\partial_\mu\Psi)(\partial^\mu\Psi^*)$ $=$ 
$\partial_\mu\Psi\partial^\mu\Psi^*$.
If we take the non-relativistic limit we recover the Lagrangian density that gives the Schr\"odinger equation,
i.e., $\lim_{c\rightarrow \infty} \mathcal{L}_{asym} =  \mathcal{L}_{Schr}$, where
\begin{equation}
\mathcal{L}_{Schr} = \frac{\hbar^2}{2m}\partial_j\Psi\partial^j\Psi^* + i\hbar \Psi^* \partial_t \Psi - V |\Psi|^2 =
-\frac{\hbar^2}{2m}\nabla\Psi\cdot \nabla \Psi^* + i\hbar \Psi^* \frac{\partial \Psi}{\partial t}  - V |\Psi|^2.
\end{equation}

Note that Lagrangian  
density (\ref{asym}) is not symmetric in the fields $\Psi$ and $\Psi^*$ due to the second term on the 
right hand side. This is
solved by working with the following Hermitian Lagrangian density,
\begin{equation}
\mathcal{L}_{sym} =  \frac{\hbar^2}{2m}\partial_\mu\Psi\partial^\mu\Psi^* + 
\frac{i\hbar c}{2} (\Psi^* \partial_0 \Psi - \Psi \partial_0 \Psi^*) - V |\Psi|^2. 
\label{symL}
\end{equation}
The Lagrangian densities (\ref{asym}) and (\ref{symL}) 
are connected by a four-divergence and as such are equivalent. Specifically,
$\mathcal{L}_{sym}=\mathcal{L}_{asym} + \partial_\mu f^\mu$, where $f^0 = -i\hbar c |\Psi|^2/2$ and
$f^j = 0$.

The advantage of working with the symmetric form is related to the fact that all Noether currents inherit that symmetry. 
This is not always the case with the asymmetric Lagrangian density although, as expected,
the conserved Noe\-ther charges are the same working with either Lagrangian density. Furthermore, after a long
but straightforward calculation, we can show that the symmetric Lagrangian density, eq.~(\ref{symL}), is Lorentz
invariant if $\Psi$ transforms according to eq.~(\ref{transformationL}) and $\partial_\mu$ transforms as a covariant
vector. The asymmetric Lagrangian density, on the other hand, is not Lorentz invariant, although the action 
$S_{asym} = \int d^4x \mathcal{L}_{asym}$ is. Obviously, the action for the symmetric Lagrangian is also Lorentz
invariant since we already have Lorentz invariance at the level of its Lagrangian density. For all these reasons
we will only work with the symmetric Lagrangian density in the rest of this work. 

Specifically, we will employ
the following symmetric Lagrangian density to describe the Lorentz covariant Schr\"o\-din\-ger fields,
akin to the usual way one writes the complex field Klein-Gordon Lagrangian density, 
\begin{equation}
\mathcal{L} =  \partial_\mu\Psi\partial^\mu\Psi^* + 
\frac{i m c}{\hbar} (\Psi^* \partial_0 \Psi - \Psi \partial_0 \Psi^*) - \frac{2mV}{\hbar^2} |\Psi|^2 
= \partial_\mu\Psi\partial^\mu\Psi^* + 
\frac{i m c}{\hbar} \Psi^* \overleftrightarrow{\partial}_{\hspace{-.15cm}0} \Psi - \frac{2mV}{\hbar^2} |\Psi|^2.
\label{symL2}
\end{equation}
Note that  
$\mathcal{L} = \frac{2m}{\hbar^2}\mathcal{L}_{sym}$. Also, if in eq.~(\ref{symL2}) 
we set $V=0$ and use eq.~(\ref{transformationKG2}), we obtain the  complex field Klein-Gordon Lagrangian density.

\subsection{The Hamiltonian density}

The Hamiltonian density of the Lorentz covariant Schr\"odinger equation is obtained from 
the Lagrangian density (\ref{symL2}) by the following Legendre transformation,
\begin{equation}
\mathcal{H} = \Pi_\Psi\partial_t\Psi + \Pi_{\Psi^{\!*}}\partial_t\Psi^* - \mathcal{L} 
=\frac{\partial\mathcal{L}}{\partial(\partial_0\Psi)}\partial_0\Psi 
+ \frac{\partial\mathcal{L}}{\partial(\partial_0\Psi^*)}\partial_0\Psi^* - \mathcal{L}, 
\label{ham}
\end{equation}
where $\partial_t = c\partial_0$. Here
\begin{eqnarray}
\Pi_\Psi &=& \frac{\partial\mathcal{L}}{\partial(\partial_t\Psi)}
=\frac{1}{c}\frac{\partial\mathcal{L}}{\partial(\partial_0\Psi)}, \\
\Pi_{\Psi^{\!*}} &=& \frac{\partial\mathcal{L}}{\partial(\partial_t\Psi^*)}
=\frac{1}{c}\frac{\partial\mathcal{L}}{\partial(\partial_0\Psi^*)},
\end{eqnarray}
are, respectively, the conjugate momenta to the fields $\Psi$ and $\Psi^*$.
Due to the symmetry of the Lagrangian density in those fields, it is not difficult to
see that $\Pi_{\Psi^{\!*}}=\Pi_{\Psi}^*$. A direct calculation gives
\begin{eqnarray}
\Pi_\Psi &=& \frac{1}{c}\partial_0\Psi^* + \frac{im}{\hbar}\Psi^*, \label{pi} \\
\Pi_{\Psi^{\!*}} &=& \frac{1}{c}\partial_0\Psi - \frac{im}{\hbar}\Psi. \label{pi*}
\end{eqnarray}
Now, inserting eqs.~(\ref{symL2}), (\ref{pi}), and (\ref{pi*}) into (\ref{ham})
we get
\begin{eqnarray}
\mathcal{H} 
& = & \partial_0\Psi \partial_0 \Psi^* + \nabla\Psi \cdot \nabla \Psi^* +  \frac{2mV}{\hbar^2}|\Psi|^2.
\label{ham2}
\end{eqnarray}

Looking at eq.~(\ref{ham2}) we see that it is clearly positive for the free field 
case ($V=0$) as well as whenever 
$mV >  0$. For $mV<0$, the positiveness of eq.~(\ref{ham2}) can be broken for a sufficiently
high value of $|mV|$. Also, using eqs.~(\ref{pi}) and (\ref{pi*}) we can express the temporal derivatives
appearing in eq.~(\ref{ham2}) as functions of the fields $\Psi$ and $\Psi^*$ and their conjugate momenta.
This leads to the following way of writing the Hamiltonian density,
\begin{eqnarray}
\mathcal{H} = c^2\Pi_{\Psi}\Pi_{\Psi^{\!*}} + \nabla\Psi \cdot \nabla \Psi^* 
+ \frac{imc^2}{\hbar}\left( \Psi \Pi_{\Psi} - \Psi^*\Pi_{\Psi^{\!*}}\right) 
+ \frac{m^2c^2}{\hbar^2}\left( 1 +  \frac{2V}{mc^2}  \right) |\Psi|^2.
\label{ham3}
\end{eqnarray}
Noting that $\Pi_{\Psi^{\!*}}=\Pi_{\Psi}^*$, eq.~(\ref{ham3}) becomes
\begin{eqnarray}
\mathcal{H} &=& c^2\Pi_{\Psi}\Pi_{\Psi}^* + \nabla\Psi \cdot \nabla \Psi^* 
+ \frac{imc^2}{\hbar}\left( \Psi \Pi_{\Psi} - \Psi^*\Pi_{\Psi}^*\right) 
+ \frac{m^2c^2}{\hbar^2}\left( 1 +  \frac{2V}{mc^2}  \right) |\Psi|^2 \label{ham4}\\
& = & c^2|\Pi_{\Psi}|^2 + |\nabla\Psi|^2 - \frac{2mc^2}{\hbar}\text{Im}(\Psi \Pi_{\Psi}) 
+ \frac{m^2c^2}{\hbar^2}\left( 1 +  \frac{2V}{mc^2}  \right) |\Psi|^2,
\label{ham5}
\end{eqnarray}
where $\text{Im}(z)$ stands for the imaginary part of the complex number $z$. 
Setting $V=0$ we can write the Hamiltonian density as 
\begin{eqnarray}
\mathcal{H} &=& c^2\Pi_{\Psi}\Pi_{\Psi}^* + \nabla\Psi \cdot \nabla \Psi^* 
  + \frac{m^2c^2}{\hbar^2} |\Psi|^2 
  + \frac{imc^2}{\hbar}\left( \Psi \Pi_{\Psi} \!-\! \Psi^*\Pi_{\Psi}^*\right).
\label{mathcalHLS}
\end{eqnarray}
Note that the first three terms at the right hand side are formally the same as those appearing in 
the complex field Klein-Gordon Hamiltonian density while the last one is particular to 
the Hamiltonian density of the Lorentz covariant Schr\"odinger equation. 
This last term couples the conjugate momenta with the fields.

\subsection{The Noether currents}
\label{noethertheorem}

Here we assume that the external potential $V$ is constant. This 
means that the Lagrangian density (\ref{symL2}) does not depend 
explicitly on the space-time coordinates, only on the fields and their
derivatives. As such, whenever the La\-gran\-gian density \cite{man86}, and more generally the action \cite{gre95},
is invariant under a continuous one-parameter set of transformations, or a symmetry transformation for short,
we get a local conserved current. This is the essence of Noether's theorem \cite{man86,gre95}. 

Calling $\delta x_\mu$ an infinitesimal variation of the space-time coordinates, $\delta \Psi_T=\Psi'(x')- \Psi(x)$ and 
$\delta\Psi_T^* = {\Psi^{*}}'(x') - \Psi^*(x)$ the total variations of the fields, the Noether four-current density of the 
Lagrangian density (\ref{symL2}) reads
\begin{equation}
j^\mu = \frac{\partial \mathcal{L}}{\partial(\partial_\mu\Psi)}\delta\Psi_T 
+\frac{\partial \mathcal{L}}{\partial(\partial_\mu\Psi^*)}\delta\Psi_T^*
- \mathcal{T}^{\mu\nu}\delta x_\nu,
\label{noetherC}
\end{equation}
where the canonical energy-momentum tensor is
\begin{equation}
\mathcal{T}^{\mu\nu} = \frac{\partial \mathcal{L}}{\partial(\partial_\mu\Psi)}\partial^\nu\Psi 
+\frac{\partial \mathcal{L}}{\partial(\partial_\mu\Psi^*)}\partial^\nu\Psi^* 
- g^{\mu\nu}\mathcal{L}.
\label{tmn}
\end{equation}
Note that to first order in the infinitesimal variation $\delta x_\mu$ we have
\begin{eqnarray}
\delta\Psi_T &=& \Psi'(x) - \Psi(x) + \delta x^\mu \partial_\mu \Psi(x) 
             =\delta \Psi(x) + \delta x^\mu \partial_\mu \Psi(x),
\end{eqnarray}
with a similar expression for $\delta\Psi_T^*$.

If under a symmetry transformation the Lagrangian density is invariant 
($\delta \mathcal{L} = 0$), we obtain
the continuity equation $\partial_\mu j^\mu = 0$. Thus, the following quantity is conserved,
\begin{equation}
\int d^3x j^0(x).
\end{equation}

\subsubsection{Invariance under space-time translations}

Under a space-time translation $\Psi'(x') = \Psi(x)$ and thus
$\delta\Psi_T$ $=$ $\delta\Psi_T^*=0$. Using this fact and assuming 
an infinitesimal constant translation $\delta x^\mu = a^\mu$ of the coordinates,
it is not difficult to see that the Lagrangian density (\ref{symL2}) is invariant under 
such operation. 

In this case the Noether current (\ref{noetherC}) becomes
\begin{equation}
 j^\mu = - \mathcal{T}^{\mu\nu} a_\nu,
\end{equation}
where
\begin{eqnarray}
\mathcal{T}^{\mu\nu} &=& \partial^\mu\Psi^*\partial^\nu\Psi +\partial^\mu\Psi\partial^\nu\Psi^* 
+\frac{imc}{\hbar}\left( \Psi^*\partial^\nu\Psi - \Psi\partial^\nu\Psi^* \right)\delta^{\mu 0}
- g^{\mu\nu}\mathcal{L}.
\label{tmn2}
\end{eqnarray}
Since each component of $a^\mu$ is arbitrary and independent of each other, we obtain that 
\begin{equation}
\partial_\mu \mathcal{T}^{\mu\nu} = 0, \hspace{.2cm}\nu = 0,1,2,3.
\end{equation}

The corresponding conserved ``charges'' are 
\begin{eqnarray}
H &=& \int d^3x \mathcal{T}^{00} = \int d^3x \mathcal{H}, \label{Hconserved}\\
cP^j &=& \int d^3x \mathcal{T}^{0j} = \int d^3x (c\mathcal{P}^{j}) , \hspace{.2cm} j = 1,2,3. \label{Pconserved}
\end{eqnarray}
Using eqs.~(\ref{symL2}) and (\ref{tmn2}), a direct computation gives
\begin{eqnarray}
\mathcal{H} &=& \partial_0\Psi \partial_0 \Psi^* + \nabla\Psi \cdot \nabla \Psi^* +  \frac{2mV}{\hbar^2}|\Psi|^2,
\label{densistyH} \\
c\mathcal{P}^j &=&  \partial^0\Psi^* \partial^j \Psi +  \partial^0\Psi \partial^j \Psi^*
+ \frac{imc}{\hbar}\left( \Psi^*\partial^j\Psi - \Psi\partial^j\Psi^* \right) \nonumber \\
& = & 2\text{Re}\left(\partial^0\Psi^* \partial^j \Psi\right) 
- \frac{2mc}{\hbar}\text{Im}\left( \Psi^*\partial^j\Psi\right),
\label{densityP}
\end{eqnarray}
where $\text{Re}(z)$ is the real part of the complex number $z$.

Comparing eq.~(\ref{densistyH}) with (\ref{ham2}) we see that, as expected, they are identical.
Equation (\ref{densistyH}) is the Hamiltonian density of the Lorentz covariant Schr\"odinger equation and
its associated 
conserved charge, eq.~(\ref{Hconserved}), is the Hamiltonian $H$ giving the total energy of the system. 
The three quantities
$\mathcal{P}^j$, $j=1,2,3$, are interpreted as the momentum density associated to the
fields $\Psi$ and $\Psi^*$ along three orthogonal spatial directions. This is true because they are derived 
from the invariance of the
Lagrangian density under spatial translations. The associated conserved charges, $P^j$, are the total
momentum of the fields projected along three orthogonal spatial directions [cf. eq.~(\ref{Pconserved})].
Note also that for $c\rightarrow \infty$ the first term on the right hand side of eq.~(\ref{densityP})
goes to zero and $\mathcal{P}^j \rightarrow \frac{2m}{\hbar}\text{Im}\left( \Psi^*\partial^j\Psi\right)$.
This latter term is proportional to the momentum density of the non-relativistic Schr\"odinger Lagrangian, which is
given by $-i\hbar\Psi^*\partial_j\Psi$ if we use the appropriate non-relativistic normalization for the wave function. 

We can also write the momentum density vector $\bm{\mathcal{P}}=(\mathcal{P}^1,\mathcal{P}^2,\mathcal{P}^3)$ as
\begin{eqnarray}
c\bm{\mathcal{P}} &=& -2\text{Re}\left(\partial^0\Psi^* \nabla \Psi\right) 
+\frac{2mc}{\hbar}\text{Im}\left( \Psi^*\nabla\Psi\right).
\end{eqnarray}
Similarly to what we did with the Hamiltonian density, if we use eqs.~(\ref{pi}) and (\ref{pi*}) we can express the 
momentum density as a function of the conjugate momenta,
\begin{equation}
\mathcal{P}^j =  \Pi_\Psi\partial^j\Psi + \Pi_{\Psi^{\!*}}\partial^j\Psi^*  
=2\text{Re}\left(\Pi_\Psi \partial^j \Psi\right),
\label{densityP2} 
\end{equation}
where the last term comes from the fact that $\Pi_{\Psi^{\!*}}=\Pi_\Psi^*$.
And noting that $\partial^j=-\partial_j$ we have
\begin{equation}
\bm{\mathcal{P}} = -2\text{Re}\left(\Pi_\Psi\nabla\Psi\right). 
\label{densityP3}
\end{equation}

\subsubsection{Invariance under spatial rotations}

Under a spatial rotation we also have $\delta\Psi_T=\delta\Psi_T^*=0$ [see eq.~(\ref{3Drotations})]. Now, however, 
an infinitesimal spatial rotation leads to $\delta x^\mu= \omega^\mu_{\;\nu}\, x^\nu$, where 
the tensor $\omega^\mu_{\;\nu}$ is antisymmetric. Using the notation $[\omega^\mu_{\;\nu}]_k$
to label about which axis we are rotating, rotations about the $x$, $y$, and
$z$ axes are, respectively, given by the tensors $[\omega^\mu_{\;\nu}]_1$, $[\omega^\mu_{\;\nu}]_2$, and
$[\omega^\mu_{\;\nu}]_3$. Calling $\epsilon$ the infinitesimal angle of rotation we have 
$[\omega^2_{\;3}]_1=-[\omega^3_{\;2}]_1=[\omega^1_{\;3}]_2=-[\omega^3_{\;1}]_2=[\omega^1_{\;2}]_3=-[\omega^2_{\;1}]_3=\epsilon$,
with all the other $[\omega^\mu_{\;\nu}]_k$ being zero.
Putting together all these results, 
and noting that the Lagrangian density (\ref{symL2}) is unchanged by this symmetry operation,
the Noether current (\ref{noetherC}) becomes
\begin{equation}
[j^\mu]_k = -\mathcal{T}^{\mu\nu}[\omega_{\nu\alpha}]_k\, x^\alpha,
\end{equation}
where $\omega_{\nu\alpha}=-\omega^\nu_{\;\alpha}$ and $\partial_\mu[j^\mu]_k=0$.

Fixing our attention to a rotation about the $z$ axis and noting that $\epsilon$ is an arbitrary constant
we have,
\begin{eqnarray}
[j^0]_3 = \left( \mathcal{T}^{01}x^2-\mathcal{T}^{02}x^1\right),
\end{eqnarray}
with the corresponding conserved charge
\begin{equation}
cM^3 = \int d^3x [j^0]_3 = \int d^3x (c\,l^3). 
\end{equation}

Computing explicitly $\mathcal{T}^{01}$ and $\mathcal{T}^{02}$ using eq.~(\ref{tmn}) 
we can express the angular momentum density along the $z$-di\-rec\-tion as
\begin{equation}
c\,l^3 = \left( \partial^0\Psi -\frac{imc}{\hbar}\Psi\right)\!\! \left( x^2\partial^1\Psi^* - x^1\partial^2\Psi^* \right) 
+ \left( \partial^0\Psi^* +\frac{imc}{\hbar}\Psi^*\right)\!\! \left( x^2\partial^1\Psi - x^1\partial^2\Psi \right)\!\!.
\end{equation}

If we repeat the above calculation for rotations about the x and y axis we get similar results, showing that the angular momentum 
densities along the $x$ and $y$-directions, $l^1$ and $l^2$, are obtained from $l^3$ 
by cyclic permutations of the spatial indexes. The angular momentum density vector $\mathbf{l}=(l^1,l^2,l^3)$ can be written
as
\begin{eqnarray}
c\mathbf{l} &=&  \left(\!\! \partial^0\Psi -\frac{imc}{\hbar}\Psi\!\!\right)\!\! \mathbf{r}\!\times\!\nabla\Psi^* + 
\left(\!\! \partial^0\Psi^* +\frac{imc}{\hbar}\Psi^*\!\!\right) \!\! \mathbf{r}\!\times\!\nabla\Psi \nonumber \\
&=& 2\text{Re}\left[ \left( \partial^0\Psi^* +\frac{imc}{\hbar}\Psi^*\right) \mathbf{r}\times\nabla\Psi\right],
\end{eqnarray}
where the symbol $\times$ above stands for the cross product. 

Using the three-dimensional Levi-Civita tensor $\epsilon^{ijk}$,
we can rewrite the angular momentum density vector as   
\begin{equation}
cl^k = 2\text{Re}\left[ \left( \partial^0\Psi^* +\frac{imc}{\hbar}\Psi^*\right) \epsilon^{ijk}x_i\partial_j\Psi\right].
\end{equation}
Here $\epsilon^{ijk}=g^{ia}g^{jb}g^{kc}\epsilon_{abc}$, where $\epsilon_{abc}=1$ for $abc=123$ and cyclic permutations,
$\epsilon_{abc}=-1$ for $abc=213$ and cyclic permutations, and $\epsilon_{abc}=0$ whenever at least two indexes are equal.

Note that the angular momentum density tends to   
$\mathbf{l} \rightarrow 
2\text{Re}\left[  \frac{im}{\hbar}\Psi^* \mathbf{r}\times\nabla\Psi\right]$
when $c\rightarrow \infty$,
which is proportional, up to a normalization constant, to the angular momentum density of the non-relativistic
Schr\"odinger Lagrangian.

Finally, using eqs.~(\ref{pi}) and (\ref{pi*}) we can write the 
angular momentum density as a function of the conjugate momenta,
\begin{equation}
\mathbf{l} =  \mathbf{r}\!\times\!\Pi_{\Psi^{\!*}}\nabla\Psi^* + 
 \mathbf{r}\!\times\!\Pi_\Psi \nabla\Psi = 
 2\mathbf{r}\times\text{Re}\left[ \Pi_\Psi \nabla\Psi\right]. 
\end{equation}
It is worth mentioning
that the ``correct'' sign for the angular momentum density does not follow from
the Noether's theorem and in order to have an angular momentum density agreeing with
the standard definition of angular momentum we need to insert a minus sign \cite{gre00}.
Using eq.~(\ref{densityP3}) this gives
\begin{equation}
\mathbf{l} =  - 2\mathbf{r}\times\text{Re}\left[ \Pi_\Psi \nabla\Psi\right]
 = \mathbf{r} \times \bm{\mathcal{P}}, 
\end{equation}
the expected standard relation between vector momentum and vector angular momentum densities. Observe that 
the total angular momentum is given by\footnote{We are running out of letters to denote
the several quantities in this work. We used the letter $M$ for the angular momentum instead of $L$
since the latter is already used to denote the Lagrangian. Sometimes we will even use the same letter to
denote different quantities in order to comply with the usual notation for those quantities.
The context will make it clear which meaning to ascribe to a given notation.}
\begin{equation}
\textbf{M} = \int d^3x \;\mathbf{l}.
\label{capitalM}
\end{equation}

\subsubsection{Invariance under global phase transformations}

The Lagrangian density (\ref{symL2}) is clearly invariant under the following
transformation of the field,
\begin{equation}
\Psi'(x) = e^{-i\theta}\Psi(x),
\end{equation}
where $\theta$ is a real constant. For an infinitesimal $\theta$ we get that
$\Psi'(x) \approx (1-i\theta)\Psi(x)$. This leads to
\begin{equation}
\delta\Psi_T=-i\theta\Psi(x)
\label{deltaPsi}
\end{equation}
and analogously to
\begin{equation}
\delta\Psi_T^*=i\theta\Psi^*(x).
\label{deltaPsi2}
\end{equation}
Since now there is no change in the coordinates 
(we are dealing
with an internal symmetry) we have that $\delta x^\mu = 0$. 
The Noether current (\ref{noetherC}) is thus 
\begin{equation}
 j^\mu =  \frac{\partial \mathcal{L}}{\partial(\partial_\mu\Psi)}\delta\Psi_T 
+\frac{\partial \mathcal{L}}{\partial(\partial_\mu\Psi^*)}\delta\Psi_T^*.
\label{jQ}
\end{equation}

Inserting eqs.~(\ref{symL2}), (\ref{deltaPsi}), and (\ref{deltaPsi2}) 
into (\ref{jQ}) and dropping the 
arbitrary constant $\theta$ we get
\begin{equation}
\mathbf{j} = (j^1,j^2,j^3) = i(\Psi\nabla\Psi^* - \Psi^*\nabla\Psi)
\label{jj}
\end{equation}
and
\begin{equation}
j^0 = -i(\Psi\partial_0\Psi^* - \Psi^*\partial_0\Psi) + \frac{2mc}{\hbar}|\Psi|^2,
\label{J0}
\end{equation}
where we used that $\partial_0=\partial^0$ and $\partial_j=-\partial^j$. The conserved charge 
is\footnote{We use $\tilde{q}$ and $\tilde{Q}$ instead of $q$ and $Q$ since we save the latter to 
represent the exact SI value of a given particle's electric charge.} 
\begin{eqnarray}
c\tilde{Q} &=& \int d^3x j^0=\int d^3x (c\,\tilde{q}).
\label{Qconserved}
\end{eqnarray}

Comparing eqs.~(\ref{jj}) and (\ref{J0}) with eqs.~(\ref{Jj}) and (\ref{j0}) we see that,
up to a multiplicative constant equal to $-\hbar/(2m)$, they are the same. 
We have thus recovered the
probability four-current density we obtained before using a different method.

As we did to the other conserved quantities, we 
can express the charge density $\tilde{q}$ in terms of the conjugate momenta of the fields. Inserting 
eqs.~(\ref{pi}) and (\ref{pi*}) into (\ref{J0}) we get
\begin{equation}
\tilde{q} =  -i\left( \Psi \Pi_{\Psi} \!-\! \Psi^*\Pi_{\Psi}^*\right) = 2\text{Im}(\Psi\Pi_\Psi).
\end{equation}

\subsection{Discrete symmetries}
\label{ds}

We now briefly discuss the three major discrete symmetries one usually encounters in 
any field theory, namely, space inversion 
(parity), time reversal, and charge conjugation. We save for later, after
we second quantize the Lorentz covariant Schr\"odinger equation, a more thorough discussion
on this subject, specially the aspects related to charge conjugation. 
However, even
at the first quantization level, we already see that the charge conjugation operator requires
a different definition in order to ``save'' the CPT theorem. Actually, it will become clear
after second quantization that the Lagrangian density (\ref{symL2}) satisfies the CPT 
theorem only if we 
define the charge conjugation operator such that it also
implements what we will call the ``mass conjugation operation''. Simply put, we must 
change the sign of the mass $m$ when applying the charge conjugation operation. 
Only in this way can we properly exchange the roles of particles with antiparticles and, at the same time, satisfy the CPT theorem.

\subsubsection{Space inversion or parity}

Space inversion is implemented by changing the sign of the space coordinates of the system under investigation,
\begin{equation}
x = (\mathbf{r},t) \longrightarrow x' = (\mathbf{r'},t') = (-\mathbf{r},t).
\label{inversion}
\end{equation}
Following the same steps detailed in ref. \cite{gre95}, it is not difficult to see that
the parity operation leads to the following transformation rule for the second quantized
field associated with the Lorentz covariant Schr\"odinger equation,
\begin{equation}
\mathcal{\hat{P}}\hat{\Psi}(\mathbf{r},t)\mathcal{\hat{P}}^\dagger = \hat{\Psi}(-\mathbf{r},t). \label{parity} 
\end{equation}
Here $\mathcal{\hat{P}}$ is a unitary operator denoting the space inversion 
operation.\footnote{The most general parity operation is 
$\mathcal{\hat{P}}\hat{\Psi}(\mathbf{r},t)\mathcal{\hat{P}}^\dagger = \eta_P\hat{\Psi}(-\mathbf{r},t)$, with $\eta_P$ a complex number such that $|\eta_P|=1$. However, 
for charged scalar fields one can always set $\eta_P=1$ without losing in generality.
This argument is still valid for the time reversal and charge conjugation
operations. See ref. \cite{gre95} for details.}
Note that we are using the symbol ``\;$\hat{\,}$\;'' 
to distinguish the second quantized fields
from the classical ones. In section \ref{qft}, we will drop the hat symbol when dealing
with operators to simplify
notation. 

Using eq.~(\ref{parity}) and its complex conjugated version, as well as eq.~(\ref{inversion})
and the fact that 
under space inversion $(\partial_0,\partial_j)\rightarrow (\partial_0,-\partial_j)$, 
we can show that 
\begin{eqnarray}
\mathcal{\hat{P}}\hat{\mathcal{L}}(\mathbf{r},t)\mathcal{\hat{P}}^\dagger &=& \hat{\mathcal{L}}(\mathbf{-r},t), \label{parity1}\\
\mathcal{\hat{P}}\hat{H}\mathcal{\hat{P}}^\dagger &=& 
\hat{H}, \\
\mathcal{\hat{P}}\hat{\mathbf{P}}\mathcal{\hat{P}}^\dagger &=& 
-\mathbf{\hat{P}}, \\
\mathcal{\hat{P}}\hat{\mathbf{M}}\mathcal{\hat{P}}^\dagger &=& 
\mathbf{\hat{M}}, \label{parity4}
\end{eqnarray}
where $\hat{\mathcal{L}}$, $\hat{H}$, $\mathbf{\hat{P}}$, and $\hat{\mathbf{M}}$ are
the second quantized versions of the Lagrangian density, the Hamiltonian, the 
linear momentum vector, and angular momentum vector of the Lorentz covariant
Schr\"odinger equation [see eqs.~(\ref{symL2}), (\ref{Hconserved}), (\ref{Pconserved}), 
and (\ref{capitalM})]. The relations given by eqs.~(\ref{parity1})-(\ref{parity4}) are
the ones expected from our classical intuition about space inversion. 

\subsubsection{Time reversal}

Time reversal is given by changing the sign of the time coordinate only,
\begin{equation}
x = (\mathbf{r},t) \longrightarrow x' = (\mathbf{r'},t') = (\mathbf{r},-t).
\label{timereflection}
\end{equation}
As before, the same techniques given in ref. \cite{gre95} to build the time reversal 
operator for the Klein-Gordon field 
remain valid here. Repeating those steps, we get an 
antiunitary operator $\hat{\mathcal{T}}$ that implements the time reversal operation.
Its action on the quantum field $\hat{\Psi}$ is
\begin{equation}
\mathcal{\hat{T}}\hat{\Psi}(\mathbf{r},t)\mathcal{\hat{T}}^\dagger = 
\hat{\Psi}(\mathbf{r},-t). \label{timereversal} 
\end{equation}
Since $\mathcal{\hat{T}}$ is antiunitary, we have that 
\begin{equation}
\hat{\mathcal{T}} z \hat{\mathcal{T}}^\dagger = z^*,
\label{antiunitary}
\end{equation}
where $z$ is a complex number.

Using eqs.~(\ref{timereflection})-(\ref{antiunitary}), and that  
$(\partial_0,\partial_j)\rightarrow (-\partial_0,\partial_j)$ under time reversal,
we now have
\begin{eqnarray}
\mathcal{\hat{T}}\hat{\mathcal{L}}(\mathbf{r},t)\mathcal{\hat{T}}^\dagger &=& \hat{\mathcal{L}}(\mathbf{r},-t), \label{time1}\\
\mathcal{\hat{T}}\hat{H}\mathcal{\hat{T}}^\dagger &=& 
\hat{H}, \\
\mathcal{\hat{T}}\hat{\mathbf{P}}\mathcal{\hat{T}}^\dagger &=& 
-\mathbf{\hat{P}}, \\
\mathcal{\hat{T}}\hat{\mathbf{M}}\mathcal{\hat{T}}^\dagger &=& 
-\mathbf{\hat{M}}. \label{time4}
\end{eqnarray}
Note that due to the explicit presence of the imaginary $i$ in the expression for the Lagrangian density, the antiunitarity  of $\mathcal{\hat{T}}$ is crucial to obtain the above relations.

\subsubsection{Charge conjugation}

Charge conjugation is not related to 
space-time coordinate transformations and since it exchanges particles with antiparticles,
it has no immediate classical analog. In fact, it is defined such that
the creation and annihilation operators for particles are transformed to 
the ones associated with antiparticles. 

If we follow the prescription defining 
the charge conjugation operator for the Klein-Gordon fields,  
we see that it is given by a unitary operator $\hat{\mathcal{C}}$ such that
\begin{equation}
\mathcal{\hat{C}}\hat{\Psi}(x)\mathcal{\hat{C}}^\dagger = \hat{\Psi}^\dagger(x). \label{chargeC}  
\end{equation}
However, when applying it to the Lagrangian density (\ref{symL2}) we realize 
that, due to the presence of the term proportional to the imaginary number $i$, 
the charge conjugation operation (\ref{chargeC})
does not leave the Lagrangian density invariant (normal ordering implied).
This fact, together with how the parity and time reversal operations  
$\mathcal{\hat{P}}$ and $\mathcal{\hat{T}}$
affect the Lagrangian density, would lead to
a violation of the CPT theorem since we are dealing with a Lorentz invariant Lagrangian.
A possible fix would be to define 
$\hat{\mathcal{C}}$ as an antiunitary operator. This will change the sign of $i$ in the Lagrangian
density, leaving it invariant under charge conjugation. 

Although an antiunitary charge
conjugation operator sol\-ves the CPT theorem violation problem, we will see in 
section \ref{qft} that either a unitary or an antiunitary
$\mathcal{\hat{C}}$ does not properly exchange the roles of particles with antiparticles.
The only acceptable solution is to extend the charge conjugation operation such that it 
anticommutes with any function of the mass. Putting it simply, we must change the sign of 
the mass $m$ when applying the extended charge conjugation operator.  

If we define the ``mass conjugation'' operator $\mathcal{\hat{M}}$ such that
\begin{equation}
\mathcal{\hat{M}} f(m) \mathcal{\hat{M}}^\dagger = f(-m), 
\label{massconjugation}
\end{equation}
where $f(m)$ is an arbitrary function of the mass $m$, the extended charge conjugation
operator we need is given by
\begin{equation}
\mathcal{\hat{C}}_m = \mathcal{\hat{M}}\mathcal{\hat{C}},
\label{Cm}
\end{equation}
with $\mathcal{\hat{C}}$ given by eq.~(\ref{chargeC}).
It is not difficult to see that $\mathcal{\hat{C}}_m$
leaves the Lagrangian density (\ref{symL2}) invariant and,
as we will see in section \ref{qft}, it
properly changes the roles of particles with 
antiparticles.\footnote{
We implicitly assumed that we were dealing with the free field ($V=0$).
However, looking at eq.~(\ref{symL2}), everything that was said above also applies to $V\neq 0$ provided
that $\mathcal{\hat{C}}_m V\mathcal{\hat{C}}_m^\dagger = -V$. If $V$ satisfies the previous transformation property
we get that eqs.~(\ref{charge1})-(\ref{charge4}) are also satisfied.
Moreover, it is not difficult to see that
the Lagrangian density (\ref{symL2}) is invariant if we simultaneously apply the following 
three transformations: $m \rightarrow -m$, $V\rightarrow -V$, and 
$\Psi \rightarrow \exp[{-i2mc^2t/\hbar}]\Psi$.}

Using eq.~(\ref{Cm}) we can show that
\begin{eqnarray}
\mathcal{\hat{C}}_m\hat{\mathcal{L}}(\mathbf{r},t)\mathcal{\hat{C}}_m^\dagger &=& \hat{\mathcal{L}}(\mathbf{r},t), \label{charge1}\\
\mathcal{\hat{C}}_m\hat{H}\mathcal{\hat{C}}_m^\dagger &=& 
\hat{H}, \\
\mathcal{\hat{C}}_m\hat{\mathbf{P}}\mathcal{\hat{C}}_m^\dagger &=& 
\mathbf{\hat{P}}, \\
\mathcal{\hat{C}}_m\hat{\mathbf{M}}\mathcal{\hat{C}}_m^\dagger &=& 
\mathbf{\hat{M}}. \label{charge4}
\end{eqnarray}

\subsubsection{The CPT theorem}

The CPT theorem applies to the Lorentz covariant Schr\"o\-din\-ger  Lagrangian
if we use the extended charge conjugation operator $\mathcal{\hat{C}}_m$ define above. 
Strictly speaking, we have
the CPTM theorem here since not only the charge but the mass must change sign to leave
the action invariant.

If we define the CPT operation by
\begin{equation}
\hat{\Theta} = \mathcal{\hat{C}}_m\mathcal{\hat{P}}\mathcal{\hat{T}} =  \mathcal{\hat{M}}\mathcal{\hat{C}}\mathcal{\hat{P}}\mathcal{\hat{T}},
\label{cptm}
\end{equation}
it is not difficult to show that 
\begin{eqnarray}
\hat{\Theta}\hat{\mathcal{L}}(x)\hat{\Theta}^\dagger &=& \hat{\mathcal{L}}(-x), \label{cpt1}\\
\hat{\Theta}\hat{H}\hat{\Theta}^\dagger &=& 
\hat{H}, \\
\hat{\Theta}\hat{\mathbf{P}}\hat{\Theta}^\dagger &=& 
\mathbf{\hat{P}}, \\
\hat{\Theta}\hat{\mathbf{M}}\hat{\Theta}^\dagger &=& 
-\mathbf{\hat{M}}. \label{cpt4}
\end{eqnarray}

Using eq.~(\ref{cpt1}) and eq.~(\ref{action}) that defines the action, we get
after changing the variables $x\rightarrow -x$,
\begin{equation}
\hat{\Theta}\hat{S}\hat{\Theta}^\dagger = \hat{S}. \label{cptm2}
\end{equation}
In other words, the action is invariant under the CPT operation
(\ref{cptm}) defined here.

\subsubsection{More on the new charge conjugation operation}
\label{moreCm}

In a static external gravitational field, we can as a first approximation model the 
interaction of a particle with inertial mass $m$ by setting $V=m_g\varphi(x)$, where
$\varphi(x)$ represents the gravitational potential associated to the field acting on the particle and
$m_g$ the particle's gravitational mass (``gravitational charge'').  
The corresponding interaction Lagrangian density is given by [see eq.~(\ref{symL2})]
\begin{equation}
\hat{\mathcal{L}}_{int} = -2m\,m_g\varphi\hat{\Psi}\hat{\Psi}^\dagger/\hbar^2. 
\end{equation}

So far we have not said how the charge conjugation operator affects the gravitational mass. Assuming the latter
is invariant under the action of $\mathcal{\hat{C}}_m$, we have that 
\begin{equation}
\mathcal{\hat{C}}_m\hat{\mathcal{L}}_{int}\mathcal{\hat{C}}_m^\dagger = 
2m\,m_g\varphi\hat{\Psi}\hat{\Psi}^\dagger/\hbar^2 = -\hat{\mathcal{L}}_{int}, 
\label{Lgrav}
\end{equation}
where normal ordering is implied and we have used eqs. (\ref{massconjugation}), (\ref{Cm}), and that 
$\mathcal{\hat{C}}_m m_g \mathcal{\hat{C}}_m^\dagger = m_g$. Note that $\mathcal{\hat{C}}_m$
commutes with $\varphi$ since the former does not act on the external source of the gravitational field.

Since the free field part of eq.~(\ref{symL2}) does not change under the action of $\mathcal{\hat{C}}_m$,
the minus sign in eq.~(\ref{Lgrav}) obtained after applying the charge conjugation operation 
implies that antiparticles will react to the external gravitational field differently when compared to 
the response of particles to the same field. 
If we had an attractive interaction when dealing with particles, we would 
now get a repulsive one for antiparticles. Note that we can make particles and antiparticles respond 
in the same way to a gravitational field if we postulate that 
the charge conjugation operation anticommutes with the gravitational mass $m_g$. This is 
equivalent to assuming that $m_g=m$ in eq.~(\ref{Lgrav}). In other words, if the 
gravitational mass and inertial mass are equal, we will always have particles and 
particles, antiparticles and antiparticles, and particles and antiparticles attracting
each other gravitationally.

What happens if we now have an external static electric field? After the minimal coupling prescription,
the interacting part of the Lagrangian density becomes [see eqs.~ (\ref{c1}) and (\ref{Lint})]
\begin{equation}
\hat{\mathcal{L}}_{int} = 
-\frac{iq}{\hbar}(\hat{\Psi}^\dagger\overleftrightarrow{\partial}_{\hspace{-.15cm}0}\hat{\Psi})A^0
+ \frac{q^2}{\hbar^2}\hat{\Psi}\hat{\Psi}^\dagger A_0 A^0  -\frac{2q\mu}{\hbar}\hat{\Psi}\hat{\Psi}^\dagger A^0, 
\label{Lem} 
\end{equation}
where $A^0$ is the source of the external electromagnetic field. Applying the charge conjugation operator
we now get (normal ordering implied),
\begin{equation}
\mathcal{\hat{C}}_m\hat{\mathcal{L}}_{int}\mathcal{\hat{C}}_m^\dagger =  
\frac{iq}{\hbar}(\hat{\Psi}^\dagger\overleftrightarrow{\partial}_{\hspace{-.15cm}0}\hat{\Psi})A^0
+ \frac{q^2}{\hbar^2}\hat{\Psi}\hat{\Psi}^\dagger A_0 A^0  +\frac{2q\mu}{\hbar}\hat{\Psi}\hat{\Psi}^\dagger A^0, 
\label{Lem2} 
\end{equation}
where we used that $\mu=mc/\hbar$ changes sign under the action of $\mathcal{\hat{C}}_m$
since it depends on $m$; the external potential $A^0$ is unaffected by the action
of the charge conjugation operator. Comparing eqs.~(\ref{Lem}) and (\ref{Lem2}) we see that
we can go from one to the other by simply changing the sign of the charge $q$. This is the
expected result when we apply the charge conjugation operation: the antiparticle behaves like a 
particle with opposite electric charge. 
Note that we would have obtained the same conclusion if instead of
working with the minimal coupling prescription we had used eq.~(\ref{symL2}) with the electrostatic interaction 
modeled by setting $V=q\varphi(x)$, with $\varphi(x)$ denoting the electrostatic potential 
of the external electric field.

\section{Second quantization of the Lorentz covariant Schr\"odinger equation}
\label{qft}

We now start the canonical quantization of 
the Lorentz covariant Schr\"odinger fields. We will try to stay as close as possible
to the notation and traditional ways of dealing with the canonical quantization of a
classical field theory \cite{gre95,man86}. After finishing the canonical quantization 
of the free field and the computation of the Feynman propagator, we will present two applications of the present formalism.
We will analyze the scattering 
cross-section for two particles whose interaction is
given by $(\lambda/4)[\Psi(x)\Psi^\dagger(x)]^2$ and we will also 
develop the scalar electrodynamics for the Lorentz covariant Schr\"odinger fields. Comparisons with the predictions
coming from the complex Klein-Gordon fields subjected to these same interactions 
will also be made as well as with interactions that break Lorentz 
invariance.

\subsection{General solution to the Lorentz covariant Schr\"odinger equation}

We can write the most general free field solution ($V=0$) of eq.~(\ref{GeneralEq7}) 
as\footnote{To simplify notation, 
we will not use the symbol ``\;$\hat{\,}$\;'' to denote the 
second quantized field operators.} 

\begin{equation}
\Psi(x) = \int \widetilde{dk}_+ a_{\mathbf{k}}e^{-i\omega_{\mathbf{k}}^+ t}
e^{i\mathbf{k}\cdot\mathbf{r}}
+ \int \widetilde{dk}_- b_{\mathbf{k}}^\dagger e^{i\omega_{\mathbf{k}}^- t} 
e^{-i\mathbf{k}\cdot\mathbf{r}},
\label{geral0}
\end{equation}
which is equivalent to
\begin{eqnarray}
\Psi(x) &=& \int \widetilde{dk}_+ a_{\mathbf{k}}e^{i\mu x^0}e^{-ikx}
+ \int \widetilde{dk}_- b_{\mathbf{k}}^\dagger e^{i\mu x^0}e^{ikx} 
=\Psi^+(x) + \Psi^-(x). \label{geral}
\end{eqnarray}
Note that we must multiply $b_{\mathbf{k}}^\dagger$  as well as $a_{\mathbf{k}}$ by 
$e^{i\mu x^0}$.
Here the expansion coefficients $a_{\mathbf{k}}$ and $b_{\mathbf{k}}^\dagger$, that 
depend on the wave number $\mathbf{k}=(k^1,k^2,k^3)$,
were promoted to operators. Also, $kx=k_\mu x^\mu$ and
\begin{eqnarray}
k^0&=&\omega_{\mathbf{k}}/c = E_{\mathbf{k}}/(\hbar c), \label{omegak}\\
E_{\mathbf{k}} &=&
\sqrt{m^2c^4+\hbar^2c^2|\mathbf{k}|^2}, \\
\omega_{\mathbf{k}}^\pm &=& \mp \mu c + \omega_{\mathbf{k}}, \\
\mu &=& mc/\hbar, \label{mi}\\
\widetilde{dk}_{\pm} &=& f_{\pm}(|\mathbf{k}|)d^3k,
\end{eqnarray}
with $f_{\pm}(|\mathbf{k}|)$ real functions that depend only on the magnitude of the 
wave number. The integration in $d^3k$ are carried out from $-\infty$ to $\infty$ and the fields are
usually assumed to be zero at the boundaries of integration.

If we want $a_{\mathbf{k}} (b_{\mathbf{k}})$ and $a_{\mathbf{k}}^\dagger
(b_{\mathbf{k}}^\dagger)$ to be bona fide particle (antiparticle) annihilation and creation
operators, we should impose the following commutation relations \cite{man86},
\begin{eqnarray}
[a_{\mathbf{k}},a_{\mathbf{k'}}^\dagger ] = [ b_{\mathbf{k}},b_{\mathbf{k'}}^\dagger ]
&=& \delta^{(3)}(\mathbf{k}-\mathbf{k'}), \label{aad}\\ \,
[ a_{\mathbf{k}},a_{\mathbf{k'}}] = [b_{\mathbf{k}},b_{\mathbf{k'}}]
&=& 0, \label{aa} \\ \,
[ a_{\mathbf{k}},b_{\mathbf{k'}}] = [a_{\mathbf{k}},b_{\mathbf{k'}}^\dagger]
&=& 0, \label{ab}
\end{eqnarray}
where $\delta^{(3)}(\mathbf{k}-\mathbf{k'})=\delta(k^1- k^{1'})
\delta(k^2- k^{2'})\delta(k^3-k^{3'})$ is the three-dimensional Dirac delta function.

With these assumptions, 
we must choose the functions $f_{\pm}(|\mathbf{k}|)$ 
such that the standard equal-time
commutation relations among the fields $\Psi(x)$ and $\Psi^\dagger(x)$ and their
conjugate momenta are satisfied. In other words, $f_{\pm}(|\mathbf{k}|)$ is fixed in order
to have the following set of commutation relations valid,
\begin{eqnarray}
[\Psi(t,\mathbf{r}),\Pi_{\Psi}(t,\mathbf{r'}) ] &=&  
i\hbar\delta^{(3)}(\mathbf{r}-\mathbf{r'}), \label{com1} \\ \,
[ \Psi^\dagger (t,\mathbf{r}),\Pi_{\Psi^\dagger}(t,\mathbf{r'})  ]
&=& i\hbar\delta^{(3)}(\mathbf{r}-\mathbf{r'}). \label{com2}
\end{eqnarray}
We also require that the remaining commutators involving any other combinations 
among the $\Psi$-fields, among their conjugate momenta, and among
them and their conjugate momenta, are zero.

As we show in the appendix \ref{apA}, using eqs.~(\ref{aad})-(\ref{ab}) 
we can only satisfy
eqs.~(\ref{com1}) and (\ref{com2}), together with the other 
field commutation relations that must be zero, if
\begin{equation}
f_{+}(|\mathbf{k}|) = f_{-}(|\mathbf{k}|) = \left(\frac{\hbar c^2}{(2\pi)^3 2\omega_{\mathbf{k}}}\right)^{1/2},
\label{f=f}
\end{equation}
where $\omega_{\mathbf{k}}$ is given by eq.~(\ref{omegak}). Using eq.~(\ref{f=f}) we can write 
eq.~(\ref{geral}) as
\begin{equation}
\Psi(x) = e^{i\mu x^0}\hspace{-.1cm} 
\int\hspace{-.1cm} d^3k
\left(\hspace{-.05cm}\frac{\hbar c^2}{(2\pi)^3 2\omega_{\mathbf{k}}}\hspace{-.05cm}\right)^{1/2} \hspace{-.2cm}(a_{\mathbf{k}}e^{-ikx}
+ b_{\mathbf{k}}^\dagger e^{ikx}). \label{geral2}
\end{equation}
Note that the standard complex Klein-Gordon field expansion is given 
by eq.~(\ref{geral2}) without the term $e^{i\mu x^0}$ \cite{man86}.

Inserting eq.~(\ref{geral2}) into the normal ordered expressions for the 
free field Hamiltonian, linear momentum, and conserved charge, eqs.~(\ref{Hconserved}), 
(\ref{Pconserved}), (\ref{Qconserved}), respectively, we get [see appendix \ref{apA}],
\begin{eqnarray}
H & = & \int d^3k (\hbar \omega_{\mathbf{k}}^+a_{\mathbf{k}}^\dagger a_{\mathbf{k}}
+ \hbar\omega_{\mathbf{k}}^-b_{\mathbf{k}}^\dagger b_{\mathbf{k}}), \label{HamQ}\\
P^j & = & \int d^3k [\hbar k^j (a_{\mathbf{k}}^\dagger a_{\mathbf{k}}
+ b_{\mathbf{k}}^\dagger b_{\mathbf{k}})], \label{PQ} \\
\tilde{Q} & = & \int d^3k [\hbar (a_{\mathbf{k}}^\dagger a_{\mathbf{k}}
- b_{\mathbf{k}}^\dagger b_{\mathbf{k}})], \label{QQ}
\end{eqnarray}
where
\begin{equation}
E^{\pm}_{\mathbf{k}} = \hbar  \omega_{\mathbf{k}}^\pm = \mp mc^2 + E_{\mathbf{k}}.
\label{pmE}
\end{equation}

Looking at eqs.~(\ref{HamQ}) and (\ref{pmE}), we see that the energies associated
with particles and antiparticles are not the same, contrary to what one gets when dealing 
with the Klein-Gor\-don Hamiltonian. These energies are, nevertheless,
always non-negative,
solving the ``negative energy problem'' that we faced when dealing
with the first quantized fields (section \ref{vzero}). Moreover, if 
$m\rightarrow -m$ we get that $E_{\mathbf{k}}^+ \rightarrow E_{\mathbf{k}}^-$. In this
sense, when it comes to their energies, the particles and antiparticles can be regarded as effectively differing by the sign of its mass. However, as can be seen looking at 
eqs.~(\ref{PQ}) and (\ref{QQ}), the linear momentum and conserved charge do not depend
explicitly on the sign of the mass $m$.
We can also understand the meaning of $E^{\pm}_{\mathbf{k}}$ by noting that 
$E^{+}_{\mathbf{k}}$ is actually the kinetic energy associated with a classical 
relativistic particle. In this way, $E^{-}_{\mathbf{k}}$ is the ``kinetic energy''
of a particle with an effective negative mass and it is in this sense that the discussion
carried out in section \ref{vzero} about negative masses should be 
understood.\footnote{Note that in the present theory 
the mass of a particle cannot be proportional to the
expectation value of the Hamiltonian in its rest 
frame, where the expectation value is computed using a single particle state 
$a^\dagger_{\mathbf{k}}|0\rangle=|1_{\mathbf{k}}\rangle$, with $\mathbf{k}=0$ since
we are in the particle's rest frame \cite{gre95}.
This is true because Eqs.~(\ref{HamQ}) and (\ref{pmE}) give
$\langle 1_{\mathbf{k}}|H|1_{\mathbf{k}}\rangle=0$. This implies that the usual 
proof stating that particles and antiparticles must have the same mass as 
defined by the previous expectation value 
cannot be applied here (see pages $331$ and $332$ of \cite{gre95}).}

One might wonder if it is not possible to force the energies of the particles and
antiparticles to be the same, in particular equal to $E_{\mathbf{k}}$, 
by properly adjusting the values of $f_{\pm}(|\mathbf{k}|)$ in eq.~(\ref{geral}).
As we show in the appendix \ref{apA}, this is possible. But the price to pay is 
a theory that is apparently no longer local. 
This feature shows up since we can only make 
particles and antiparticles have the same energy if 
$[\Psi(t,\mathbf{r}),\Psi^\dagger(t,\mathbf{r'})]\neq 0$, which leads to a
violation of the microcausality condition. Also, the fact that
the latter commutator is not null implies a modification to eqs.~(\ref{com1})
and (\ref{com2}), i.e., we no longer have a canonically second quantized theory
[see appendix \ref{apA}].
Unless stated otherwise, we will not  
deal with this scenario
in the remaining of this section.

\subsection{Continuous and discrete symmetries}

The conserved Noether ``charges'' obtained for the classical fields in 
section \ref{noethertheorem} manifest themselves here, for instance, 
in the following commutators being zero [see eqs.~(\ref{HamQ})-(\ref{QQ})],
\begin{equation}
[H,H] = [H,P^j] = [H,\tilde{Q}] = 0.
\end{equation}
This means that the energy, the momentum, and the charge
are all conserved for the free field.

Moving on to the discrete symmetries, we can now finish the discussion initiated in
section \ref{ds}. Starting with space inversion (parity), we can now insert 
eq.~(\ref{geral2}) into (\ref{parity}) to obtain the following transformation rules for 
the creation (annihilation) operators,
\begin{eqnarray}
\mathcal{\hat{P}}a_{\mathbf{k}}\mathcal{\hat{P}}^\dagger = a_{-\mathbf{k}},&
\mathcal{\hat{P}}a^\dagger_{\mathbf{k}}\mathcal{\hat{P}}^\dagger = a^\dagger_{-\mathbf{k}},
\label{tfirst}\\
\mathcal{\hat{P}}b_{\mathbf{k}}\mathcal{\hat{P}}^\dagger = b_{-\mathbf{k}},&
\hspace{.1cm}
\mathcal{\hat{P}}b^\dagger_{\mathbf{k}}\mathcal{\hat{P}}^\dagger = b^\dagger_{-\mathbf{k}}.
\end{eqnarray}
Proceeding similarly with the time reversal operator we get
\begin{eqnarray}
\mathcal{\hat{T}}a_{\mathbf{k}}\mathcal{\hat{T}}^\dagger = a_{-\mathbf{k}},&
\mathcal{\hat{T}}a^\dagger_{\mathbf{k}}\mathcal{\hat{T}}^\dagger = a^\dagger_{-\mathbf{k}},
\\
\mathcal{\hat{T}}b_{\mathbf{k}}\mathcal{\hat{T}}^\dagger = b_{-\mathbf{k}},&
\hspace{.1cm}
\mathcal{\hat{T}}b^\dagger_{\mathbf{k}}\mathcal{\hat{T}}^\dagger = b^\dagger_{-\mathbf{k}}.
\label{tfourth}
\end{eqnarray}

The analysis of charge conjugation, as anticipated in section \ref{ds}, is a little more 
involved. If we apply the standard charge conjugation operator $\mathcal{\hat{C}}$,
i.e. eq.~(\ref{chargeC}), to
eq.~(\ref{geral2}) we get,
\begin{equation}
\mathcal{\hat{C}}\Psi(x)\mathcal{\hat{C}}^\dagger = \int \widetilde{dk} 
( \mathcal{\hat{C}}a_{\mathbf{k}}\mathcal{\hat{C}}^\dagger e^{i\mu x^0}e^{-ikx}
+ \mathcal{\hat{C}}b_{\mathbf{k}}^\dagger\mathcal{\hat{C}}^\dagger e^{i\mu x^0}e^{ikx}). \label{geral2C1}
\end{equation}
where $\widetilde{dk}=\sqrt{\hbar c^2/[(2\pi)^32\omega_{\mathbf{k}}]}d^3k$.
Comparing eq.~(\ref{geral2C1}) wi\-th the complex conjugate of eq.~(\ref{geral2}),
\begin{equation}
\Psi^\dagger (x)= \int \widetilde{dk} 
(b_{\mathbf{k}} e^{-i\mu x^0}e^{-ikx} + 
a_{\mathbf{k}}^\dagger e^{-i\mu x^0}e^{ikx}), \label{geral2C2}
\end{equation}
we realize that is not possible to identify 
$\mathcal{\hat{C}}a_{\mathbf{k}}\mathcal{\hat{C}}^\dagger$ with $b_{\mathbf{k}}$ and
$\mathcal{\hat{C}}b_{\mathbf{k}}^\dagger\mathcal{\hat{C}}^\dagger$ with 
$a_{\mathbf{k}}^\dagger$ due to the presence of the term $e^{i\mu x^0}$ in
eq.~(\ref{geral2C1}) and $e^{-i\mu x^0}$ in eq.~(\ref{geral2C2}). Defining 
$\mathcal{\hat{C}}$ antiunitary will not do the job either since in this case eq.~(\ref{geral2C1})
becomes
\begin{equation}
\mathcal{\hat{C}}\Psi(x)\mathcal{\hat{C}}^\dagger = \int \widetilde{dk} 
( \mathcal{\hat{C}}a_{\mathbf{k}}\mathcal{\hat{C}}^\dagger e^{-i\mu x^0}e^{ikx}
+ \mathcal{\hat{C}}b_{\mathbf{k}}^\dagger\mathcal{\hat{C}}^\dagger e^{-i\mu x^0}e^{-ikx}). \label{geral2C3}
\end{equation}
Comparing eq.~(\ref{geral2C3}) with (\ref{geral2C2}) we see that we do not get the expected interchange 
of particles with antiparticles. Actually we get 
$\mathcal{\hat{C}}a_{\mathbf{k}}\mathcal{\hat{C}}^\dagger=a_{\mathbf{k}}^\dagger$
and $\mathcal{\hat{C}}b_{\mathbf{k}}^\dagger\mathcal{\hat{C}}^\dagger=b_{\mathbf{k}}$.

However, we can get a bona fide charge conjugation operator if 
we use $\mathcal{\hat{C}}_m$ as given by eq.~(\ref{Cm}). The operator
$\mathcal{\hat{C}}_m$ incorporates the ``mass conjugation''
operation as defined in eq.~(\ref{massconjugation}). This means that 
$\mathcal{\hat{C}}_m$ anticommutes with any function of the mass $m$.
In this case we have,
\begin{eqnarray}
\mathcal{\hat{C}}_m\Psi(x)\mathcal{\hat{C}}_m^\dagger &=& \int \widetilde{dk} 
( \mathcal{\hat{C}}_ma_{\mathbf{k}}\mathcal{\hat{C}}_m^\dagger e^{-i\mu x^0}e^{-ikx}
+ \mathcal{\hat{C}}_mb_{\mathbf{k}}^\dagger\mathcal{\hat{C}}_m^\dagger e^{-i\mu x^0}e^{ikx}), \label{geral2CA}
\end{eqnarray}
since $\mu$ is linear in $m$ and $k^0$ and $\widetilde{dk}$ are functions of $m^2$.
Comparing eq.~(\ref{geral2CA}) with (\ref{geral2C2}) we immediately get
\begin{eqnarray}
\mathcal{\hat{C}}_ma_{\mathbf{k}}\mathcal{\hat{C}}_m^\dagger = b_{\mathbf{k}},&
\mathcal{\hat{C}}_ma^\dagger_{\mathbf{k}}\mathcal{\hat{C}}_m^\dagger = b^\dagger_{\mathbf{k}}, \label{tfifth}
\\
\mathcal{\hat{C}}_mb_{\mathbf{k}}\mathcal{\hat{C}}_m^\dagger = a_{\mathbf{k}},&
\hspace{.1cm}
\mathcal{\hat{C}}_mb^\dagger_{\mathbf{k}}\mathcal{\hat{C}}_m^\dagger = a^\dagger_{\mathbf{k}}, \label{tsixth}
\end{eqnarray}
the desired properties of a good charge conjugation operator.

We should also mention that although we used eq.~(\ref{geral2}) to arrive at 
the previous transformation properties for the creation and annihilation operators, 
all the results remain valid had we used
eq.~(\ref{geral}) and the fact that 
$\mathcal{\hat{C}}_m\widetilde{dk}_{\pm}\mathcal{\hat{C}}_m^\dagger =\widetilde{dk}_{\mp}$.
In particular, this means that all the previous results apply to 
the creation and annihilation operators of the non-canonically 
second quantized theory given 
in the appendix \ref{apA}.

\subsection{Microcausality and the Lorentz covariance of the field commutators}
\label{microSec}

Our first task will be the explicit calculation of the commutator $[\Psi(x),\Psi^\dagger(y)]$ 
at arbitrary space-time points $x=(x^0,x^1,$ $x^2,x^3)$ and 
$y=(y^0,y^1,y^2,y^3)$. This will allow us to clearly see its Lorentz covariance
and prove that the Lorentz covariant Schr\"odinger fields respect the microcausality
condition \cite{man86,gre95}.
Then, we will present a general argument connecting the commutators between the
Lorentz covariant Schr\"odinger fields with the commutators of the Klein-Gordon fields,
allowing us to readily extend the commutation properties of the latter 
to the former fields.

Let us start introducing the following shorthand notation for the field operators,
\begin{eqnarray}
\Psi(x) = \Psi_x,& \Psi^\dagger(x) = \Psi_x^\dagger, \label{short1}\\
\Pi_\Psi(x)=\Pi_x,& \Pi_{\Psi^\dagger}(x)=\Pi^\dagger_x, \label{short2}
\end{eqnarray}
where the last equality causes no confusion since for the Lorentz covariant 
Schr\"odinger fields we always have $\Pi_{\Psi^\dagger}(x)$ $=$ $\Pi_{\Psi}^\dagger(x)$.
Now, using eqs.~(\ref{geral}), (\ref{short1}), and (\ref{short2}) we get 
\begin{eqnarray}
[\Psi(x),\Psi^\dagger(y)] = [\Psi^+_x,(\Psi^+_y)^\dagger] + [\Psi^-_x,(\Psi^-_y)^\dagger],
\label{psipsidagger}
\end{eqnarray}
where we used that $[\Psi^+_x,(\Psi^-_y)^\dagger]=[\Psi^-_x,(\Psi^+_y)^\dagger]=0$ to arrive 
at the right hand side of 
eq.~(\ref{psipsidagger}).\footnote{Note that $(\Psi^+_x)^\dagger$ is one thing
and $(\Psi^\dagger_x)^+$ is another. The latter is the positive frequency part of 
$\Psi^\dagger_x$, associated with the destruction operator $b_\mathbf{k}$,
while the former is the adjoint of $\Psi^+_x$.} 
This is true since $\Psi^+_x$ is proportional to the 
operator $a_\mathbf{k}$, $(\Psi^-_y)^\dagger$ to $b_\mathbf{k}$, 
and 
$[a_\mathbf{k},b_\mathbf{k'}]=0$.

Using the explicit plane wave expansions for $\Psi_x^\pm$ as given in eq.~(\ref{geral}), 
and the commutation relations (\ref{aad}), we obtain
\begin{equation}
[\Psi^\pm_x,(\Psi^\pm_y)^\dagger] = i\hbar c \Delta^\pm_{LS}(x-y),
\label{lspm}
\end{equation}
where
\begin{eqnarray}
\Delta_{LS}^\pm(x) &=& e^{i\mu x^0}\Delta^\pm(x), \label{lsdeltapm} \\
\Delta^\pm(x) &=& \frac{1}{\pm i\hbar c}\int d^3k \; [f^\pm (\mathbf{k})]^2
e^{\mp ikx}. \label{deltapm}
\end{eqnarray}
The subscript ``LS'' reminds us that the quantities above are the ones
for the Lorentz covariant Schr\"odinger fields. 
The same quantities without the subscript,
as we will see shortly if we use eq.~(\ref{f=f}), 
are the ones we obtain dealing with the complex 
Klein-Gordon fields.

Using eqs.~(\ref{lspm}) and (\ref{lsdeltapm}) we can write eq.~(\ref{psipsidagger}) as
\begin{equation}
[\Psi(x),\Psi^\dagger(y)] = i\hbar c\, \Delta_{LS}(x-y), \label{commutatorLS}
\end{equation}
where 
\begin{eqnarray}
\Delta_{LS}(x) &=& \Delta_{LS}^+(x)+\Delta_{LS}^-(x) 
=e^{i\mu x^0} [\Delta^+(x)+\Delta^-(x)] 
=e^{i\mu x^0} \Delta(x). \label{deltalsdelta}
\end{eqnarray}

If we now use eq.~(\ref{f=f}) we get
\begin{equation}
\Delta^{\pm}(x) = \frac{\mp ic}{2(2\pi)^3}\int d^3k\, 
\frac{e^{\mp ikx}}{\omega_{\mathbf{k}}}
=-\Delta^{\mp}(-x) \label{deltapm2}
\end{equation}
and
\begin{equation}
\Delta(x) = \Delta^+(x)+\Delta^-(x)= \frac{-c}{(2\pi)^3}\int d^3k\, 
\frac{\sin (kx)}{\omega_{\mathbf{k}}}.
\label{deltakg}
\end{equation}
As usual, we are employing the definitions given in eqs. (\ref{omegak})-(\ref{mi}).
Equations (\ref{deltapm2}) and (\ref{deltakg}) are exactly the ones we get when 
working with the complex Klein-Gordon Lagrangian \cite{man86,gre95}. Note that
$\Delta(x)$ and $\Delta^{\pm}(x)$ satisfy the Klein-Gordon equation \cite{man86,gre95} and
a direct calculation shows that $\Delta_{LS}(x)$ and $\Delta^{\pm}_{LS}(x)$
satisfy the Lorentz covariant Schr\"odinger equation.

The Lorentz invariant form of eq.~(\ref{deltakg}) is \cite{man86}
\begin{equation}
\Delta(x) = \frac{-i}{(2\pi)^3}\int d^4k\, 
\delta(k^2-\mu^2)\varepsilon(k^0)e^{-ikx},
\label{deltakg2}
\end{equation}
where $d^4k=dk^0dk^1dk^2dk^3$, $k^2=k_{\nu}k^{\nu}$, $\mu$, and $kx=k_{\nu}x^{\nu}$ 
are Lorentz invariants and the integrals run from $-\infty$ to $\infty$. Here
$\varepsilon(k^0)$ gives the sign of $k^0$, which is also Lorentz
invariant.\footnote{The sign of $k^0$ is invariant under proper Lorentz 
transformations for time-like vectors, which is what we have here for the four-vector $k$
due to the mass shell condition $\delta(k^2-\mu^2)$. By a proper Lorentz transformation
we mean spatial rotations and Lorentz boosts. Time reversal
and space inversion are excluded.}

Looking at eqs.~(\ref{commutatorLS}) and (\ref{deltalsdelta}), 
it is clear that the commutator (\ref{commutatorLS}) is covariant under
spatial rotations. Let us see what happens under a Lorentz boost. 
Assuming, without loss of generality, a Lorentz boost along the $x^1$-direction we have
\begin{eqnarray}
\Delta(x) &\longrightarrow& \Delta(x), \label{d-d}\\
\Psi(x)  &\longrightarrow& e^{i\theta(x)/\hbar}\Psi(x), \label{psi-psi}\\
x^0 &\longrightarrow& \gamma (x^0+\beta x^1), \label{x-x}
\end{eqnarray}
with $\theta(x)$ given by eq.~(\ref{thetaL}). 
Using eqs.~(\ref{d-d})-(\ref{x-x}), a direct calculation gives
\begin{equation}
[\Psi_x,\!\Psi_y^\dagger] = i\hbar c \Delta_{LS}(x\!-\!y)\! \longrightarrow\!
[\Psi_x,\!\Psi_y^\dagger] = i\hbar c \Delta_{LS}(x\!-\!y).
\label{LIcommutator}
\end{equation}
This last result, together with the covariance of the commutator under spatial rotations, proves the covariance of 
$[\Psi(x),$ $\!\Psi^\dagger(y)]$ under proper Lorentz transformations.

If we compute the previous commutator at equal times  
[see eq.~(\ref{f1}) of the appendix \ref{apA}] we 
get
\begin{equation}
[\Psi(t,\mathbf{x}),\!\Psi^\dagger(t,\mathbf{y})] = 0,
\end{equation}
which implies according to eq.~(\ref{commutatorLS}) that
\begin{equation}
\Delta_{LS}(0,\mathbf{x-y}) =0.
\label{dls0}
\end{equation}
When the time argument is zero, eq.~(\ref{deltalsdelta}) gives
\begin{equation}
\Delta_{LS}(0,\mathbf{x-y}) = \Delta(0,\mathbf{x-y}) 
\end{equation}
and thus according to eq.~(\ref{dls0}), 
\begin{equation}
\Delta(0,\mathbf{x-y}) = 0.
\end{equation}

But since $\Delta(x-y)$ is invariant under a proper Lorentz transformation, we have that 
$\Delta(x-y)=0$ whenever $(x-y)^2<0$. This is true because any two space-like vectors are connected to each
other via a proper Lorentz transformation. Also, 
the fact $\Delta(x-y)=0$ when $(x-y)^2<0$ and eq.~(\ref{deltalsdelta}) 
imply that $\Delta_{LS}(x-y)=0$ whenever $(x-y)^2<0$. This last result
combined with eq.~(\ref{commutatorLS}) and with eq.~(\ref{LIcommutator}),
which shows the Lorentz
covariance of the commutator, lead to
\begin{equation}
[\Psi(x),\Psi^\dagger(y)] = 0, \hspace{.25cm}\mbox{if}\hspace{.25cm} (x-y)^2<0. \label{micro}
\end{equation}
Equation (\ref{micro}), telling us that fields with space-like separation commute,
is the microcausality condition. This guarantees that the 
second quantized Lorentz covariant Schr\"odin\-ger Lagrangian leads to a local
quantum field theory \cite{man86,gre95}. 

We can also connect the field commutators of  
the Lorentz covariant Schr\"odinger theory with the 
ones of the Klein-Gordon theory as follows. First, if we use
eq.~(\ref{transformationKG2}), i.e., the relation connecting the Lorentz
covariant Schr\"odinger Lagrangian with the Klein-Gordon one, we get  
\begin{equation}
[\Psi(x),\Psi^\dagger(y)] = e^{i\mu (x^0-y^0)}[\Phi(x),\Phi^\dagger(y)],
\label{phiphikg}
\end{equation}
where $\Phi(x)$ and $\Phi^\dagger(x)$ are the complex Klein-Gordon fields.
Moreover, applying eq.~(\ref{transformationKG2}) to (\ref{pi}) we obtain
\begin{equation}
\Pi_{\Psi}(x) = e^{-i\mu x^0} \Pi_{\Phi}(x),
\label{pikg}
\end{equation}
where $\Pi_{\Phi}(x)$ is the conjugate momentum to $\Phi(x)$. Thus, using 
eqs.~(\ref{transformationKG2}) and (\ref{pikg}), the remaining non-trivial (not obviously 
zero) commutators can be written as
\begin{eqnarray}
[\Pi_{\Psi}(x),\Pi_{\Psi^\dagger}(y)] \!\!&\!=\!&\!\! e^{-i\mu (x^0-y^0)}
[\Pi_{\Phi}(x),\Pi_{\Phi^\dagger}(y)], \label{pipikg}\\ \,
[\Psi(x),\Pi_{\Psi}(y)] &=& e^{i\mu (x^0-y^0)}
[\Phi(x),\Pi_{\Phi}(y)]. \label{phipikg}
\end{eqnarray}

The first thing worth noting is that the equal time commutation relations of the Lorentz
covariant Schr\"odinger and Klein-Gordon fields are the same. Indeed, for equal
times we have $x^0=y^0$ and according to eqs.~(\ref{phiphikg}), (\ref{pipikg}),
and (\ref{phipikg}) we immediately get
\begin{eqnarray}
[\Psi(t,\mathbf{x}),\Psi^\dagger(t,\mathbf{y})] &=& 
[\Phi(t,\mathbf{x}),\Phi^\dagger(t,\mathbf{y})], \\ \,
[\Pi_{\Psi}(t,\mathbf{x}),\Pi_{\Psi^\dagger}(t,\mathbf{y})] \!\!&\!=\!&\!\! 
[\Pi_{\Phi}(t,\mathbf{x}),\Pi_{\Phi^\dagger}(t,\mathbf{y})], \\ \,
[\Psi(t,\mathbf{x}),\Pi_{\Psi}(t,\mathbf{y})] &=& 
[\Phi(t,\mathbf{x}),\Pi_{\Phi}(t,\mathbf{y})]. 
\end{eqnarray}
As such, we can understand eqs.~(\ref{transformationKG2}) and (\ref{pikg}) as a canonical
transformation since they preserve the commutation relations as we move to the new ``variables'', namely, fields. It is also important to remark that eqs.~(\ref{phiphikg}),
(\ref{pipikg}), and (\ref{phipikg}) are only valid if eq.~(\ref{f=f}) is true, since 
this is the only way to guarantee a canonical quantization of the Lorentz covariant
Schr\"odinger fields. Therefore, eqs.~(\ref{phiphikg}),
(\ref{pipikg}), and (\ref{phipikg}) do not apply to the non-canonically quantized 
Lorentz covariant Schr\"odinger fields given in the appendix \ref{apA}, where,
for example, $[\Psi(t,\mathbf{x}),\Psi^\dagger(t,\mathbf{y})]\neq 0$, in clear contradiction
to eq.~(\ref{phiphikg}) and the fact that 
$[\Phi(t,\mathbf{x}),\Phi^\dagger(t,\mathbf{y})]= 0$.

Before we move on to the computation of the Feynman propagator, we note that we can
write $\Delta^\pm(x)$ and $\Delta(x)$ as the following integrals,
where $k^0$ is considered a complex variable \cite{man86},
\begin{eqnarray}
\Delta^\pm(x) &=& -\frac{1}{(2\pi)^4}\int_{C^{\pm}}d^4k\,
\frac{e^{-ikx}}{k^2-\mu^2}, \label{contourCpm}\\
\Delta(x) &=& -\frac{1}{(2\pi)^4}\int_Cd^4k\,
\frac{e^{-ikx}}{k^2-\mu^2}. \label{contourC}
\end{eqnarray}
The complex contour integral in $k_0$ is such that
$C^\pm$ is any counterclockwise closed
path encircling only $\pm \omega_{\mathbf{k}}/c$ and
$C$ is any counterclockwise closed
path encircling $\omega_{\mathbf{k}}/c$ and $-\omega_{\mathbf{k}}/c$,
the two simple poles of the integrand.

\subsection{The Feynman propagator}

The first step towards a perturbative 
treatment of any interaction involving the Lorentz covariant Schr\"odinger fields,  
or involving them and other types of fields, 
is the calculation of the Feynman propagator. The Feynman propagator 
$\Delta_{F_{\!_{LS}}}(x-x')$ is defined as \cite{man86,gre95}
\begin{equation}
i\hbar c\, \Delta_{F_{\!_{LS}}}(x-x') = \langle 0| T\left\{\Psi(x)\Psi^\dagger(x')\right\} 
\!|0\rangle, \label{FPLS}
\end{equation}
where $|0\rangle$ is the vacuum state and  
we use the subscript ``LS'' to differentiate the Feynman propagator associated with
the Lorentz covariant Schr\"odinger fields from $\Delta_F(x)$, the Feynman propagator
related to the Klein-Gordon fields. 

In eq.~(\ref{FPLS}) the symbol $T$ denotes time ordering, giving rise to the 
T-product (time-ordered product),
\begin{eqnarray}
T\left\{\Psi(x)\Psi^\dagger(x')\right\} &=& h(x^0-x^{0'})\Psi(x)\Psi^\dagger(x')
+h(x^{0'}-x^0)\Psi^\dagger(x')\Psi(x), \label{Tproduct}
\end{eqnarray}
with $h$ being the Heaviside step function,
\begin{eqnarray}
h(x^0) &=& 1, \;\mbox{if}\; x^0>0, \nonumber \\
h(x^0) &=& 0, \;\mbox{if}\; x^0<0.
\label{hfunction}
\end{eqnarray}
Using eq.~(\ref{lspm}) and 
remembering that $\Psi(x) =\Psi^+(x) + \Psi^-(x)$,  with $\Psi^+(x)$ and $\Psi^-(x)$
respectively functions of annihilation and creation operators, we have 
\begin{eqnarray}
\langle 0|\Psi(x)\Psi^\dagger(x')|0\rangle &=& i\hbar c\, \Delta_{LS}^+(x-x'), 
\label{ppd}\\
\langle 0|\Psi^\dagger(x')\Psi(x)|0\rangle &=& -i\hbar c\, \Delta_{LS}^-(x-x').
\label{pdp}
\end{eqnarray}
If we now insert eqs.~(\ref{Tproduct}), (\ref{ppd}), and (\ref{pdp}) into
(\ref{FPLS}) we get
\begin{equation}
\Delta_{F_{\!_{LS}}}(x) =   h(x^0)\Delta_{LS}^+(x) 
- h(-x^0)\Delta_{LS}^-(x) \label{FPLS2}
\end{equation}
and also
\begin{equation}
\Delta_{F_{\!_{LS}}}(x) = e^{i\mu x^0}\Delta_F(x),  \label{FPLS3}
\end{equation}
if we employ eq.~(\ref{lsdeltapm}). Here 
$\Delta_F(x)= h(x^0)\Delta^+(x)
- h(-x^0)$ $\Delta^-(x)$ is the Feynman propagator associated with the complex field
Klein-Gordon Lagrangian, also known as the char\-ged meson propagator \cite{man86,gre95}.

Under a proper Lorentz transformation $\Psi(x)$ changes according to 
eqs.~(\ref{transformationL}) and (\ref{thetaL}) while $h(x^0)$ is an invariant.
The latter is true because the sign of $x^0$ for 
time-like vectors is not altered under a proper Lorentz transformation. 
These two facts
lead to the following 
transformation rule for eq.~(\ref{FPLS}),
\begin{equation}
\Delta_{F_{\!_{LS}}}(x) \longrightarrow 
e^{i\theta(x)/\hbar}\Delta_{F_{\!_{LS}}}(x).
\label{dLorentz}
\end{equation}
To arrive at eq.~(\ref{dLorentz}), which tells us how the Feynman propagator of the Lorentz covariant Schr\"odinger fields transforms under a proper Lorentz transformation,
we have also assumed that the latter changes the vacuum state $|0\rangle$ 
by at most a global phase. Equation (\ref{dLorentz}), 
combined with the transformation rule for $x^0$ 
and the invariance of $\Delta_F(x)$
under a proper Lorentz transformation, allow us to show that 
eq.~(\ref{FPLS3}) is covariant under a proper Lorentz transformation, i.e.,
\begin{equation}
\Delta_{F_{\!_{LS}}}(x) = e^{i\mu x^0}\Delta_F(x) \longrightarrow 
\Delta_{F_{\!_{LS}}}(x) = e^{i\mu x^0}\Delta_F(x). \label{dCovariant}
\end{equation}

\begin{figure}[!ht]
\centering
\includegraphics[width=8.5cm]{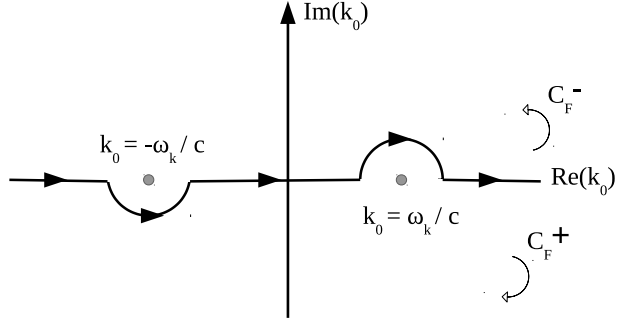}
\caption{\label{fig2} The contours of integration leading to the 
Feynman propagator for the Lorentz covariant Schr\"odinger fields.}
\end{figure}

Due to the connection between the Klein-Gordon and the Lorentz covariant 
Schr\"odin\-ger propagators as given by eq. (\ref{FPLS3}), we can carry over the integral 
representations of the former to the latter. Following ref.~\cite{man86} we have
\begin{equation}
\Delta_{F_{\!_{LS}}}(x) = \frac{e^{i\mu x^0}}{(2\pi)^4}\int_{C_F}d^4k\,
\frac{e^{-ikx}}{k^2-\mu^2},
\end{equation}
where the contour of integration $C_F$ is given in figure \ref{fig2} 
and should be understood in the following sense. 
For $x^0>0$ the contour of integration is closed in the lower half of the 
complex $k^0$-plane,
leading to a clockwise contour integration whose path we denote by $C_F^+$. 
On the other hand, for $x^0<0$ we use the upper half plane to close
the path of integration, which gives a counterclockwise integration around path 
$C_F^-$. In this way, computing the complex integral
in $k^0$ using the residue theorem, we obtain $\Delta_{LS}^+(x)$ for the former
contour of integration 
and $-\Delta_{LS}^-(x)$ for the latter, the 
expected results for the Feynman propagator [see eq.~(\ref{FPLS2})].

We can also write the Feynman propagator with all integrals being real variable
integrals if we slightly displace the poles off the real axis, 
\begin{equation}
\pm \frac{\omega_{\mathbf{k}}}{c} \longrightarrow 
\pm \frac{\omega_{\mathbf{k}}}{c} \mp i\eta,
\end{equation}
with $0<\eta\ll 1$, and use the following integral representation,
\begin{eqnarray}
\Delta_{F_{\!_{LS}}}(x)
&=& \frac{e^{i\mu x^0}}{(2\pi)^4}\int d^4k\,
\frac{e^{-ikx}}{(k^0)^2-(\omega_{\mathbf{k}}/c-i\eta)^2} 
=\frac{e^{i\mu x^0}}{(2\pi)^4}\int d^4k\,
\frac{e^{-ikx}}{k^2-\mu^2+i\epsilon}.
\label{integralDFLS}
\end{eqnarray}
To arrive at the last equality we have neglected the $\eta^2$ term since 
$\eta \ll 1$ and made the following identification,
\begin{equation}
\epsilon = \frac{2\eta\omega_{\mathbf{k}}}{c}.
\end{equation}

With this definition the $k^0$ variable is also integrated from 
$-\infty$ to $\infty$. After the integration in $k^0$ 
we take the limit $\eta\rightarrow 0$, or equivalently $\epsilon \rightarrow 0$, 
obtaining the corresponding expression 
for the Feynman propagator, namely, $\Delta_{LS}^+(x)$ for $x^0>0$ and 
$-\Delta_{LS}^-(x)$ for $x^0<0$. 

Using eq.~(\ref{integralDFLS}) we can show that the Feynman propagator 
$\Delta_{F_{\!_{LS}}}(x)$ is the Green's function of the free Lorentz covariant
Schr\"odinger equation. Setting $V=0$ in eq.~(\ref{GeneralEq7}) and replacing 
$\Psi(x)$ for $\Delta_{F_{\!_{LS}}}(x)$ as given in eq.~(\ref{integralDFLS}), we
obtain after a little calculation and after taking the limit 
$\epsilon \rightarrow 0$ that 
\begin{equation}
\partial_\nu\partial^\nu\Delta_{F_{\!_{LS}}}(x) -i2\mu\partial_0 \Delta_{F_{\!_{LS}}}(x)
= -\frac{e^{i\mu x^0}}{(2\pi)^4}\int d^4k\, e^{-ikx}.
\end{equation}
Noting that the 
four dimensional Dirac delta function can be written as 
$(2\pi)^4\delta^{(4)}(x)=\int d^4k\, e^{-ikx}$ we get  
\begin{eqnarray}
\partial_\nu\partial^\nu\Delta_{F_{\!_{LS}}}(x) -i2\mu\partial_0 \Delta_{F_{\!_{LS}}}(x)
&=& -e^{i\mu x^0}\delta^{(4)}(x) 
= -\delta^{(4)}(x).
\label{green1}
\end{eqnarray}
We have used that $e^{i\mu x^0}\delta^{(4)}(x)=\delta^{(4)}(x)=0$ for 
$x^0\neq 0$ and that $\lim_{x^0\rightarrow 0 }e^{i\mu x^0}\delta^{(4)}(x)=
\lim_{x^0\rightarrow 0 }\delta^{(4)}(x)$ to arrive at the last equality. 
If we equate the left hand side of eq.~(\ref{green1}) to zero we get 
the free Lorentz covariant Schr\"odinger equation. Equating it to $-\delta^{(4)}(x)$ 
we obtain the definition of its Green's function,  
proving that $\Delta_{F_{\!_{LS}}}(x)$ is indeed the Green's function of 
the free Lorentz covariant Schr\"odinger equation. For the Klein-Gordon equation,
the analogous expression to eq.~(\ref{green1}) is \cite{gre95}
\begin{displaymath}\partial_\nu\partial^\nu\Delta_{F}(x) + \mu^2\Delta_{F}(x) = -\delta^{(4)}(x). 
\end{displaymath}

\subsection{Turning on the interaction among the fields}

The techniques built to perturbatively handle scattering problems among interacting 
quantum fields, in particular those suited to the Klein-Gordon fields \cite{man86,gre95}, 
remain valid for the Lorentz covariant Schr\"odinger fields. 
Working in the interaction picture,\footnote{In previous sections of this work, 
specially when dealing with the second quantization of the free fields, 
we were working in the Heisenberg picture.} it is not difficult to see that the 
Lorentz covariant Schr\"odinger fields 
evolve according to the same dynamical equations and satisfy the same commutation relations of the free fields. On the other hand, the evolution of the quantum state describing the initial particle configuration evolves according to the Schr\"odinger equation, whose Hamiltonian is given by the interaction part of the total Hamiltonian describing the system being investigated.
Moreover, a simple inspection on the proofs leading to the Dyson series and to Wick's theorem shows that they remain valid here too. 

As such, the probability amplitude for a system initially in the state 
$|\alpha_{in}\rangle$ to be found after the collision (scattering) in
the final state $|\alpha_{out}\rangle$ is 
\begin{equation}
S_{fi} = \langle \alpha_{out}| S | \alpha_{in}\rangle.
\label{sfi}
\end{equation}
Here, invoking the adiabatic hypothesis, $|\alpha_{in(out)}\rangle$ are eigenstates of
the free-field Hamiltonian, with $|\alpha_{in}\rangle$ the initial state in the remote 
past ($t\rightarrow -\infty$) and $|\alpha_{out}\rangle$ the final state in the far 
future ($t\rightarrow \infty$). The operator describing this transition is called the 
S-matrix and its manifestly covariant perturbative 
expansion is given by the Dyson series,
\begin{eqnarray}
S &=& T \exp\left\{-\frac{i}{\hbar c}\int d^4x \mathcal{H}^I_{int}(x)\right\} \nonumber \\
&=& \mathbb{1} + \sum_{n=1}^{\infty}\frac{[-i/(\hbar c)]^n}{n!}
\int d^4x_1\ldots d^4x_n T\left\{\mathcal{H}^I_{int}(x_1) \ldots 
\mathcal{H}^I_{int}(x_n)\right\}, \label{dyson}
\end{eqnarray}
where $T$ is the time ordering operator and $\mathcal{H}^I_{int}$ is the 
Hamiltonian density describing the interaction among the fields 
expressed in the interaction picture (the superscript $I$ denotes interaction
picture). Similarly to the Klein-Gordon case, the time-ordered term in the Dyson series
is expanded using Wick's theorem in order to carry out transition amplitude 
calculations. It is in this step that the Feynman propagator
(\ref{FPLS}) appears, being usually called 
in this context a ``contraction''. 

An important property of the Lorentz covariant Schr\"o\-din\-ger fields, valid 
when they are canonically quantized, is the simple relation connecting them to the complex 
Klein-Gordon fields as given by eqs.~(\ref{transformationKG2}) and (\ref{pikg}). 
These relations, 
originally shown to be true for the free fields being described in the Heisenberg picture,
remain valid in the interaction picture. This is easily seen by either inspecting the 
time evolution of these fields in the interaction picture or by noting that these transformations connecting the Lorentz covariant Schr\"odinger fields to the Klein-Gordon
ones are given by c-numbers and, as such, commute with the unitary transformation taking 
us from the Heisenberg to the interaction picture. Written in the interaction picture,
eqs.~(\ref{transformationKG2}) and (\ref{pikg}) become
\begin{eqnarray}
\Psi^I(x) &=& e^{i\mu x^0}\Phi^I(x), \label{ikg} \\
\Pi^I_{\Psi}(x) &=& e^{-i\mu x^0}\Pi^I_{\Phi}(x). \label{ipikg}
\end{eqnarray}

As can be seen explicitly in the appendix \ref{apC}, 
these relations are crucial to prove that to any order in perturbation theory and for Lorentz covariant interactions, the Klein-Gordon and the Lorentz covariant
Schr\"odinger theories are essentially equivalent. However, for interactions
violating Lorentz invariance, the predictions stemming from both theories are drastically 
different.\footnote{The non-canonically quantized theory given in the 
appendix \ref{apA}
is not equivalent to the Klein-Gordon theory either, even for Lorentz
covariant interactions. This is true because, 
contrary to the canonically quantized case, the propagator of the non-canonically
quantized theory is not trivially related to the Klein-Gordon one.}

\subsection{Breaking Lorentz invariance}
\label{bli}

The equivalence between the Klein-Gordon and the Lorentz covariant Schr\"odinger theories
established above is only valid for Lorentz invariant Lagrangian densities 
(actually, being more general, we just need Lorentz invariant actions). If we introduce,
nevertheless, interaction terms breaking Lorentz invariance, we can have certain processes 
that only occur in the Lorentz covariant Schr\"odinger theory. See, however, 
Refs. \cite{edw18,and04} for interesting and alternative ways of introducing 
interaction terms that break
Lorentz invariance in the framework of the standard Klein-Gordon theory and
in the Higgs sector.

For example, consider the following interaction term (as always, normal ordering is implied, and here it is also implied an adjoint term in order to have 
a Hermitian Hamiltonian),
\begin{equation}
\mathcal{H}_{int} \propto \Psi(\Psi^\dagger)^3. 
\label{breakingLI1}
\end{equation}
This term is not Lorentz invariant since after a proper Lorentz transformation we have,
according to eq.~(\ref{transformationL}),
\begin{equation}
\Psi(x)[\Psi^\dagger(x)]^3 \longrightarrow e^{-i2\theta(x)/\hbar}\Psi(x)[\Psi^\dagger(x)]^3. 
\end{equation}
The analogous term for 
the Klein-Gordon theory, namely, $\Phi(\Phi^\dagger)^3$, is invariant under a proper 
Lorentz transformation since in this case $\Phi \rightarrow \Phi$.

\begin{figure}[!ht]
\centering
\includegraphics[width=12cm]{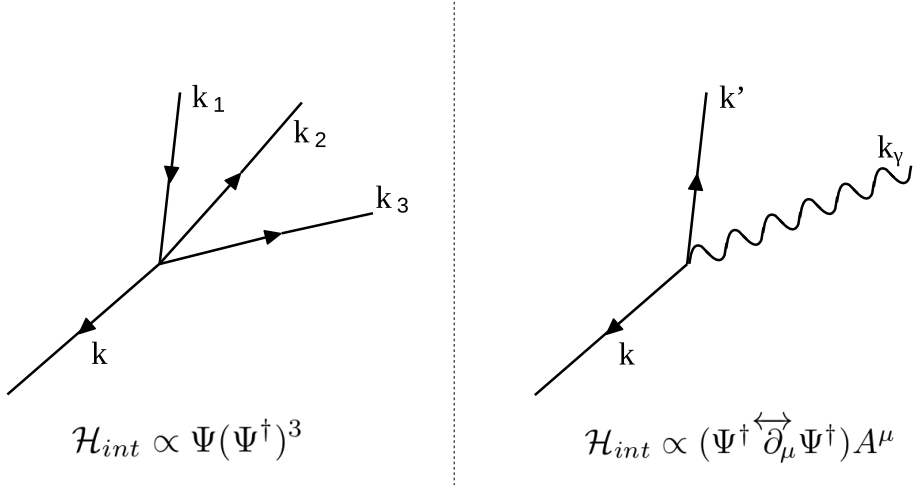}
\caption{\label{fig3} Feynman diagrams in momentum space showing 
two possible processes for the Lorentz covariant Schr\"odinger theory when we introduce  interaction terms that break Lorentz invariance. These two processes
are kinematically forbidden for the Klein-Gordon theory subjected to analogous
interaction terms. 
Left panel: An antiparticle ($b$-particle) with momentum $\mathbf{k}$ and energy 
$E^-_{\mathbf{k}}$ decays into another antiparticle and two particles ($a$-particles). 
The energies of the particles are $E^+_{\mathbf{k_3}}$ and
$E^+_{\mathbf{k_2}}$. We can represent this process by the following equation, 
$b\rightarrow baa$. Note that the process (not shown in the picture)
in which particles and antiparticles exchange roles, 
$a\rightarrow abb$, is kinematically forbidden. Right panel: An antiparticle 
decays into a photon and a particle, $b\rightarrow a\gamma$. 
The processes (not shown in the picture)
$a\rightarrow b\gamma$, $a\rightarrow a\gamma$, and $b\rightarrow b\gamma$ are all
kinematically forbidden.}
\end{figure}

An interaction term such as (\ref{breakingLI1}) allows, among other processes, for an
antiparticle  at rest to decay into two particles 
and one antiparticle, all of
them at rest too (see the left panel of figure \ref{fig3}). 
This process is kinematically forbidden for the Klein-Gordon theory,
violating conservation of energy, while for the Lorentz covariant Schr\"odinger theory
the total energy and momentum can be made equal before and after the decay. 

This can be seen by noting that to first order in perturbation theory (at the tree level),
this process is associated to the initial state 
$|\alpha_{in}\rangle=b^\dagger_\mathbf{k}|0\rangle$ 
and to the final state 
$|\alpha_{out}\rangle=
b^\dagger_\mathbf{k_1}a^\dagger_\mathbf{k_2}a^\dagger_\mathbf{k_3}|0\rangle$. 
A non-null probability amplitude is obtained using the following term coming from (\ref{breakingLI1}), 
\begin{equation}
\Psi^-[(\Psi^\dagger)^-]^2(\Psi^\dagger)^+,
\label{intBLI}
\end{equation}
where $(\Psi^\dagger)^+$ destroys the antiparticle, $\Psi^-$ creates another
antiparticle, and $[(\Psi^\dagger)^-]^2$ creates the remaining two particles. 
The previous field operators are defined  by noting 
that we can write $\Psi = \Psi^+ + \Psi^-$ and 
$\Psi^\dagger = (\Psi^\dagger)^+ +$ $(\Psi^\dagger)^-$, which after  
eq.~(\ref{geral2}) can be explicitly written as 
\begin{eqnarray}
\Psi^+(x) &=& \int\hspace{-.1cm} d^3k
f_{\mathbf{k}}
e^{i\mu x^0}e^{-ikx}a_{\mathbf{k}}, \label{psi+}\\
\Psi^-(x) &=& \hspace{-.1cm} 
\int\hspace{-.1cm} d^3k
f_{\mathbf{k}}
e^{i\mu x^0} e^{ikx}b_{\mathbf{k}}^\dagger, \\
\![\Psi^\dagger(x)]^- &=& \int\hspace{-.1cm} d^3k
f_{\mathbf{k}}
e^{-i\mu x^0}e^{ikx}a_{\mathbf{k}}^\dagger, \\
\![\Psi^\dagger(x)]^+ &=& \hspace{-.1cm} 
\int\hspace{-.1cm} d^3k
f_{\mathbf{k}}
e^{-i\mu x^0} e^{-ikx}b_{\mathbf{k}}, \label{psid-}
\end{eqnarray}
where
\begin{eqnarray}
f_{\mathbf{k}} &=& \left[\frac{\hbar c^2}{(2\pi)^3 2\omega_{\mathbf{k}}}\right]^{1/2}, 
\\
k^0 &=& \omega_{\mathbf{k}}/c = E_{\mathbf{k}}/(\hbar c), \\
E_{\mathbf{k}}&=&\sqrt{m^2c^4+\hbar^2c^2|\mathbf{k}|^2}.
\end{eqnarray}

Suppressing the superscript $I$ that indicates operators in the interaction picture,
eqs. (\ref{sfi}), (\ref{dyson}), and (\ref{intBLI}) give to first order in perturbation theory
\begin{equation}
S_{fi}^{(1)} =  \langle \alpha_{out}| S^{(1)} | \alpha_{in}\rangle,
\end{equation}
where
\begin{eqnarray}
S^{(1)} \!&\!=\!&\! -\frac{i}{\hbar c}\int d^4x \mathcal{H}_{int}  
=-\frac{ig}{\hbar c}\int d^4x 
\Psi^-(x)\{[\Psi^\dagger(x)]^-\}^2[\Psi^\dagger(x)]^+,   
\end{eqnarray}
with $g$ being the effective coupling constant associated to this process.
If we now use eqs.~(\ref{psi+})-(\ref{psid-}) and the explicit expansion for 
$|\alpha_{in(out)}\rangle$ we get
%
\begin{eqnarray}    
 S_{fi}^{(1)} &=&  -\frac{ig}{\hbar c}\int d^4x d^3q_1\ldots d^3q_4 f_{\mathbf{q_1}}
\ldots f_{\mathbf{q_4}} e^{-i[2\mu x^0+(q_4-q_1-q_2-q_3)x]} \nonumber \\
&& \times \langle 0| 
a_{\mathbf{k_2}} a_{\mathbf{k_3}} a_{\mathbf{q_2}}^\dagger a_{\mathbf{q_3}}^\dagger
b_{\mathbf{k_1}} b_{\mathbf{q_1}}^\dagger b_{\mathbf{q_4}} b_{\mathbf{k}}^\dagger
|0\rangle. 
\label{sfi1}
\end{eqnarray}
A direct calculation using the commutation relations given by eq.~(\ref{aad})
leads to
\begin{eqnarray}
 \langle 0| 
a_{\mathbf{k_2}} a_{\mathbf{k_3}} a_{\mathbf{q_2}}^\dagger a_{\mathbf{q_3}}^\dagger
b_{\mathbf{k_1}} b_{\mathbf{q_1}}^\dagger b_{\mathbf{q_4}} b_{\mathbf{k}}^\dagger
|0\rangle &=&  \delta^{(3)}(\mathbf{k-q_4})
\delta^{(3)}(\mathbf{k_1-q_1})\delta^{(3)}(\mathbf{k_2-q_3})\delta^{(3)}(\mathbf{k_3-q_2})
\nonumber \\
&+& \delta^{(3)}(\mathbf{k-q_4}) \delta^{(3)}(\mathbf{k_1-q_1})
\delta^{(3)}(\mathbf{k_2-q_2})
\delta^{(3)}(\mathbf{k_3-q_3}).\nonumber \\
&& \,
\label{d3k}
\end{eqnarray}
Inserting eq.~(\ref{d3k}) into (\ref{sfi1}), and noting that by relabeling the 
integration variables the two terms in 
eq.~(\ref{d3k}) give the same contribution, we get 
\begin{eqnarray}    
 S_{fi}^{(1)} &=&  -\frac{2ig}{\hbar c}\int d^4x f_{\mathbf{k}}
f_{\mathbf{k_1}}f_{\mathbf{k_2}}f_{\mathbf{k_3}} 
e^{-i(2\mu +k^0-k_1^0-k_2^0-k_3^0)x^0}e^{i(\mathbf{k-k_1-k_2-k_3})\cdot \mathbf{r}}
\nonumber \\
&=&-\frac{2ig}{\hbar c}f_{\mathbf{k}}
f_{\mathbf{k_1}}f_{\mathbf{k_2}}f_{\mathbf{k_3}}(2\pi)^4 
\delta(2\mu +k^0-k_1^0-k_2^0-k_3^0)\delta^{(3)}(\mathbf{k-k_1-k_2-k_3}). \nonumber \\
&& \,
\label{conservationEP}
\end{eqnarray}
To arrive at the last line we employed 
the integral representation of the Dirac delta function.


The two Dirac delta functions in eq.~(\ref{conservationEP}) are related to the
conservation of energy and linear momentum for the present process, i.e.,
$S_{fi}^{(1)}$ is different from zero only if the arguments of the delta functions are zero.
Considering the decaying particle at rest,
we have $\hbar c k^0= mc^2$ and $\mathbf{k}=0$. 
Thus, conservation of energy leads to
\begin{eqnarray}
2\mu +k^0-k_1^0-k_2^0-k_3^0 &=& 0, \nonumber \\
2\hbar c \mu  +\hbar c k^0- \hbar c (k_1^0-k_2^0-k_3^0) &=& 0, \nonumber \\
2mc^2 + mc^2 -E_{\mathbf{k_1}}-E_{\mathbf{k_2}}-E_{\mathbf{k_3}} &=& 0.
\end{eqnarray}
But $E_{\mathbf{k_j}} = \sqrt{m^2c^4 + h^2c^2|\mathbf{k_j}|^2}\leq mc^2$, for $j=1,2,3$.
Therefore, we can only satisfy 
\begin{equation}
 E_{\mathbf{k_1}}+E_{\mathbf{k_2}}+E_{\mathbf{k_3}} = 3mc^2
\end{equation}
if $\mathbf{k_j}=0$, i.e., all particles are at rest after the decay. Note that 
the momentum conservation equation are automatically satisfied if 
$\mathbf{k} = \mathbf{k_j}=0$.

We could have anticipated the possibility of this decay process 
by noting that for the
Lorentz covariant Schr\"odinger theory, the particle and antiparticles have
different energies, $E^+_{\mathbf{k}} = -mc^2+E_{\mathbf{k}}$ and 
$E^-_{\mathbf{k}} = mc^2+E_{\mathbf{k}}$, respectively [see eqs. (\ref{geral0}),
(\ref{HamQ}), and (\ref{pmE})].
Hence, if we look at the Feynman diagram for this process (left panel of figure \ref{fig3}),
energy conservation leads to
\begin{equation}
E_{\mathbf{k}}^- = E_{\mathbf{k_1}}^- + E_{\mathbf{k_3}}^+ + E_{\mathbf{k_3}}^+,
\label{Ek+-}
\end{equation}
which is exactly what we obtain according to eq.~(\ref{conservationEP}) if we use the definition of $E_{\mathbf{k}}^\pm$.\footnote{For the Klein-Gordon theory,
the analogue of eq.~(\ref{Ek+-}) is 
$E_{\mathbf{k}} = E_{\mathbf{k_1}} + E_{\mathbf{k_3}} + E_{\mathbf{k_3}}$, which
can never be satisfied. Indeed, in the rest frame of the antiparticle 
with momentum $\mathbf{k}$ we have $E_{\mathbf{k}}=mc^2$ while the right hand
side is always greater or equal to $3mc^2$. Also, to obtain
the same predictions associated with the Lorentz covariant Schr\"odinger theory
when the fields interact according to 
$\mathcal{H}_{int} \propto \Psi(\Psi^\dagger)^3$,
we need for the Klein-Gordon theory an interaction given by  
$\mathcal{H}_{int} \propto e^{-i2\mu x^0}\Phi(\Phi^\dagger)^3$, a very unnatural
interaction term.}

For Lorentz invariant interactions, which we have
seen must always be built with bilinear terms such as
$\Psi\Psi^\dagger$, this gap of energy between particles
and antiparticles gets 
``averaged out''. This happens
because we always have an equal number of 
$e^{-i\mu x^0}$ and $e^{i\mu x^0}$ 
when we employ the field expansions to do any calculation
[see, for instance, eqs.~(\ref{psi+})-(\ref{psid-})].
However, when we have an interaction term breaking Lorentz invariance, we no longer
have a balanced number of $e^{\pm i\mu x^0}$ in the field expansions. This is
why the previous process does not violate energy-momentum conservation and why we 
have $S_{fi}^{(1)}$ as given by eq.~(\ref{conservationEP}). 

\textit{Remark.} The above interaction that breaks Lorentz invariance should be considered
with caution. It is best to understand it as a ``toy model'' and not as a fundamental
interaction. The same should apply to the interaction we will study next. 
They are simple examples illustrating a kinematically forbidden
interaction in the framework of the Klein-Gordon theory that is kinematically allowed
in the framework of the Lorentz covariant Schr\"odinger theory. 
The interaction term given by eq. (\ref{breakingLI1}), leading to the left Feynman
diagram of figure \ref{fig3}, when taken alone brings to the table an ``unstable vacuum''.
Since an antiparticle decays into two particles and an antiparticle, the decay process will
never cease because we will always have the same antiparticle as a product of the decay.
This unstoppable proliferation of decays is not currently observed. We could have
avoided this problem working with the following interaction, $\mathcal{H}_{int} \propto (\Psi^\dagger)^3$, which also breaks Lorentz invariance but leads to one antiparticle
decaying into only two particles. We can alternatively understand the interaction
(\ref{breakingLI1}) as being modulated by a time dependent coupling constant 
that decreases as time goes by. In primordial times it was relevant but today it is completely suppressed. 

We can also have processes involving photons that 
are not seen in the Klein-Gordon theory but that are
present in the Lorentz covariant Schr\"odinger theory if we introduce
appropriate Lorentz invariance-breaking terms in the interaction Hamiltonian density. 
For instance, interactions such as
\begin{equation}
(\Psi^\dagger \overleftrightarrow{\partial}_{\hspace{-.15cm}\mu}
\Psi^\dagger) A^\mu \hspace{.25cm}  \mbox{or} \hspace{.25cm} 
\Psi^\dagger \Psi^\dagger A^\mu  
\end{equation}
allow for the possibility of an antiparticle decaying into a particle and a photon
(see the right panel of figure \ref{fig3}). 

The equation expressing the conservation of energy for this process is 
\begin{equation}
E_{\mathbf{k}}^- = E_{\mathbf{k'}}^+ + E_{\mathbf{k_\gamma}},
\label{Ekgamma}
\end{equation}
which in the rest frame of the decaying particle ($\mathbf{k}=0$) becomes
\begin{eqnarray}
mc^2 + E_{\mathbf{k}} &=& - mc^2 + E_{\mathbf{k'}} + E_{\mathbf{k_\gamma}}, \nonumber \\
2mc^2 &=& - mc^2 + E_{\mathbf{k'}} + E_{\mathbf{k_\gamma}}.
\label{Ekgamma2}
\end{eqnarray}
Equation (\ref{Ekgamma2}) tells us that we must have
\begin{equation}
E_{\mathbf{k'}} + E_{\mathbf{k_\gamma}} = 3mc^2,
\end{equation}
which can be satisfied since any process such that  
$E_{\mathbf{k'}} + E_{\mathbf{k_\gamma}} > mc^2$ and that conserves momentum
($\mathbf{k'+k_\gamma}=0$) is kinematically  possible. The solution to this problem
is $E_{\mathbf{k_\gamma}} = 4mc^2/3$, $E_{\mathbf{k}}^+ = 2mc^2/3$, and
$|\mathbf{p_\gamma}|=|\mathbf{p'}|=\hbar c |\mathbf{k_{\gamma}}|
=\hbar c |\mathbf{k'}|=4mc/3$.

For the Klein-Gordon theory we would have instead of eqs.~(\ref{Ekgamma}) and 
(\ref{Ekgamma2}),
\begin{eqnarray}
E_{\mathbf{k}} &=& E_{\mathbf{k'}} + E_{\mathbf{k_\gamma}}, \nonumber \\
mc^2 &=&  E_{\mathbf{k'}} + E_{\mathbf{k_\gamma}},
\label{Ekgamma3}
\end{eqnarray}
which cannot be satisfied because the right hand side must be greater than $mc^2$.

\textit{Remark.} The two interactions that break Lorentz invariance studied in this
section when taken alone lead to the violation of charge conservation as well. 
For the interaction
given by the right Feynman diagram of figure \ref{fig3} we have, for instance,
a positive antiparticle decaying into a neutral photon and a negative particle. 
If we want to save charge conservation we can either say that those processes are 
forbidden despite being kinematically allowed 
or postulate the existence of a negative antiparticle decaying 
in an analogous way into a positive particle and a photon. Those two processes taken
together would lead to an overall charge conservation in our universe. A similar 
reasoning can restore overall charge conservation for the interaction (\ref{breakingLI1}).

Before we move on it is worth mentioning two points concerning the previous results.
First, the Lorentz covariant Schr\"odinger theory supplemented with appropriate 
interaction terms that break Lorentz invariance might lead to simple and  
useful effective field theories to describe condensed
matter physical process. Indeed, in a 
condensed matter system there is a ``privileged'' inertial 
frame, namely, the condensed matter system itself, and the introduction of 
Lorentz invariance-breaking interactions 
is not a serious threat to the modeling of the 
interactions among pseudo-particles within it. 

Second, the asymmetry in the decay rates
associated with particles and antiparticles, due to the introduction of 
interaction terms breaking Lorentz invariance, 
points to a possible way to understand the asymmetry 
between matter and antimatter in our universe \cite{din04}. As we previously remarked, 
the introduction of an interaction term such as $\Psi(\Psi^\dagger)^3$ implies that
only the decay $b \rightarrow baa$ is possible (see left panel of figure \ref{fig3}),
while the corresponding process obtained by exchanging the roles of particles with antiparticles, $a\rightarrow abb$, is kinematically forbidden. Therefore, for an 
initially symmetric distribution of matter and antimatter, the asymmetry in the previous 
two decay channels will eventually lead to an asymmetry in the matter and antimatter
distribution (baryogenesis). 
And if this asymmetry in the decay channels is always present and 
not counterbalanced by any other process, in the long
run there will be practically no antimatter in the universe.

Note that the underlying reason for the asymmetry in the two decay channels above
is the gap in 
the mass between particle and antiparticle naturally 
present in the Lorentz covariant Schr\"odinger
theory and no extra fields or interactions are needed to obtain an asymmetric 
behavior between matter and antimatter \cite{ber97,col98,car06,ces15,sak17,edw18,and04}. 
On one hand, for Lorentz invariant interactions, this mass difference 
is effectively suppressed and no asymmetry between matter and antimatter is observed. 
On the other hand, 
for interaction terms violating Lorentz invariance, the mass gap becomes relevant,
leading to an asymmetry between matter and antimatter that is testable.  
With that in mind, two
important issues arise: (1) Is there a fundamental physical  
process leading to the violation of the Lorentz invariance
of the Lorentz covariant Schr\"odinger theory? 
What causes it? 
Can it be traced back to the presence of a background gravitational field?
Is this Lorentz invariance-breaking process a feature of the present day universe or
it was relevant in its beginning, being suppressed during its evolution?
(2) Can we find a Lorentz invariant theory that gives 
\textit{different masses} for particles and antiparticles and, at the same time, leads to 
an asymmetry in the decay of particles and antiparticles as described above?
Can it be done without violating the CPT-theorem, or at least without violating
its extension, the CPTM-theorem as given in this work?

\section{Generalizing the Lorentz covariant Schr\"odinger Equation}
\label{glcse}

If we look at the second term of the Lagrangian density (\ref{symL2}), we realize that 
there is no derivatives in the spatial variables and the following question naturally
arises: Is it possible to come up with a Lorentz invariant Lagrangian density where the 
four space-time derivatives appear on an equal footing? Furthermore, and similarly
to the Lorentz covariant Schr\"odinger fields of the previous sections, 
can this be done in such a way that this Lagrangian density yields the same
predictions of the Klein-Gordon one 
if we restrict ourselves to Lorentz invariant interactions? 
Our goal in this section is to show that the answers to those questions are
affirmative.

\subsection{Obtaining the Lagrangian density and the transformation law for 
$\mathbf{\Psi(x)}$}

We start with the following Hermitian Lagrangian density ansatz, 
\begin{equation}
\mathcal{L} = \partial_\mu \Psi \partial^\mu \Psi^* 
+ i\kappa^\mu \Psi^* \partial_\mu \Psi -i(\kappa^\mu)^*\Psi \partial_\mu \Psi^*,
\label{GLS}
\end{equation}
where $\kappa^\mu$ \textit{does not} transform like a four-vector under a 
Lo\-ren\-tz transformation. It is just a shorthand notation for the four \textit{constants}
$\kappa^0,\kappa^1,\kappa^2,\kappa^3$. Being more specific, we postulate that 
$\kappa^0,\kappa^1,\kappa^2,\kappa^3$ are Lorentz invariants
under proper Lorentz transformations. These constants are the analog in the present theory
of the mass ``$m$'' that appears in the Klein-Gordon Lagrangian density. 
We will see later that 
$\kappa^\mu$ must be a real quantity to guarantee the Lorentz invariance of the Lagrangian
density (\ref{GLS}) and that
$\kappa^2=\kappa_\mu\kappa^\mu=\mu^2=m^2c^2/\hbar^2$, with $m$ being the rest mass of 
a Klein-Gordon scalar particle, if we want the present theory to be  
equivalent to the Klein-Gordon theory. Also, if $\kappa^0=\mu$ and
$\kappa^j=0$, we recover the Lorentz covariant Schr\"odinger Lagrangian density
previously studied.

Under an infinitesimal proper Lorentz transformation (s\-pa\-tial rotations or boosts), 
the space-time coordinates in the rest frames $S$ and $S'$ are connected by the following
relation,
\begin{equation}
x^\mu = x^{\mu'} - \epsilon^{\mu\nu}x_{\nu'},
\end{equation}
where $|\epsilon^{\mu\nu}|\ll 1$ and $\epsilon^{\mu\nu}=-\epsilon^{\nu\mu}$. The 
antisymmetry of $\epsilon^{\mu\nu}$ is a consequence of the invariance
of the norm of the four-vector $x^\mu$ under a proper Lorentz transformation. The 
derivative transforms according to
\begin{equation}
\partial_\mu = \partial_{\mu'} - \epsilon_{\mu\nu}\partial^{\nu'}.
\label{delmuinf}
\end{equation}
We also 
assume that under an infinitesimal proper Lorentz transformation, the wave function 
$\Psi(x)$ and $\Psi'(x')$ are connected by
\begin{equation}
\Psi(x) = e^{\frac{i}{\hbar}b_\mu x^{\mu'}}\Psi'(x'),
\label{psiinf}
\end{equation}
where $b_\mu$ is real and of the order of $\epsilon_{\mu\nu}$.

Inserting eqs.~(\ref{delmuinf}) and (\ref{psiinf}) into (\ref{GLS}), we get to first order
in $\epsilon_{\mu\nu}$,
\begin{eqnarray}
\mathcal{L} &=& \partial_{\mu'} \Psi' \partial^{\mu'} \Psi'^* 
+ i\kappa^\mu \Psi'^* \partial_{\mu'} \Psi' -i(\kappa^\mu)^*\Psi' \partial_{\mu'} \Psi'^*
-\frac{|\Psi'|^2}{\hbar}[\kappa^\mu + (\kappa^\mu)^*]b_\mu \nonumber \\
&&-i\left(\frac{b^\mu}{\hbar} + \epsilon^{\nu\mu}\kappa_\nu\right)
\Psi'^* \partial_{\mu'} \Psi' 
+i\left(\frac{b^\mu}{\hbar} + \epsilon^{\nu\mu}(\kappa_\nu)^*\right)
\Psi' \partial_{\mu'} \Psi'^* + \mathcal{O}(|\epsilon_{\mu\nu}|^2).
\label{GLS2}
\end{eqnarray}

Comparing eqs.~(\ref{GLS}) and (\ref{GLS2}), we see that they have the same form if
\begin{eqnarray}
[\kappa^\mu + (\kappa^\mu)^*]b_\mu & = & 0, \label{firstoftheset}\\
\frac{b^\mu}{\hbar} + \epsilon^{\nu\mu}\kappa_\nu &=& 0, \\
\frac{b^\mu}{\hbar} + \epsilon^{\nu\mu}(\kappa_\nu)^* &=& 0 \label{thirdoftheset}.
\end{eqnarray}
The last two equations imply that $(\kappa_\nu)^*=\kappa_\nu$, i.e., $\kappa_\nu$ is 
real, and that
\begin{equation}
b^\mu = \hbar \epsilon^{\mu\nu}\kappa_\nu,
\label{bmu}
\end{equation}
where we used the antisymmetry of $\epsilon^{\mu\nu}$ to express the right hand side
as shown above.
Inserting eq.~(\ref{bmu}) into the left hand side of 
(\ref{firstoftheset}), and noting that $\kappa^\mu$ is real and that  
$\epsilon_{\mu\nu}$ is antisymmetric, we get
\begin{equation}
2\hbar\epsilon_{\mu\nu}\kappa^\mu\kappa^\nu =
\hbar(\epsilon_{\mu\nu}+\epsilon_{\nu\mu})\kappa^\mu\kappa^\nu = 0.
\end{equation}
Thus,  
eqs.~(\ref{firstoftheset})-(\ref{thirdoftheset}) are all satisfied if 
$b^\mu$ is given by eq.~(\ref{bmu}).

This means that under an infinitesimal proper Lorentz transformation, 
the Lagrangian density\footnote{It is worth noting that we cannot have Lorentz 
invariance if 
$\mathcal{L} = 
i\kappa^\mu \Psi^* \partial_\mu \Psi -i(\kappa^\mu)^*\Psi \partial_\mu \Psi^*$.
The term $\partial_\mu \Psi \partial^\mu \Psi^*$ is crucial 
to guarantee a Lorentz
invariant Lagrangian density.} 
\begin{equation}
\mathcal{L} = \partial_\mu \Psi \partial^\mu \Psi^* 
+ i\kappa^\mu\Psi^* \overleftrightarrow{\partial}_{\hspace{-.15cm}\mu}
\Psi,
\label{GLS3}
\end{equation}
with $\kappa^\mu$ real and constant, 
is invariant if $\Psi$ transforms as
\begin{equation}
\Psi(x) \rightarrow e^{-i\epsilon_{\mu\nu}\kappa^\mu x^\nu}\Psi(x).
\label{PsiTransf}
\end{equation}

To arrive at the transformation law for $\Psi(x)$ under finite proper Lorentz 
transformations, we follow the usual prescription of implementing $N$ infinitesimal
transformations and then letting $N\rightarrow \infty$. For instance, a counterclockwise 
rotation of $\phi$ radians about the $x^3$-axis gives [see appendix \ref{apB}]
\begin{eqnarray}
\Psi(x) &\rightarrow& e^{i[\kappa^1(1-\cos\phi)-\kappa^2\sin\phi]x^1} 
e^{i[\kappa^1\sin\phi + \kappa^2(1-\cos\phi)]x^2}
\Psi(x) \nonumber \\
&\rightarrow& e^{i[(\kappa^1x^1+\kappa^2x^2)(1-\cos\phi)
+(\kappa^1x^2-\kappa^2x^1)\sin\phi]}\Psi(x).\nonumber \\
&&\! \label{rotationx3}
\end{eqnarray}

Similarly, for a rotation of $\phi$ about the $x^1$-axis we get
\begin{equation}
\Psi(x) \rightarrow e^{i[(\kappa^2x^2+\kappa^3x^3)(1-\cos\phi)
+(\kappa^2x^3-\kappa^3x^2)\sin\phi]}\Psi(x). \label{rotationx1}
\end{equation}
The transformation law for a rotation about the $x^2$-axis is obtained from
the previous one by relabeling the superscript indexes: $2\rightarrow 3$ and 
$3\rightarrow 1$.\footnote{From the above results it is clear that the transformation
law for $\Psi(x)$ when we have a rotation of $\phi$ 
about an arbitrary axis $\mathbf{n}$ is
$\Psi(x) \!\rightarrow\! \exp{
\{i[\mathbf{r}-(\mathbf{r\cdot \hat{n}})\mathbf{\hat{n}}]\cdot\bm{\kappa}(1-\cos\phi)
\!+\!i(\bm{\kappa}\times\mathbf{r})\!\cdot\!\mathbf{\hat{n}}\sin\phi\}}\Psi(x)$.
}

Since from spatial rotations about the $x^3$ and $x^1$ axes we can build an arbitrary spatial rotation, 
it is sufficiently general to give only how $\Psi(x)$ transforms after a boost 
in a specified direction. For a boost along the $x^1$-axis characterized by $\beta$ 
we have [see appendix \ref{apB}],
\begin{equation}
\Psi(x) \rightarrow e^{i[(\gamma - 1)\kappa^0-\gamma\beta\kappa^1]x^0
+i[\gamma\beta\kappa^0-(\gamma - 1)\kappa^1 ]x^1}
\Psi(x),  \label{boostx1}
\end{equation}
where $\gamma=1/\sqrt{1-\beta^2}$ is the Lorentz factor. To check the consistency
of the calculations given in the appendix \ref{apB}, a direct calculation  
using eqs.~(\ref{rotationx3}), (\ref{rotationx1}), and (\ref{boostx1}), with
the corresponding transformation laws for $\partial_\mu$, shows that 
the Lagrangian (\ref{GLS3}) is indeed invariant under those finite proper Lorentz 
transformations.

\subsection{The wave equation and its solution}

The Lagrangian density (\ref{GLS3}) 
leads to the following wave equation if we insert it into
the Euler-Lagrange equation (\ref{euler2}),
\begin{equation}
\boxed{\partial_\mu\partial^\mu\Psi - 2i\kappa^\mu\partial_\mu\Psi = 0.}
\label{we}
\end{equation}

Inserting the ansatz 
\begin{equation}
\Psi(x) = e^{i(\kappa_\mu x^\mu \pm k_\mu x^\mu)} = e^{i(\kappa \pm k)x}
\label{ansatzwe}
\end{equation}
into eq.~(\ref{we}) we get
\begin{equation}
(\kappa^2 - k^2)\Psi = 0.
\label{ceq}
\end{equation}
We can only satisfy eq.~(\ref{ceq}), and thus raise the ansatz (\ref{ansatzwe}) 
to the status
of a solution to the wave equation, if 
\begin{equation}
k^2 = \kappa^2  = (\kappa^0)^2-|\bm{\kappa}|^2.  
\label{disprelGLS}
\end{equation}
Due to the linearity of the wave equation (\ref{we}), 
its general solution is a linear 
combination of (\ref{ansatzwe}) comprising all values of $k$
compatible with the boundary conditions of the problem being solved.

So far the four real constants $\kappa^\mu$ are arbitrary. If we set 
$\kappa^j=0$ and $\kappa^0=\mu=mc/\hbar$, we get back to the 
Lorentz covariant Schr\"odinger 
equation. This shows that the latter equation 
is a particular case of the general wave equation
we are now dealing with. As we show next, if 
$\kappa^2  = \mu^2$, with $m$ interpreted as the rest mass of a scalar particle,
we can connect the present wave equation with the Klein-Gordon equation in
a straightforward way.

\subsection{Connection with the Klein-Gordon equation}

Building on our previous experience with the Lorentz covariant Schr\"odinger equation,
we expect that the following relation will take us from the present wave equation to 
the Klein-Gordon one,
\begin{equation}
\Psi(x) = e^{i\kappa x}\Phi(x)=e^{i\kappa_\mu x^\mu}\Phi(x).
\label{GLStoKG}
\end{equation}

Using eq.~(\ref{GLStoKG}) we get that
\begin{eqnarray}
\partial_\mu\partial^\mu \Psi &=& (\partial_\mu\partial^\mu \Phi + 2i\kappa^\mu\partial_\mu\Phi -\kappa^2\Phi)e^{i\kappa x}, \\
-2i\kappa^\mu\partial_\mu\Psi &=& (2\kappa^2\Phi-2i\kappa^\mu\partial_\mu\Phi)e^{i\kappa x}, 
\end{eqnarray}
and thus the wave equation (\ref{we}) becomes
\begin{equation}
\partial_\mu\partial^\mu\Phi + \kappa^2\Phi = 0.
\label{weKG}
\end{equation}

The wave equation (\ref{weKG}) is formally equivalent to the Klein-Gordon one 
and we have an exact match if 
\begin{equation}
\kappa^2=\kappa_\mu\kappa^\mu = \mu^2=(mc/\hbar)^2,
\label{dispKG}
\end{equation}
where $m$ is the rest mass of the scalar particle described by the Klein-Gordon equation. 
We can also show that eq.~(\ref{GLStoKG}) when inserted into the Lagrangian density
(\ref{GLS3}) leads to the Klein-Gordon Lagrangian density if eq.~(\ref{dispKG}) is
satisfied. 

It is worth mentioning that we just need $\kappa^2=\mu^2$ to identify the transformed
wave equation with the Klein-Gordon one. In other words, out of the four real constants
$\kappa^\mu$, only one is fixed by eq.~(\ref{dispKG}). Three of them are still free to
be set to any value we wish. For an isotropic three-dimensional space, we expect  
all $\kappa^j$ to be equal. This reduces the free parameters to just one real constant.
However, for condensed matter systems, where anisotropic physical systems are commonplace,
the freedom to choose the values of $\kappa^j$ 
might be an important asset \cite{saf93,zha19}. 

\subsection{Canonical quantization}

We now start the canonical quantization of the generalized Lorentz covariant Schr\"odinger 
fields. We have two main go\-als in carrying out this task. First, we want to show that
it is possible to implement this second quantization program 
to its logical completion without any contradiction. Then, for Lorentz covariant interactions, we want to show the equivalence
between the just developed second quantized theory and Klein-Gordon's.  

\subsubsection{Equal time commutation relations}

The most general solution to the wave equation (\ref{we}), now understood as an operator,
can be written as 
\begin{equation}
\Psi(x) = \Psi^+(x) + \Psi^-(x), 
\label{psiGLS}
\end{equation}
where
\begin{eqnarray}
\Psi^+(x) & = & \int \tilde{dk_+}a_{\mathbf{k}}e^{i\kappa x}e^{-ikx}, \\ 
\Psi^-(x) & = & \int \tilde{dk_-}b^\dagger_{\mathbf{k}}e^{i\kappa x}e^{ikx},
\end{eqnarray}
and
\begin{eqnarray}
k^0 &=& \sqrt{\kappa^2 + |\mathbf{k}|^2}, \\
\tilde{dk_\pm} &=& f_\pm(|\mathbf{k}|)d^3k. 
\end{eqnarray}

Requiring that the commutation relations for the 
creation and annihilation operators are given by eqs.~(\ref{aad})-(\ref{ab}),
the fields $\Psi$ and $\Psi^\dagger$ satisfy the equal time canonical commutation 
relations (\ref{com1}) and (\ref{com2}) if
\begin{equation}
f_+(|\mathbf{k}|) = f_-(|\mathbf{k}|) = f(|\mathbf{k}|) = 
\left(\frac{\hbar c^2}{(2\pi)^32\omega_{\mathbf{k}}}\right)^{1/2},
\end{equation}
where
\begin{equation}
\hbar \omega_{\mathbf{k}} = E_\mathbf{k} = \hbar c k^0.
\end{equation}
The conjugate momenta to the fields are given by
\begin{eqnarray}
\Pi_\Psi = \frac{1}{c}\partial_0\Psi^\dagger + \frac{i\kappa^0}{c}\Psi^\dagger, \\
\Pi_{\Psi^\dagger} = \frac{1}{c}\partial_0\Psi - \frac{i\kappa^0}{c}\Psi,
\end{eqnarray}
and they are connected to the Klein-Gordon ones as follows,
\begin{eqnarray}
\Pi_\Psi &=& e^{-i\kappa x} \Pi_\Phi, \label{piiGLS}\\
\Pi_{\Psi^\dagger} &=& e^{i\kappa x} \Pi_{\Phi^\dagger}.
\end{eqnarray}

\subsubsection{Conserved quantities}

Following the prescription given in section \ref{cft}, we now have the following expressions
for the Hamiltonian, momentum, and charge densities (normal ordering always implied),
\begin{eqnarray}
\mathcal{H} & = & \partial_0\Psi\partial_0\Psi^\dagger + \nabla\Psi \cdot \nabla\Psi^\dagger 
- i\bm{\kappa}\cdot\Psi^\dagger\overleftrightarrow{\nabla}
\Psi, \label{hGLS}\\
c\mathcal{P}^j & = &  -\partial_0\Psi\partial_j\Psi^\dagger 
-\partial_0\Psi^\dagger\partial_j\Psi
- i\kappa^0\Psi^\dagger \overleftrightarrow{\partial}_{\hspace{-.15cm}j}
\Psi, \label{pGLS} \\
c\tilde{q} & = & i\Psi^\dagger \overleftrightarrow{\partial}_{\hspace{-.15cm}0}
\Psi + 2\kappa^0\Psi^\dagger\Psi. \label{cGLS}
\end{eqnarray}

Note that now the charge current is given by
\begin{equation}
c\mathbf{\tilde{j}} = i\Psi^\dagger\overleftrightarrow{\nabla}
\Psi + 2\bm{\kappa}\Psi^\dagger\Psi. \label{jGLS}
\end{equation}

Similarly to what we did for the Lorentz covariant Schr\"o\-din\-ger fields, by inserting eq.~(\ref{psiGLS}) into eqs.~(\ref{hGLS})-(\ref{cGLS}), and then into 
eqs.~(\ref{Hconserved}), (\ref{Pconserved}), and (\ref{Qconserved}), we obtain
\begin{eqnarray}
H & = & \int d^3k (\hbar \omega_{\mathbf{k}}^+a_{\mathbf{k}}^\dagger a_{\mathbf{k}}
+ \hbar\omega_{\mathbf{k}}^-b_{\mathbf{k}}^\dagger b_{\mathbf{k}}), \label{HamQ2}\\
P^j & = & \int d^3k (\hbar k^j_+ a_{\mathbf{k}}^\dagger a_{\mathbf{k}}
+ \hbar k^j_- b_{\mathbf{k}}^\dagger b_{\mathbf{k}}), \label{PQ2} \\
\tilde{Q} & = & \int d^3k [\hbar (a_{\mathbf{k}}^\dagger a_{\mathbf{k}}
- b_{\mathbf{k}}^\dagger b_{\mathbf{k}})], \label{QQ2}
\end{eqnarray}
where
\begin{eqnarray}
E^{\pm}_{\mathbf{k}} = \hbar  \omega_{\mathbf{k}}^\pm 
&=& \mp \hbar c \kappa^0 + E_\mathbf{k} \\
p^j_{\pm} = \hbar k^j_\pm &=& \mp\hbar\kappa^j + \hbar k^j.
\label{pmE2}
\end{eqnarray}

In addition to the energy gap associated with a pair of
particle and antiparticle characterized by the vector $\mathbf{k}$,
we also have a momentum gap. 
In contrast to the Lorentz covariant Schr\"odinger equation, 
we now have a complete symmetry between energy and momentum, which allows us to
define the two quantities below,
\begin{equation}
p_\pm^\mu = (E^{\pm}_{\mathbf{k}},c\mathbf{p}_{\pm}), \label{two4vector}
\end{equation}
where $\mathbf{p}_{\pm}=(p^1_{\pm},p^2_{\pm},p^3_{\pm})$.
Equation (\ref{two4vector}) gives the energy and momentum of a particle ($p_+^\mu$) 
and antiparticle ($p_-^\mu$)
created, respectively, by $a_\mathbf{k}^\dagger$ and $b_\mathbf{k}^\dagger$ acting on
the vacuum state.

When it comes to the discrete symmetries, we need to modify the operators that implement
the parity, time reversal, and charge conjugation operations as follows. The parity,
time reversal, and charge conjugation operators 
are now defined as\footnote{Note that to simplify notation we are not using the symbol 
``$\hat{\;\;\;}$'' to denote operators
as we did in section \ref{ds}. } 
\begin{eqnarray}
\mathcal{P}_\kappa\Psi(\mathbf{r},t)\mathcal{P}_\kappa^\dagger &=& \Psi(-\mathbf{r},t),  \label{parity2} \\
\mathcal{T}_\kappa\Psi(\mathbf{r},t)\mathcal{T}_\kappa^\dagger &=& \Psi(\mathbf{r},-t),  \label{time2} \\
\mathcal{C}_\kappa\Psi(\mathbf{r},t)\mathcal{C}_\kappa^\dagger &=& \Psi^\dagger(\mathbf{r},t). \label{charge2} 
\end{eqnarray}

In eqs.~(\ref{parity2})-(\ref{charge2}) we have 
\begin{eqnarray}
\mathcal{P}_\kappa &=& \mathcal{K}_1\mathcal{P},   \label{parity2a} \\
\mathcal{T}_\kappa &=& \mathcal{K}_1\mathcal{T},  \label{time2a} \\
\mathcal{C}_\kappa &=& \mathcal{K}_2\mathcal{C}, \label{charge2a} 
\end{eqnarray}
where
\begin{eqnarray}
\mathcal{K}_1f(\kappa^\mu)\mathcal{K}_1^\dagger &=& f(\kappa_\mu),  \label{ka1} \\
\mathcal{K}_2f(\kappa^\mu)\mathcal{K}_2^\dagger &=& f(-\kappa^\mu). \label{ka2} 
\end{eqnarray}
Here $f(\kappa^\mu)$ is an arbitrary function of $\kappa^\mu$ and $\mathcal{P},
\mathcal{T}$, and $\mathcal{C}$ are the standard parity, time reversal, and charge
conjugation operators of the Klein-Gordon theory [see also section \ref{ds}]. We should
remark that $\mathcal{P}$ and $\mathcal{C}$ are unitary operators while  
$\mathcal{T}$ is antiunitary. 

Equation (\ref{ka2}) generalizes  the ``mass conjugation''
operation, eq.~(\ref{massconjugation}), 
which is needed to correctly define a charge conjugation
operation for the Lorentz covariant Schr\"o\-din\-ger equation. 
The operation defined in eq.~(\ref{ka1}), on the other hand, only changes the sign of the spatial 
components of $\kappa^\mu$, i.e., 
$\mathcal{K}_1f(\kappa^0,\bm{\kappa})\mathcal{K}_1^\dagger = f(\kappa^0,-\bm{\kappa})$.

Using eqs.~(\ref{parity2})-(\ref{charge2}), we can show that 
we get the expected behavior for properly defined parity, time reversal, and
charge conjugation operations. Specifically, eqs.~(\ref{parity1})-(\ref{parity4}), 
eqs.~(\ref{time1})-(\ref{time4}), and eqs.~(\ref{charge1})-(\ref{charge4}) 
are all satisfied. Furthermore, eqs.~(\ref{parity2})-(\ref{charge2}) imply that 
the creation and annihilation operators satisfy the correct transformation laws,
namely, eqs.~(\ref{tfirst})-(\ref{tfourth}) and (\ref{tfifth})-(\ref{tsixth}). 

Similarly, the CPT-theorem, which in the present context should be  
more properly called the CPTM-theorem, is once more ``saved'' if we define the 
CPT operation as
\begin{equation}
\Theta = \mathcal{C}_\kappa\mathcal{P}_\kappa\mathcal{T}_\kappa 
= \mathcal{K}_2\mathcal{C}\mathcal{P}\mathcal{T}.
\end{equation}
To arrive at the last 
equality we used that $\mathcal{K}_{j}$ commutes with 
$\mathcal{C}, \mathcal{P}$, and $\mathcal{T}$ and that $\mathcal{K}_{j}^2$
is the identity operator.
With this definition for the CPT operation, 
eqs.~(\ref{cpt1})-(\ref{cptm2}) continue to hold for
the generalized Lorentz covariant Schr\"odinger theory.

\subsubsection{Arbitrary time commutators}

A direct calculation gives
\begin{eqnarray}
[\Psi(x),\Psi^\dagger(x')] &=& \int d^3k[f(|\mathbf{k}|)]^2[e^{-i(k-\kappa)(x-x')}
-e^{i(k+\kappa)(x-x')}] \nonumber \\
&=&  \int d^3k[f(|\mathbf{k}|)]^2e^{i(\mathbf{k}\bm{-\kappa})\cdot\mathbf{(r-r')}}
[e^{-i\omega_{\mathbf{k}}^+(t-t')}-e^{i\omega_{\mathbf{k}}^-(t-t')}].
\end{eqnarray}

Similarly to the Lorentz covariant Schr\"odinger fields, we can write the previous
commutator as
\begin{equation}
[\Psi(x),\Psi^\dagger(x')] = i\hbar c \Delta_{GLS}(x-x'),
\end{equation}
where
\begin{equation}
\Delta_{GLS}(x) = e^{i\kappa x}\Delta(x).
\end{equation}
Here the Lorentz invariant $\Delta(x)$ is given by
\begin{equation}
\Delta(x) = \frac{-i}{2\pi^3}\int d^4k\delta(k^2-\kappa^2)\varepsilon(k^0)e^{-ikx},
\end{equation}
where all variables are real and integrated from $-\infty$ to $\infty$.
Note that contrary to eq.~(\ref{deltakg2}), 
we have $\kappa^2$ instead of $\mu^2$ inside
the delta function.

The other non-null commutators are
\begin{eqnarray}
[\Pi_\Psi(x),\Pi_{\Psi^\dagger}(x')]  
&=& e^{-i\kappa(x-x')}\int d^3k[f(|\mathbf{k}|)]^2
\left(\frac{\omega_{\mathbf{k}}^2}{c^4}\right)
[e^{-ik(x-x')} -e^{ik(x-x')}] \nonumber \\
&=&  \int d^3k[f(|\mathbf{k}|)]^2
\left(\frac{\omega_{\mathbf{k}}^2}{c^4}\right)e^{i(\mathbf{k+}\bm{\kappa})
\cdot (\mathbf{r-r'})}
[e^{-i\omega_{\mathbf{k}}^-(t-t')}-e^{i\omega_{\mathbf{k}}^+(t-t')}]
\end{eqnarray}
and
\begin{eqnarray}
[\Psi(x),\Pi_\Psi(x')] 
&=& e^{i\kappa(x-x')}\int d^3k[f(|\mathbf{k}|)]^2
\left(\frac{i\omega_{\mathbf{k}}}{c^2}\right)
[e^{-ik(x-x')} +e^{ik(x-x')}] \nonumber \\
&=&  \int d^3k[f(|\mathbf{k}|)]^2
\left(\frac{i\omega_{\mathbf{k}}}{c^2}\right)e^{i(\mathbf{k-}\bm{\kappa})
\cdot (\mathbf{r-r'})}
[e^{-i\omega_{\mathbf{k}}^+(t-t')}+e^{i\omega_{\mathbf{k}}^-(t-t')}].
\end{eqnarray}

\subsubsection{The Feynman propagator}

The Feynman propagator, which we call $\Delta_{F_{\!_{GLS}}}(x)$, is given by
\begin{equation}
\Delta_{F_{\!_{GLS}}}(x) = h(x^0)\Delta_{GLS}^+(x) - h(-x^0)\Delta_{GLS}^-(x),
\label{dfgls}
\end{equation}
where
\begin{equation}
\Delta_{GLS}^\pm(x) = e^{i\kappa x}\Delta^\pm(x).
\label{glspm}
\end{equation}
Here $\Delta^\pm(x)$ are the standard delta functions appearing in the 
analysis of the Klein-Gordon propagator, with all quantities that are 
functions of $\mu^2$ changed to functions of $\kappa^2$ [eq.~(\ref{deltapm2})
with $k^0$ and $\omega_{\mathbf{k}}$ expressed as functions of $\kappa^2$
instead of $\mu^2$].

Alternatively, using eqs.~(\ref{dfgls}) and (\ref{glspm}) we have
\begin{equation}
\Delta_{F_{\!_{GLS}}}(x) = e^{i\kappa x}\Delta_F(x),
\end{equation}
with
\begin{equation}
\Delta_F(x) = \frac{1}{(2\pi)^4}\int d^4k\,
\frac{e^{-ikx}}{k^2-\kappa^2+i\epsilon}.
\label{integralDFGLS}
\end{equation}
Here $\eta \ll 1$, $\epsilon = 2\eta\omega_{\mathbf{k}}/c$, and
all variables are real and integrated from $-\infty$ to $\infty$. 
As usual, the limit $\epsilon \rightarrow 0$ is implied  
after the integration is made.

\subsection{Connection with the Klein-Gordon theory}

We divide the forthcoming analysis in two parts. First, we show that the 
S-matrix describing a given process (scattering or decay)
involving the generalized Lorentz covariant Schr\"odinger 
fields is equal to the S-matrix one obtains modeling the same process
using Klein-Gordon fields. This only happens for interactions that are
invariant under a proper Lorentz transformation.

Second, we show that the same S-matrix leads to the same scattering cross section (or decay rate) for both theories. This apparently trivial result should be properly discussed since
the generalized Lorentz covariant Schr\"odinger Lagrangian and the Klein-Gordon Lagrangian lead to formally different expressions for the current density vector (flux of particles),
an important quantity employed in the definition of a scattering cross section.

\subsubsection{Equivalence of the S-matrices}

For interactions of the type given by eq.~(\ref{lambdaNi}), with the 
$(\lambda/4)(\Psi\Psi^\dagger)^2$ interaction being a particular example, 
the equivalence of the S-matrices is proved using exactly the same arguments and
steps employed in the corresponding proof given before 
for the Lorentz covariant 
Schr\"odinger fields.
The only difference, a formal one that does not change the arguments used in the previous
proof, is that instead of eqs.~(\ref{ikg}) and (\ref{ipikg}) we now have according to
eqs.~(\ref{GLStoKG}) and (\ref{piiGLS}),
\begin{eqnarray}
\Psi^I(x) = e^{i\kappa x}\Phi^I(x), \label{psiIGLStoKG} \\
\Pi^I_\Psi(x) = e^{-i\kappa x}\Pi^I_\Phi(x).
\end{eqnarray}

Furthermore, since in the present case the conjugate momenta to the fields
are the same as the ones associated with the Lorentz covariant Schr\"odinger fields,
the proof of the equivalence of the S-matrices when we include 
electromagnetic interactions is almost 
the same as the one given in section \ref{sedsection}.
Note that the conjugate momenta do not change because 
the new extra terms appearing in the Lagrangian density describing 
the generalized Lorentz covariant Schr\"odinger fields do not depend on time derivatives.

Repeating the same calculations given in section \ref{sedsection}, we obtain in the 
interaction picture the following
Hamiltonian density describing the electromagnetic interaction among the fields,
\begin{eqnarray}
\mathcal{H}_{int}^I &=& -\mathcal{L}_{int}^I 
+ \frac{q^2}{\hbar^2}(\Psi^I)^\dagger\Psi^I A^I_0 (A^I)^0,
\label{HintSED3GLS}
\end{eqnarray}
where 
\begin{eqnarray}
\mathcal{L}_{int}^I & = & 
-\frac{iq}{\hbar}[(\Psi^I)^\dagger \overleftrightarrow{\partial}_{\hspace{-.15cm}\mu}
\Psi^I](A^I)^\mu
+\frac{q^2}{\hbar^2}(\Psi^I)^\dagger\Psi^I A^I_\mu (A^I)^\mu 
-\frac{2 q}{\hbar}\kappa^\mu(\Psi^I)^\dagger\Psi^I A^I_\mu. \label{LintSEDGLS}
\end{eqnarray}

If we now use eq.~(\ref{psiIGLStoKG}), a direct calculation shows that 
eq.~(\ref{HintSED3GLS}) is transformed to eq.~(\ref{HintSED5}), 
the interaction Hamiltonian density for the Klein-Gordon theory.
Using the same arguments given in section \ref{sedsection}, 
this fact is enough to establish
the equivalence between the 
S-matrix of the present theory and the S-matrix coming from the Klein-Gordon theory.

\subsubsection{Equivalence of the scattering cross sections}

If we analyze the standard way to define  
a differential cross section for a given process \cite{man86,gre95}, 
we realize that
the following three distinct steps are taken to arrive at an experimentally
meaningful  quantity:
\begin{itemize}
\item[(1)] Terms like $[(2\pi)^4\delta^{(4)}(\sum k'_{final} - \sum k_{initial})]^2$
are identified with $VT(2\pi)^4\delta^{(4)}(\sum k'_{final} - \sum k_{initial})$, where
$V$ and $T$ represent, respectively, a finite volume (box normalization for plane waves) 
and the duration of the experiment. Here 
$\sum k'_{final} - \sum k_{initial}$ denotes the conservation of energy and momentum
written in terms of the four-wave vector. The most common situation is the one where
we have $k_1$ and $k_2$, two initial particles, and two or more final ones.
\item[(2)] The calculation of the transition probability per unit time, 
$w = |S_{fi}|^2/T$, where $S_{fi}$ is the shorthand notation for the probability amplitude
describing a given process.
\item[(3)] The definition of the differential scattering cross section as 
\begin{equation}
d \sigma = \frac{w}{|\mathbf{j}|}\prod_{final}
\frac{Vd^3k'_{final}}{(2\pi)^3},
\end{equation}
where $|\mathbf{j}|$ is the magnitude of the flux of incoming particles and
$\prod_{final}Vd^3k'_{final}/(2\pi)^3$ represents a group of final particles with
wave numbers in the interval $\mathbf{k'}_{final}$ and
$\mathbf{k'}_{final} + d\mathbf{k'}_{final}$. Also, the normalization
adopted for the wave functions is one particle per volume $V$. 
\end{itemize}

Since we have proved the equivalence of the S-matrices,
items (1) and (2) above are easily seen to be the same 
whe\-ther we deal with the Klein-Gordon or with
the generalized Lo\-ren\-tz covariant Schr\"odinger fields. Item (3),
however, deserves a little more thought. To prove that it is equivalent
for both theories, we have to show the equivalence of the 
incident particle flux $\mathbf{j}$. 

The flux associated with the generalized Lorentz covariant Schr\"odinger
equation, given by eq.~(\ref{jGLS}), can be written as
\begin{equation}
\mathbf{j} = -\frac{i\hbar}{2m}(
\Psi^\dagger\overleftrightarrow{\nabla}
\Psi + 2i\bm{\kappa}\Psi^\dagger\Psi). \label{jGLS2}
\end{equation}
In the above expression 
we are using the first quantization non-relativistic normalization 
$-i\hbar/(2m)$ for the flux for reasons that will become clear in a moment. If we now use
eq.~(\ref{GLStoKG}), eq.~(\ref{jGLS2}) becomes
\begin{equation}
\mathbf{j} = -\frac{i\hbar}{2m}
\Phi^\dagger \overleftrightarrow{\nabla}
\Phi, \label{jGLS3}
\end{equation}
which is the incident particle flux we obtain working directly with the Klein-Gordon equation.

To better appreciate the equivalence of both fluxes, we explicitly compute $\mathbf{j}$
for the two cases of interest here. For simplicity,
we will work  at the first quantization level. 

For the Klein-Gordon equation, and assuming 
the target at rest, 
the plane wave normalized to one particle per volume representing an incident flux of particles with four-wave number $k^\mu$ is given by
\begin{equation}
\Phi(x) = \frac{1}{\sqrt{V}}e^{-ikx}.
\end{equation}
A direct calculation using eq.~(\ref{jGLS3}) gives 
\begin{equation}
\mathbf{j} =\frac{\hbar \mathbf{k}}{m}|\Phi|^2=\frac{\mathbf{p}}{m}\frac{1}{V}
=\frac{\mathbf{v}}{V},
\end{equation}
where we have made the identification of $\hbar\mathbf{k}$ with the 
Klein-Gordon particle's
momentum and $\mathbf{p}/m$ with its velocity $\mathbf{v}$. 

On the other hand, the solution to the generalized Lorentz covariant Schr\"odinger
equation representing a particle with four-wave number $k^\mu$ is
\begin{equation}
\Psi(x) = \frac{1}{\sqrt{V}}e^{i\kappa x}e^{-ikx}.
\end{equation}
Using eq.~(\ref{jGLS2}) we get
\begin{equation}
\mathbf{j} =\frac{\hbar \mathbf{k}}{m}|\Psi|^2=\frac{\hbar \mathbf{k}}{m}|\Phi|^2=
\frac{\hbar \mathbf{k}}{mV},
\end{equation}
which is clearly the same flux $\mathbf{j}$ 
we obtain working with the Klein-Gordon fields, proving thus the complete
equivalence between both theories.

Before closing this section, we would like to reinforce once more 
that the above analysis,
in particular the equivalence of the S-matrices, is only true for a Lorentz invariant
Lagrangian. The same arguments given in section \ref{bli} apply here. Therefore,
for interaction terms breaking Lorentz invariance,
we can have certain types of decay and scattering processes that are impossible 
to happen in the Klein-Gordon theory.\footnote{Note also that the previous equivalence
proof between the two theories assumed two types of possible interactions: 
self-interactions and electromagnetic interactions. It is possible
that whenever we have a consistent quantum field theory of gravitation,
the Klein-Gordon and the Lorentz covariant Schr\"odinger theories will show different predictions even if a fully
Lorentz invariant interaction is present. The reason for such a guess is related to
the fact that the mass(energy) and momentum gap may not be trivially ``averaged out''
if a consistent quantum gravity theory is built. More intuitively, ``negative''
masses will reduce the usual contribution of positive masses in generating a 
gravitational field. However, as we showed in section \ref{moreCm}, the mathematical framework of the 
Lorentz covariant Schr\"odinger Lagrangian can be adjusted to accommodate particles and
antiparticles attracting or repelling each other gravitationally, at least at the level
of Newtonian static gravitational fields. Therefore, at our present level of understanding,
the issue of how particles and antiparticles interact gravitationally in the framework of the Lorentz covariant
Schr\"odinger theory is left as an open problem.}

\section{Conclusion}

The original motivation leading to this work, on one hand, 
stems from the fact that the great majority of
physical observables in the framework of quantum field theory are bilinear functions of the fields. 
In non-relativistic quantum mechanics we also have that the experimentally relevant quantities are
almost always functions of 
the probability density $\Psi(x)\Psi^*(x)$, a bilinear function of 
the wave function. 
Therefore, working either with $\Psi(x)$ or $e^{\frac{i}{\hbar}f(x)}\Psi(x)$, where $f(x)$ is an arbitrary function of the spacetime coordinates, 
we obtain the same predictions. 

On the other hand, 
in classical and quantum field theories one usually defines a complex scalar field 
as a quantity that is itself invariant under a given symmetry transformation:
$\Psi(x)\rightarrow \Psi(x)$. However, it is clear that whenever bilinear functions
of the fields are associated with observable quantities, we only need those bilinears 
to be invariant under that symmetry operation to obtain invariant physical results. 
In other words, by assuming the more general transformation
rule for the fields, $\Psi(x)\rightarrow e^{\frac{i}{\hbar}f(x)}\Psi(x)$, 
we should get in principle an equivalent description for 
a physical system whose observables are bilinear functions of those
fields. 
Therefore, we revisited the classical and
quantum field theories of complex scalar fields assuming from the start 
the more general transformation law for the fields when they are subjected to a symmetry operation. 
We wanted to check if we could consistently develop
those field theories to their logical conclusion, comparing the final products with current
complex scalar field theories based on the more simple transformation law, where
it is assumed that $f(x)$ is either zero or a non-null constant.

As we showed in this work, we can indeed build logically consistent classical and quantum
field theories assuming the more general transformation law for the fields under a symmetry
operation. In particular, we showed that it is possible to have a Lorentz covariant theory
if we assume that the complex scalar field $\Psi(x)$ transforms 
according to the more general
prescription above under proper Lorentz transformations (boosts or spatial rotations).
With the aid of very natural auxiliary assumptions, namely, linearity and a wave equation with at
most second order derivatives,  
we obtained the Lagrangian, the corresponding wave equation, and the function $f(x)$ 
that give the most general Lorentz covariant theory in this scenario.
We also determined under what conditions the complex scalar field theories here developed
match the Klein-Gordon theory and under what conditions we may have different predictions.
It turned out that for Lorentz invariant self-interactions and for electromagnetic interactions, we can make the present theories equivalent to the 
Klein-Gordon theory. For interaction terms that violate Lorentz invariance, however,
we showed by giving explicit examples that the present theories 
are no longer equivalent to Klein-Gordon's.

In addition to the formal development of the present complex scalar 
field theories, we applied them to describe several physical systems. 
This helped us to become familiarized with the new concepts introduced along the logical 
construction of the theory. Of the many physical systems we studied, we would like to
call attention to our investigations 
about the bound states and the scattering cross sections associated with two interacting charged particles, 
where both the electromagnetic and gravitational interactions were simultaneously 
included to model the interaction between them. 
The electromagnetic interaction
entered via the minimal coupling prescription while gravity was introduced 
via an external potential. We obtained the exact solution for the bound state
problem and computed perturbatively the differential scattering cross section when dealing with 
the corresponding scattering problem. 
We also estimated the order of magnitude of the charges and masses of the
particles in which gravitational effects can no longer be discarded.

During our studies a few unexpected results emerged. The first one showed up already at 
the first quantization level. We observed that the eigenenergies associated with the plane wave
solutions to the Lorentz covariant Schr\"odinger equation implied that particles and 
antiparticles no longer had degenerated relativistic energies and, intriguingly,  
we observed that we could only exchange the roles of particles with
antiparticles if, in addition to changing the sign of the charge,
we also changed the sign of the mass for all physical quantities depending 
on it. This suggested
that we could assign ``negative'' masses to the antiparticles. A full understanding of this 
fact was possible when we second quantized the theory, where it was noted that this interpretation for
the mass sign of the antiparticle is actually necessary for the logical consistency of the theory. 
This became apparent when we dealt with the charge conjugation operation. It could only
be consistently defined if, and only if,
we extended its standard definition such that the charge conjugation operator anticommuted with functions of the mass of the particle.
In other words, the charge conjugation operation has to change the sign of the masses when acting upon 
a given field operator to be properly defined. 

Despite the ``negative'' masses for antiparticles, we show\-ed that 
both particles and antiparticles have non-negative energies and
the same momentum for a given wave number $\mathbf{k}$. Later, at the last part of this work, 
when we generalized the Lorentz covariant Schr\"odinger equation, we showed that it is also possible to break
the degeneracy in the value of the momentum. For a given value of $\mathbf{k}$, particles and antiparticles also
have different momenta.

As we already remarked above, the present theories were shown to be equivalent to 
the Klein-Gordon one for Lorentz invariant self-interactions and electromagnetic interactions.
However, by including interaction terms in the Lagrangian density that are 
not Lorentz invariant, we showed that this equivalence with the Klein-Gordon theory is no longer valid. 
We showed that certain types of
interactions that violate Lorentz invariance imply that an
antiparticle can decay into two particles and one antiparticle or into a particle and a photon. 
The equivalent processes for particles decaying into more antiparticles than particles were shown
to be kinematically forbidden. This pointed to a second unexpected result, namely, 
that very simple Lorentz invariance-breaking 
interactions could explain the asymmetry in the abundance of matter and antimatter in the present day
universe. 

In order to better understand the non-degeneracy of the energies of particles and 
antiparticles, we also second quantized the Lorentz covariant Schr\"odinger Lagrangian 
imposing that the energies of particles and antiparticles were the same and given by
the standard relativistic expression. We observed that this could only be achieved
if the fields did not satisfy the canonical commutation relations anymore. 
This led to the violation of the microcausality condition as well. 
Albeit this non-local character of the non-canonically quantized theory, 
we noted that the specific way in which the microcausality condition was violated
resulted in the emergence of an instantaneous gravitational-like interaction between the particles of the theory. 
In other words,
particles and antiparticles could be brought to have the same relativistic energy if they
interacted similarly to what Newton's gravitational law prescribes. 
The emergence of this gravitational-like potential may be 
just a particularity of the present theory but it suggests that 
the assumption that antiparticles have negative masses with positive energies 
might shed a new light in our quest for a consistent quantum theory of gravity. 

Summing up, let us distill and write down the two major messages that we tried to convey by writing this work.
First, it is possible to build logically consistent and Lorentz covariant
classical and quantum field theories assuming a more general transformation law for complex 
scalar fields under a symmetry operation. These theories can be adjusted to reproduce exactly 
the Klein-Gordon theory when we have Lorentz invariant self-interactions and when we introduce
electromagnetic interactions via the minimal coupling prescription. Second, for logical
consistency we have to assume that antiparticles possess negative masses while still having
positive energies. These two points when analyzed together tell us that it is perfectly legitimate,
at least at the level of electromagnetic interactions, to assume that particles and antiparticles have masses 
with different signs without contradicting any known experimental fact. 
When we include gravitational interactions, it remains an open problem to rigorously 
analyze whether matter and antimatter attract or repel each other within the present
framework (at the level of static Newtonian fields, we showed that it can 
be adjusted to accommodate both possibilities). 
To solve this issue for ordinary matter and antimatter, 
we need to measure with high accuracy 
how a particle and an antiparticle 
interact gravitationally. We need to know, for example, 
how an antihydrogen atom responds 
to the Earth's gravitational field \cite{saf18}. 
If they repel each other,  
the Lorentz covariant Schr\"odinger Lagrangian is another possible starting
point to theoretically understand more fully 
what is going on \cite{bon57,kow96,vil11,far18}. If they attract each other,
the present theory is as good as the Klein-Gordon one to describe this 
experimental result.

\appendix

\section*{Appendix}

\section{Proof of the main results of section \ref{qft}}
\label{apA}

\subsection{Canonical versus non-canonical quantization}

Our main goal here is to show how eqs.~(\ref{com1}) and (\ref{com2})
follow from eqs.~(\ref{aad})-(\ref{ab}) with the appropriate choice of
$f_{\pm}(|\mathbf{k}|)$ for eq.~(\ref{geral}). We also want to determine 
what values for $f_{\pm}(|\mathbf{k}|)$ make particles and antiparticles 
have the same energies and, then, we want to explore the main 
consequences of the theory built on this
particular choice for $f_{\pm}(|\mathbf{k}|)$. 
From now on we assume we are dealing with the free
field ($V=0$).

Let us start listing the mathematical identities employed routinely 
in all calculations that follow. In the calculations of either 
the Hamiltonian, linear momentum, and conserved charge or 
in the computations of the several commutators below, the following
representation of the Dirac delta function is useful,
\begin{equation}
\delta^{(3)}(\mathbf{k-k'}) = \frac{1}{(2\pi)^3}\int d^3x \;
e^{\pm i(\mathbf{k-k'})\cdot 
\mathbf{r}}. \label{diracdelta}
\end{equation}
With the help of eq.~(\ref{diracdelta}) it is not difficult to arrive at the following 
identities,
\begin{eqnarray}
\frac{1}{(2\pi)^3}\!\int\!\! d^3x\; e^{\pm i (k-k')x} \!\!&=&\!\!
\delta^{(3)}(\mathbf{k-k'}), \label{diracM}\\
\frac{1}{(2\pi)^3}\!\int\!\! d^3x\; e^{\pm i (k+k')x} \!\!&=&\!\! 
e^{\pm i(k^0+k^{0'})x^0} \delta^{(3)}(\mathbf{k+k'}), \label{diracP} 
\end{eqnarray}
where $k^0$ and $k^{0'}$ are given by eq.~(\ref{omegak}). Another pair of useful
identities is
\begin{equation}
\omega^-_{\mathbf{k}}\omega^+_{\mathbf{k}} = c^2|\mathbf{k}|^2 \hspace{.25cm}\mbox{and}
\hspace{.25cm}
\sqrt{\frac{\omega^-_{\mathbf{k}}}{\omega^+_{\mathbf{k}}}} - 
\sqrt{\frac{\omega^+_{\mathbf{k}}}{\omega^-_{\mathbf{k}}}} = \frac{2\mu}{|\mathbf{k}|},
\label{ide1}
\end{equation}
where $\mu$ is given in eq.~(\ref{mi}) and 
$\omega^{\pm}_{\mathbf{k}}$ are defined in eq.~(\ref{pmE}).

Inserting eq.~(\ref{geral}) into the normal ordered expressions of 
eqs.~(\ref{Hconserved}), (\ref{Pconserved}), and (\ref{Qconserved}), we obtain
with the help of eqs.~(\ref{diracdelta})-(\ref{ide1}),
%
\begin{eqnarray}
H & = & \int d^3k \frac{2(2\pi)^3}{c^2}
\left\{
[f_+(\mathbf{k})]^2 \omega_{\mathbf{k}}^+\omega_{\mathbf{k}}
a_{\mathbf{k}}^\dagger a_{\mathbf{k}}
+ [f_-(\mathbf{k})]^2 \omega_{\mathbf{k}}^-\omega_{\mathbf{k}}
b_{\mathbf{k}}^\dagger b_{\mathbf{k}}
\right\}, \label{HamG}\\
P^j & = & \int d^3k \frac{2(2\pi)^3}{c^2}
\left\{
[f_+(\mathbf{k})]^2  \omega_{\mathbf{k}} k^j 
a_{\mathbf{k}}^\dagger a_{\mathbf{k}}
+ [f_-(\mathbf{k})]^2\omega_{\mathbf{k}} k^j
b_{\mathbf{k}}^\dagger b_{\mathbf{k}}
\right\}, \label{PG} \\
\tilde{Q} & = & \int d^3k \frac{2(2\pi)^3}{c^2}
\left\{
[f_+(\mathbf{k})]^2  \omega_{\mathbf{k}}  
a_{\mathbf{k}}^\dagger a_{\mathbf{k}}
- [f_-(\mathbf{k})]^2\omega_{\mathbf{k}} 
b_{\mathbf{k}}^\dagger b_{\mathbf{k}}
\right\}. \label{QG}
\end{eqnarray}
Similarly, using the commutation relations for the creation and annihilation
operators given by eqs.~(\ref{aad})-(\ref{ab}), 
the equal time commutation relations involving the fields and their conjugate 
momenta [eqs.~(\ref{pi}) and (\ref{geral})] become
\begin{eqnarray}
[\Psi(t,\mathbf{r}),\Psi(t,\mathbf{r'}) ] = 
[\Psi^\dagger(t,\mathbf{r}),\Psi^\dagger(t,\mathbf{r'}) ]
= 
[\Pi_{\Psi}(t,\mathbf{r}), \Pi_{\Psi}(t,\mathbf{r'})  ] &=& 0, 
\\ \, 
[\Pi_{\Psi^\dagger}(t,\mathbf{r}), \Pi_{\Psi^\dagger}(t,\mathbf{r'})  ]
= 
[\Psi(t,\mathbf{r}), \Pi_{\Psi^\dagger}(t,\mathbf{r'})  ] = 
[\Psi^\dagger(t,\mathbf{r}), \Pi_{\Psi}(t,\mathbf{r'})  ]
&=& 0, 
\end{eqnarray}
with the non-trivial ones being
\begin{eqnarray}
[\Psi(t,\mathbf{r}),\Psi^\dagger(t,\mathbf{r'}) ] &=&
-[\Psi^\dagger(t,\mathbf{r}),\Psi(t,\mathbf{r'}) ]^\dagger \nonumber \\
&=& \int d^3k\; e^{i\mathbf{k\cdot (r-r')}}
\left\{
[f_+(\mathbf{k})]^2-[f_-(\mathbf{k})]^2]
\right\}, \label{f1}\\ \,
[\Pi_{\Psi}(t,\mathbf{r}), \Pi_{\Psi^\dagger}(t,\mathbf{r'}) ] &=&
-[\Pi_{\Psi^\dagger}(t,\mathbf{r}), \Pi_{\Psi}(t,\mathbf{r'}) ]^\dagger \nonumber \\
&=&\int d^3k \;e^{i\mathbf{k\cdot (r-r')}}\left(\frac{\omega_{\mathbf{k}}^2}{c^4}\right)
\left\{
[f_-(\mathbf{k})]^2-[f_+(\mathbf{k})]^2]
\right\}, \label{f2} \\ \,
[\Psi(t,\mathbf{r}),\Pi_{\Psi}(t,\mathbf{r'}) ] &=&
-[\Psi^\dagger(t,\mathbf{r}),\Pi_{\Psi^\dagger}(t,\mathbf{r'}) ]^\dagger \nonumber \\
&=&\int d^3k \;e^{i\mathbf{k\cdot (r-r')}}\left(\frac{i\omega_{\mathbf{k}}}{c^2}\right)
\left\{
[f_+(\mathbf{k})]^2+[f_-(\mathbf{k})]^2]
\right\}. \label{f3}
\end{eqnarray}

Repeating the above calculations for two arbitrary space-time points $x$ and $x'$,
and using that $x^0=ct$,
we get
\begin{equation}
[\Psi(x),\Psi(x') ] = 
[\Psi^\dagger(x),\Psi^\dagger(x') ]
= 
[\Pi_{\Psi}(x), \Pi_{\Psi}(x')  ]= 0,
\end{equation}
\begin{equation}
[\Pi_{\Psi^\dagger}(x), \Pi_{\Psi^\dagger}(x')  ]
= 
[\Psi(x), \Pi_{\Psi^\dagger}(x')  ] = 
[\Psi^\dagger(x), \Pi_{\Psi}(x')  ]
= 0 
\end{equation}
and
\begin{eqnarray}
[\Psi(x),\Psi^\dagger(x') ] &=&
-[\Psi^\dagger(x),\Psi(x') ]^\dagger
\nonumber \\
&=&\int d^3k\; e^{i\mathbf{k\cdot (r-r')}}
\hspace{-.1cm}\left\{
e^{-i\omega^+_{\mathbf{k}}(t-t')}[f_+(\mathbf{k})]^2
-e^{i\omega^-_{\mathbf{k}}(t-t')}[f_-(\mathbf{k})]^2
\right\}\!\!, \label{f1c} \nonumber \\ \,
[\Pi_{\Psi}(x), \Pi_{\Psi^\dagger}(x') ] &=&
-[\Pi_{\Psi^\dagger}(x), \Pi_{\Psi}(x') ]^\dagger 
\nonumber \\
&=&\int d^3k \;e^{i\mathbf{k\cdot (r-r')}}\hspace{-.1cm}\left(\hspace{-.05cm}\frac{\omega_{\mathbf{k}}^2}{c^4}\hspace{-.05cm}\right)
\hspace{-.1cm}\left\{
e^{-i\omega^-_{\mathbf{k}}(t-t')}[f_-(\mathbf{k})]^2
-e^{i\omega^+_{\mathbf{k}}(t-t')}[f_+(\mathbf{k})]^2
\right\}\!\!, \nonumber \\ \,
[\Psi(x),\Pi_{\Psi}(x') ] &=&
-[\Psi^\dagger(x),\Pi_{\Psi^\dagger}(x') ]^\dagger
\nonumber \\
&=&\int d^3k \;e^{i\mathbf{k\cdot (r-r')}}
\hspace{-.1cm}\left(\hspace{-.1cm}\frac{i\omega_{\mathbf{k}}}{c^2}\hspace{-.1cm}\right)
\hspace{-.1cm}\left\{
e^{-i\omega^+_{\mathbf{k}}(t-t')}[f_+(\mathbf{k})]^2
+e^{i\omega^-_{\mathbf{k}}(t-t')}[f_-(\mathbf{k})]^2
\right\}\!\!. \nonumber
\end{eqnarray}
%

Looking at eqs.~(\ref{f1}) and (\ref{f2}), we see that they can only be zero if 
$f_+(|\mathbf{k}|)$ $=\pm f_-(|\mathbf{k}|)$. The only remaining non zero commutators
with this choice are given by eq.~(\ref{f3}) and they become equal to 
$i\hbar\delta^{(3)}(\mathbf{r-r'})=(i\hbar)/[(2\pi)^3]$ $\int d^3k$ $\;
e^{i(\mathbf{r-r'})\cdot \mathbf{k}}$
if
\begin{equation}
\left(\frac{i2\omega_{\mathbf{k}}}{c^2}\right)
[f_\pm(\mathbf{k})]^2 = \frac{i\hbar}{(2\pi)^3}.
\end{equation}
Solving for $f_\pm(\mathbf{k})$ we obtain eq.~(\ref{f=f}) and inserting it into eq.~(\ref{f3})
we recover eqs.~(\ref{com1}) and (\ref{com2}) of the main text.
Note that strictly speaking $f_+(\mathbf{k})=-f_-(\mathbf{k})$ is also a possible
solution. However, all relevant quantities depend only on the square of these
functions and we choose $f_+(\mathbf{k})=f_-(\mathbf{k})$, eq.~(\ref{f=f}), for 
simplicity. Moreover, inserting eq.~(\ref{f=f}) into eqs.~(\ref{HamG})-(\ref{QG})
we obtain eqs.~(\ref{HamQ})-(\ref{QQ}) of the main text. 

We have just seen that the choice $f_+(\mathbf{k})=f_-(\mathbf{k})=$  
$[(\hbar c^2)$ $/((2\pi)^3 2\omega_{\mathbf{k}})]^{1/2}$ leads to the 
canonical commutation relations for the Lorentz covariant Schr\"odinger fields.
In this scenario the energies of the particles and antiparticles are not the same.
We now look for the values of 
$f_+(\mathbf{k})$ and $f_-(\mathbf{k})$ that lead to particles and antiparticles 
having the same energy. As we will see, this can only be achieved if the fields
no longer satisfy the canonical commutation relations and if the microcausality
condition \cite{man86} is violated. 

If we set 
\begin{equation}
[f_{\pm}(\mathbf{k})]^2 = \frac{\hbar c^2}{(2\pi)^3 2\omega_{\mathbf{k}}^{\pm}},
\label{fpm2}
\end{equation}
we can write the normal ordered eqs.~(\ref{HamG})-(\ref{QG}) as follows
\begin{eqnarray}
H & = & \int d^3k 
\hbar \omega_{\mathbf{k}}\left(
a_{\mathbf{k}}^\dagger a_{\mathbf{k}}
+ b_{\mathbf{k}}^\dagger b_{\mathbf{k}}
\right), \label{HamG2}\\
P^j & = & \int d^3k\; \hbar k^j 
\left(
\frac{\omega_{\mathbf{k}}}{\omega_{\mathbf{k}}^+} a_{\mathbf{k}}^\dagger a_{\mathbf{k}}
+ \frac{\omega_{\mathbf{k}}}{\omega_{\mathbf{k}}^-} b_{\mathbf{k}}^\dagger b_{\mathbf{k}}
\right), \label{PG2} \\
\tilde{Q} & = & \int d^3k\; \hbar 
\left(
\frac{\omega_{\mathbf{k}}}{\omega_{\mathbf{k}}^+} a_{\mathbf{k}}^\dagger a_{\mathbf{k}}
- \frac{\omega_{\mathbf{k}}}{\omega_{\mathbf{k}}^-} b_{\mathbf{k}}^\dagger b_{\mathbf{k}}
\right). \label{QG2}
\end{eqnarray}
Looking at eq.~(\ref{HamG2}) we see that the particle and the antiparticle have the same
energy $E_{\mathbf{k}}=\hbar\omega_{\mathbf{k}}$.
If we use eqs.~(\ref{omegak}) and (\ref{pmE}) we get
\begin{equation}
\frac{\omega_{\mathbf{k}}}{\omega_{\mathbf{k}}^{\pm}} 
= \left( 1 \pm \frac{mc^2}{\hbar\omega^{\pm}_{\mathbf{k}}}\right), 
\end{equation}
which gives us the correction to the momentum and charge as compared to what one 
traditionally expect for these quantities. For particles created by 
$a^\dagger_{\mathbf{k}}$, their momentum and charge 
is increased
by $(mc^2)/(\hbar\omega^{+}_{\mathbf{k}})$ while for antiparticles created by
$b^\dagger_{\mathbf{k}}$ the same quantities are reduced by 
$(mc^2)/(\hbar\omega^{-}_{\mathbf{k}})$. This is consistent with considering 
particles with mass $m$ and antiparticles with mass $-m$. Also, we can understand
$(mc^2)/$ $(\hbar\omega^{+}_{\mathbf{k}})$ as the ratio between the rest energy of
a particle with mass $m$ to its kinetic energy $\hbar\omega^{+}_{\mathbf{k}}$. Now,
if $m\rightarrow -m$ we get $(mc^2)/(\hbar\omega^{+}_{\mathbf{k}})\rightarrow 
(-mc^2)/(\hbar\omega^{-}_{\mathbf{k}})$. This means that we can interpret 
the latter as the ratio between the rest energy of
an antiparticle with mass $-m$ and kinetic energy $\hbar\omega^{-}_{\mathbf{k}}$.
Note that we can 
interchange the roles of particles with antiparticles in the expressions for the momentum and charge, leaving them invariant,
if $a_{\mathbf{k}} \leftrightarrow b_{\mathbf{k}}$ and $m \rightarrow -m$, where the 
latter operation implies $\omega_{\mathbf{k}}^{\pm}
\rightarrow \omega_{\mathbf{k}}^{\mp}$. 

Let us see now what happens to the equal time commutation relations of the fields. 
Inserting eq.~(\ref{fpm2}) into  
eqs.~(\ref{f1})-(\ref{f3}), the no-trivial commutation relations become
\begin{eqnarray}
[\Psi(t,\mathbf{r}),\!\Psi^\dagger(t,\mathbf{r'}) ] &=&
-\frac{mc^2}{\nabla_{\mathbf{r}}^2}\delta^{(3)}(\mathbf{r-r'}), \label{f1a}\\ \,
[\Pi_{\Psi}(t,\mathbf{r}),\! \Pi_{\Psi^\dagger}(t,\mathbf{r'}) ] &=&
-m\!\!\left(\!\!1 \!-\!\frac{\mu^2}{\nabla_{\mathbf{r}}^2}\!\right)\!\!\delta^{(3)}\!(\mathbf{r-r'})\!, \label{f2a} \\ \,
[\Psi(t,\mathbf{r}),\!\Pi_{\Psi}(t,\mathbf{r'}) ] &=&
i\hbar \!\left(\!\!1 -\frac{\mu^2}{\nabla_{\mathbf{r}}^2}\!\right)\!\!\delta^{(3)}\!(\mathbf{r-r'}), \label{f3a}
\end{eqnarray}
where $1/\nabla_{\mathbf{r}}^2$ is the inverse Laplacian,
\begin{equation}
\frac{1}{\nabla_{\mathbf{r}}^2}e^{i\mathbf{k\cdot r}} = -\frac{e^{i\mathbf{k\cdot r}}}
{|\mathbf{k}|^2}.
\end{equation}
Note that for this particular choice of $f_{\pm}(\mathbf{k})$ we have that 
$[\Pi_{\Psi}(t,\mathbf{r}),\! \Pi_{\Psi^\dagger}(t,\mathbf{r'})]=(im/\hbar)$ $[\Psi(t,\mathbf{r}),\!\Pi_{\Psi}(t,\mathbf{r'})]$.

If we use the integral representation of the Dirac delta function, eq.~(\ref{diracdelta}), and apply the inverse Laplacian,
we get the following integral during the calculations leading to the commutators (\ref{f1a})-(\ref{f3a}),
\begin{equation}
\int d^3k\; \frac{e^{i\mathbf{k\cdot (r - r')}}}{|\mathbf{k}|^2} = \frac{4\pi}{|\mathbf{r-r'}|}\int_0^\infty du \frac{\sin u}{u} = 
\frac{2\pi^2}{|\mathbf{r-r'}|}. \label{integral1}
\end{equation}
With the aid of eq.~(\ref{integral1}) we get for eqs.~(\ref{f1a})-(\ref{f3a}),
\begin{eqnarray}
[\Psi(t,\mathbf{r}),\!\Psi^\dagger(t,\mathbf{r'}) ] &=&
\frac{mc^2}{4\pi}\frac{1}{|\mathbf{r-r'}|}, \label{f1b}\\ \, 
[\Pi_{\Psi}(t,\mathbf{r}),\! \Pi_{\Psi^\dagger}(t,\mathbf{r'}) ] &=&
-m\!\left(\!\delta^{(3)}\!(\mathbf{r-r'})\!+\!\frac{\mu^2}{4\pi|\mathbf{r-r'}|}\right)\!, \label{f2b} \nonumber \\
\\ \,
[\Psi(t,\mathbf{r}),\!\Pi_{\Psi}(t,\mathbf{r'}) ] &=&
i\hbar\! \left(\!\delta^{(3)}\!(\mathbf{r-r'})\! +\!\frac{\mu^2}{4\pi|\mathbf{r-r'}|}\right)\!. \nonumber \\
&&\, \label{f3b}
\end{eqnarray}

The important messages are contained in eqs.~(\ref{f1b}) and (\ref{f3b}). First, eq.~(\ref{f1b}) 
tells us that $[\Psi(t,\mathbf{r}),\!\Psi^\dagger(t,\mathbf{r'}) ]\neq 0$, which
implies the violation of microcausality \cite{man86,gre95}. On the other hand,
eq.~(\ref{f3b}) shows that the field and its conjugate momentum do not satisfy the canonical
commutation relation. In addition to the canonical term, we have a correction given by
$i\hbar \mu^2/(4\pi|\mathbf{r-r'}|)$. This term, whose origin stems from eq.~(\ref{f1b}),
can be seen as giving origin to an
instantaneous Coulomb-like potential. 
Contrary to the instantaneous Coulomb potential
coming from the photon propagator \cite{man86}, here the ``charges'' 
associated with this 
interaction are the masses of the particles. 
What is intriguing here, however, is that we are dealing with scalar matter fields not 
vector fields like the photons. The emergence of this Coulomb-like potential may be 
just a coincidence but, pushing further
the speculation, it may be that a more general quantum field theory incorporating 
positive and negative masses might naturally 
lead to the emergence of the gravitational interaction.  

\subsection{More on the commutators for the non-canonically quantized fields}

We can repeat the steps followed in section \ref{microSec} using eqs.~(\ref{f1c})
and (\ref{fpm2}) to obtain that
\begin{eqnarray}
[\Psi(x),\Psi^\dagger(y)] = i\hbar c\, \tilde{\Delta}_{LS}(x-y),
\label{moreC}
\end{eqnarray}
where
\begin{eqnarray}
\tilde{\Delta}_{LS}(x) &=& e^{i\mu x^0}\tilde{\Delta}(x), \\
\tilde{\Delta}(x) &=& \tilde{\Delta}^+(x)+\tilde{\Delta}^-(x), \\ 
\tilde{\Delta}^+(x) &=& \frac{-ic}{2(2\pi)^3}\int d^3k\, 
\frac{e^{-ikx}}{\omega_{\mathbf{k}}^+}, \\
\tilde{\Delta}^-(x) &=& \frac{ic}{2(2\pi)^3}\int d^3k\, 
\frac{e^{ikx}}{\omega_{\mathbf{k}}^-},
\end{eqnarray}
with $k^0=\omega_{\mathbf{k}}/c$. We are putting the tilde ``$\;\tilde{\,}\;$'' on top
of any quantity coming from the non-canonically quantized theory in order to
differentiate it from the ones associated with the canonically
quantized theory in the main text. 

If we use the Heaviside step function $h(k^0)$, in which $h(k^0)=1$ for $k^0>0$ and 
$h(k^0)=0$ for $k^0<0$, we have
\begin{eqnarray}
\tilde{\Delta}^{\!\pm}(x) &\!=\!& 
\frac{-i}{(2\pi)^3}\int d^4k \,
h(\pm k^0)\varepsilon(k^0)\delta(k^2\pm2\mu\omega_{\mathbf{k}}^{\pm}/c)e^{-ikx}\!
\nonumber \\
&\!=\!& \frac{-i}{(2\pi)^3}\int d^4k \frac{k^0}{k^0-\mu}
h(\pm k^0)\varepsilon(k^0)\delta(k^2-\mu^2)e^{-ikx}\!,\nonumber \\
&& \, \label{a36}
\end{eqnarray}
where the other quantities above were defined in section \ref{qft}.

Noting that $h(k^0)+h(-k^0)=1$ for any $k^0\neq 0$ we get
\begin{equation}
\tilde{\Delta}(x) \!=\!\frac{-i}{(2\pi)^3}\int d^4k \frac{k^0}{k^0-\mu}
\varepsilon(k^0)\delta(k^2-\mu^2)e^{-ikx}\!.
\end{equation}
If inside the integral sign we add and subtract the term $-\mu/$ $(k^0-\mu)$, we can write
$\tilde{\Delta}(x)$ in the following illustrative way
\begin{equation}
\tilde{\Delta}(x) \!=\! \Delta(x) + \tilde{\delta}(x),
\label{a38}
\end{equation}
where
\begin{equation}
\tilde{\delta}(x) = \frac{-i}{(2\pi)^3}\int d^4k \frac{\mu}{k^0-\mu}
\varepsilon(k^0)\delta(\!k^2-\mu^2)e^{-ikx}
\label{a39}
\end{equation}
and $\Delta(x)$ is given by eq.~(\ref{deltakg2}). 
We can thus understand $\tilde{\delta}(x)$ as the correction to $\Delta(x)$ when
we deviate from the canonical quantization path of section \ref{qft}. 

In analogy to eq.~(\ref{contourCpm}), we can write 
$\tilde{\Delta}^\pm(x)$ as the following contour integral,
where $k^0$ is considered a complex variable,
\begin{eqnarray}
\tilde{\Delta}^\pm(x) &=& -\frac{1}{(2\pi)^4}\int_{\tilde{C}^{\pm}}d^4k\,
\frac{e^{-i\mu x^0}e^{-ikx}}{k_0^2-(\omega_{\mathbf{k}}^\pm/c)^2} 
=-\frac{1}{(2\pi)^4}\int_{\tilde{C}^{\pm}}d^4k\,
\frac{e^{-i\mu x^0}e^{-ikx}}{k^2\pm (2\mu\omega_{\mathbf{k}}^\pm/c)}.
\label{contourCtildepm}
\end{eqnarray}
Here $\tilde{C}^+$ is any counterclockwise closed
path encircling only $\omega_{\mathbf{k}}^+/c$ and
$\tilde{C}^-$ is any counterclockwise closed
path encircling $-\omega_{\mathbf{k}}^-/c$. Obviously,
$\tilde{\Delta}(x)=\tilde{\Delta}^+(x)+\tilde{\Delta}^-(x)$. We can have an integral representation in which $\tilde{\Delta}^+(x)$ and 
$\tilde{\Delta}^-(x)$ have the same integrand, differing only in the path of integration,
\begin{equation}
\tilde{\Delta}^\pm(x) = -\frac{1}{(2\pi)^4}\int_{C^{\pm}}d^4k\,
\frac{k^0}{k^0-\mu}
\frac{e^{-ikx}}{k^2-\mu^2}. \label{contourCpm2}
\end{equation}
The above paths of integration are defined in eq.~(\ref{contourCpm}), i.e.,
$C^{\pm}$ encircles counterclockwise only $\omega_{\mathbf{k}}^{\pm}/c$.
Note that we cannot express $\tilde{\Delta}(x)$ with the integral representation
above and using the path $C$, as we did for $\Delta(x)$ in eq.~(\ref{contourC}), 
since now we have a pole in $k^0=\mu$. If we insist using the path $C$ of 
eq.~(\ref{contourC}) we must subtract the contribution of the latter pole to the integral,
\begin{equation}
\tilde{\Delta}(x) = -\frac{1}{(2\pi)^4}\int_{C}d^4k\,
\left[\frac{k^0}{k^0-\mu}
\frac{e^{-ikx}}{k^2-\mu^2}\right]
-\frac{i\mu e^{-i\mu x^0}}{2(2\pi)|\mathbf{x}|}. \label{contourC2}
\end{equation}

Let us return to the second line of eq.~(\ref{a36}), which is conveniently written to 
prove the invariance  of $\tilde{\Delta}^\pm(x)$ under proper
Lorentz transformations. Its invariance under spatial rotations is obvious. The 
subtle point lies in proving its invariance under a Lorentz boost. 
To prove that we first note that the Dirac delta guarantees 
that we are on mass shell and therefore  
$h(\pm k^0)$ and $\epsilon(k^0)$ are invariant under a Lorentz boost. 
The other invariants in eq.~(\ref{a36}) are $d^4k,k^2, \mu^2$, and $kx$. 
Thus, after a Lorentz boost from reference frame $S$ to $S'$ 
in the $x^1$-direction we have
\begin{eqnarray}
[\tilde{\Delta}^{\!\pm}]'(x') \!&\!=\!&\!
\frac{-i}{(2\pi)^3}\int d^4k' 
\frac{\gamma(k^{0'}+\beta k^{1'})}{\gamma(k^{0'}+\beta k^{1'})-\mu} 
h(\pm k^{0'})\varepsilon(k^{0'})\delta[(k')^2-\mu^2]e^{-ik'x'}\!\!.
\label{a36b}
\end{eqnarray}

If we now make the following change of variables (this is not another Lorentz boost),
\begin{eqnarray}
k^{0''} = \gamma(k^{0'}+\beta k^{1'}), \label{cv1}\\
k^{1''} = \gamma(k^{1'}+\beta k^{0'}), \label{cv2}
\end{eqnarray}
we get for the right hand side of eq.~(\ref{a36b}),
\begin{eqnarray}
[\tilde{\Delta}^{\!\pm}]'(x')\!&\!=\!&\!
\frac{-i}{(2\pi)^3}\int d^4k'' 
\frac{k^{0''}}{k^{0''}-\mu} h(\pm k^{0''})\varepsilon(k^{0''}) 
\delta[(k'')^2-\mu^2] e^{-ik^{0''}[\gamma(x^{0'}+\beta x^{1'})]}
\nonumber \\
\!&\!\times\!&\! e^{
ik^{1''}[\gamma(x^{1'}+\beta x^{0'})]
+ik^{2''}x^{2'}+ik^{3''}x^{3'}}\!.
\label{a36c}
\end{eqnarray}
Note that we still have $x'$ in eq.~(\ref{a36c})
since we are just changing variables in the 
integral and not performing a Lo\-ren\-tz boost. Moreover, the change of variables 
(\ref{cv1})-(\ref{cv2}) are formally equivalent to a Lorentz boost from the point of 
view of $k'$ and this is why
$\int d^4k' h(\pm k^{0'})\varepsilon(k^{0'})\delta[\!(k')^2-\mu^2]$ is invariant when
we go to $k''$.

Finally, if we remember that the variables in $S'$ are connected to the ones in $S$
by the following relation,
\begin{eqnarray}
x^0 &=& \gamma (x^{0'} + \beta x^{1'}), \\
x^1 &=& \gamma (x^{1'} + \beta x^{0'}), \\
x^2 &=& x^{2'}, \\
x^3 &=& x^{3'},
\end{eqnarray}
eq.~(\ref{a36c}) becomes after we rename the integration variable $k''$ to $k$, 
\begin{eqnarray}
[\tilde{\Delta}^{\!\pm}]'\!(x')\!\!&\!=\!&\!\!
\frac{-i}{(2\pi)^3}\int d^4k 
\frac{k^{0}}{k^{0}-\mu} h(\pm k^{0})\varepsilon(k^{0})\delta(k^2\!-\!\mu^2) e^{-ikx}. 
\nonumber \\
&& \,
\label{a36d}
\end{eqnarray}
Comparing eq.~(\ref{a36d}) with (\ref{a36}) we see that they are the same, which 
proves that under a proper Lorentz transformation 
$\tilde{\Delta}^{\!\pm}(x)$ is invariant. The invariance of 
$\tilde{\Delta}^{\!\pm}(x)$ and $\Delta^{\!\pm}(x)$ also imply that 
$\tilde{\Delta}(x)$ and $\tilde{\delta}(x)$ are invariant
under proper Lorentz transformations and that the commutator 
(\ref{moreC}) and the Feynman propagator (\ref{FPNC2}) 
are Lorentz covariant quantities.

\subsection{The Feynman propagator for the non-canonically quantized fields}

In the present case eq.~(\ref{FPLS2}) becomes
\begin{equation}
\tilde{\Delta}_{F_{\!_{LS}}}(x) =   h(x^0)\tilde{\Delta}_{LS}^+(x) 
- h(-x^0)\tilde{\Delta}_{LS}^-(x), \label{FPNC2}
\end{equation}
with
\begin{equation}
\tilde{\Delta}^{\pm}_{LS}(x) = e^{i\mu x^0}\tilde{\Delta}^{\pm}(x).
\label{a42}
\end{equation}
Using eq.~(\ref{contourCtildepm}) we get
\begin{equation}
\tilde{\Delta}^{\pm}_{LS}(x) 
= -\frac{1}{(2\pi)^4}\int_{\tilde{C}^{\pm}}d^4k\,
\frac{e^{-ikx}}{k^2\pm (2\mu\omega_{\mathbf{k}}^\pm/c)}.
\label{contourCtildepm2}
\end{equation}

It is not as simple as in the canonically quantized case to 
write the Feynman propagator as a single complex integral. The reason for this
difficulty can be seen looking at the denominators of eq.~(\ref{contourCtildepm}).
Instead of the two simple poles we met when dealing with the canonically quantized theory,
we now have four simple poles, namely,
$k^0=\pm \omega_{\mathbf{k}}^+/c$ and $k^0=\pm \omega_{\mathbf{k}}^-/c$. 
This means that the 
path along the real axis 
of the complex contour integral 
defining the propagator is different whether 
we have $x^0>0$ or $x^0<0$. For $x^0>0$, the clockwise path $\tilde{C}_F^+$
must enclose only the pole $k^0=\omega_{\mathbf{k}}^+/c$, skirting counterclockwise the other three poles. For $x^0<0$, the counterclockwise  path $\tilde{C}_F^-$
must enclose only the pole $k^0=-\omega_{\mathbf{k}}^-/c$, skirting 
clockwise the remaining three poles. With this understanding for the path
$\tilde{C}_F$ we have
\begin{eqnarray}
\tilde{\Delta}_{F_{\!_{LS}}}(x) \!&\!=\!&\!  \frac{1}{(2\pi)^4}\int_{\tilde{C}_F}\!\!d^4k\!
\left[\frac{1}{k^2 + (2\mu\omega_{\mathbf{k}}^+/c)} 
+\frac{1}{k^2 - (2\mu\omega_{\mathbf{k}}^-/c)}\right] e^{-ikx} \nonumber \\
\!&\!=\!&\!  \frac{1}{(2\pi)^4}\!\!\int_{\tilde{C}_F}\!\!\!\!d^4k\!
\left[\! \frac{2k^2-4\mu^2}{k^4-4\mu^2(k^0)^2}\!\right]\!\!e^{-ikx}.
\end{eqnarray}

The same difficulty above manifests itself when writing the propagator considering 
the four variables $k^\nu$ real, with all integrals running from $-\infty$ to
$\infty$. In this case we can circumvent the four poles problem using the Heaviside
step function, which allows us to write
\begin{eqnarray}
\tilde{\Delta}_{F_{\!_{LS}}}(x) \!&\!=\!&\!  \frac{1}{(2\pi)^4}\int \!\!d^4k\!
\left[\frac{h(x^0)}{k^2 + (2\mu\omega_{\mathbf{k}}^+/c) + i\epsilon^+} 
+\frac{h(-x^0)}{k^2 - (2\mu\omega_{\mathbf{k}}^-/c)+ i\epsilon^-}\right] 
e^{-ikx}. \label{FPNC3}
\end{eqnarray}
In eq.~(\ref{FPNC3}) the limit $\epsilon^{\pm}\rightarrow 0$ is understood. Also, in analogy
to the canonically quantized theory, 
\begin{equation}
\epsilon^{\pm} = \frac{2\eta\omega_{\mathbf{k}}^{\pm}}{c},
\end{equation}
with $0<\eta \ll 1$, and
\begin{equation}
k^2 \pm (2\mu\omega_{\mathbf{k}}^{\pm}/c) + i\epsilon^{\pm} = 
(k^0)^2 -(\omega_{\mathbf{k}}^{\pm}/c - i\eta)^2,
\end{equation}
where second order terms in $\eta^2$ were discarded.

Using eqs.~(\ref{a38})-(\ref{a39}), (\ref{FPNC2})-(\ref{a42}), and
(\ref{FPNC3}), we have
\begin{equation}
\tilde{\Delta}_{F_{\!_{LS}}}(x) = \Delta_{F_{\!_{LS}}}(x) + 
\tilde{\delta}_{F_{\!_{LS}}}(x),
\end{equation}
where $\Delta_{F_{\!_{LS}}}(x)$ is the propagator for the canonically quantized theory
and 
\begin{eqnarray}
\tilde{\delta}_{F_{\!_{LS}}}(x) &=& \frac{-ie^{i\mu x^0}}{(2\pi)^3}\int d^4k
\left[ 
\frac{\mu}{k^0-\mu}
\frac{1+\epsilon(k^0)\epsilon(x^0)}{2}
\delta(k^2-\mu^2)\right]e^{-ikx},
\end{eqnarray}
where $\epsilon(x^0)$ is the sign function (do not confuse it with the small
$\epsilon$ appearing in the denominator of the propagator). 
A similar interpretation attributed to
$\tilde{\delta}(x)$ applies here, namely, $\tilde{\delta}_{F_{\!_{LS}}}(x)$ is the 
correction to the Feynman propagator due to our departure from the canonical quantization
of section \ref{qft}.

Building on eq.~(\ref{contourC2}) we also have
\begin{eqnarray}
\tilde{\Delta}_{F_{\!_{LS}}}(x) &=& \frac{e^{i\mu x^0}}{(2\pi)^4}\int d^4k\,
\left[\frac{k^0}{k^0-(\mu-i\alpha)}
\frac{e^{-ikx}}{k^2-\mu^2+i\epsilon}\right] 
-\frac{i\mu h(x^0)}{2(2\pi)|\mathbf{x}|},
\label{dfls}
\end{eqnarray}
where $\alpha>0, \epsilon >0$, and the limit 
$(\alpha, \epsilon) \rightarrow (0,0)$ is implied 
after integration in $k^0$. If we use that $\lim_{x^0\rightarrow 0}h(x^0)=1/2$,
we realize that in eq.~(\ref{dfls}) the last
term is related to the instantaneous gravitational-like potential that arises
when we non-canonically quantize the Lorentz covariant Schr\"odinger equation.

Furthermore, if we use the integral representation for the Heaviside step function,
\begin{equation}
h(x^0) = -\frac{1}{2\pi i}\int dk^0 \frac{e^{-ik^0x^0}}{k^0+i\alpha},
\end{equation}
with $\alpha>0$ and $\alpha \rightarrow 0$ implied, and that
\begin{equation}
\frac{1}{|\mathbf{x}|} = \frac{2}{(2\pi)^2}\int d^3k 
\frac{e^{i\mathbf{k}\cdot\mathbf{x}}}{|\mathbf{k}|^2},
\end{equation}
we have after changing the variable of integration $k^0\rightarrow k^0 -\mu$,
\begin{equation}
\frac{-i\mu h(x^0)}{2(2\pi)|\mathbf{x}|} = \frac{\mu e^{i\mu x^0}}{(2\pi)^4}
\int d^4k \frac{e^{-ikx}}{[k^0-(\mu -i\alpha)]|\mathbf{k}|^2}. \label{inth}
\end{equation}
Thus, using eq.~(\ref{inth}) we can rewrite eq.~(\ref{dfls}) as
\begin{eqnarray}
\tilde{\Delta}_{F_{\!_{LS}}}(x) &=& \frac{e^{i\mu x^0}}{(2\pi)^4}\!\int\!\! d^4k\!
\left\{ \frac{k^0}{[k^0-(\mu-i\alpha)][k^2-\mu^2+i\epsilon]} 
+\frac{\mu}{[k^0-(\mu -i\alpha)]|\mathbf{k}|^2} \right\}e^{-ikx} \nonumber \\
&=& \frac{e^{i\mu x^0}}{(2\pi)^4}\!\int\!\! d^4k\, e^{-ikx}
\left\{\frac{1}{k^0-(\mu -i\alpha)}\left[ 
\frac{k^0}{k^2-\mu^2+i\epsilon} +\frac{\mu}{|\mathbf{k}|^2}\right] \right\}.
\end{eqnarray}

\section{Equivalence with the Klein-Gordon theory}
\label{apC}

\subsection{The $\mathbf{(\lambda/4)[\Psi(x)\Psi^\dagger(x)]^2}$ interaction}
\label{lambda4b}

The interaction Lagrangian density  
\begin{displaymath}
\mathcal{L}_{int}=-(\lambda/4)[\Psi(x)\Psi^\dagger(x)]^2
\end{displaymath}
does not depend on time derivatives 
and thus \cite{man86,gre95}
\begin{equation}
\mathcal{H}_{int} = -\mathcal{L}_{int}=(\lambda/4)[\Psi(x)\Psi^\dagger(x)]^2.
\label{lambda4}
\end{equation}
Note that $\mathcal{H}_{int}$ is clearly Lorentz covariant since the phase change
induced in $\Psi(x)$ by a proper Lorentz transformation is compensated by the one 
induced in $\Psi^\dagger(x)$ [See eq.~(\ref{transformationL})].
Also, since the interaction picture is connected to the Heisenberg picture via a 
unitary transformation we obtain
\begin{equation}
\mathcal{H}_{int}^I = -\mathcal{L}_{int}^I=(\lambda/4)\{\Psi^I(x)[\Psi^I(x)]^\dagger\}^2.
\label{lambda4i}
\end{equation}

Inserting eq.~(\ref{lambda4i}) into (\ref{dyson}) we get
\begin{equation}
S = T \exp\left\{-\frac{i\lambda}{4\hbar c}\int d^4x \{\Psi^I(x)[\Psi^I(x)]^\dagger\}^2\right\}. \label{lsint}
\end{equation}
If we now use eq.~(\ref{ikg}) we arrive at
\begin{equation}
S = T \exp\left\{-\frac{i\lambda}{4\hbar c}\int d^4x \{\Phi^I(x)[\Phi^I(x)]^\dagger\}^2\right\}. \label{kgint}
\end{equation}

Looking at eq.~(\ref{kgint}), we realize that it is the S-matrix for the 
complex Klein-Gordon fields if they interact according to
\begin{equation}
\mathcal{H}_{int}^I = (\lambda/4)\{\Phi^I(x)[\Phi^I(x)]^\dagger\}^2.
\label{lambda4ikg}
\end{equation}

This tells us that the S-matrix for the Klein-Gordon and Lorentz covariant Schr\"odinger
fields are identical if they are subjected to the same type of interaction
[eqs.~(\ref{lambda4i}) and (\ref{lambda4ikg})]. Furthermore, 
the flux and linear momentum  
associated to a particle are formally the same in both theories, 
as can be seen comparing eqs.~(\ref{Jj}) and (\ref{PQ})   
with the respective ones coming from 
the Klein-Gordon theory. Therefore, any scattering cross section (decay rate) 
computed for both theories have the same value if the colliding (decaying) particles
are assumed to have the same initial linear momentum.

The preceding analysis can be easily extended to any interaction of the type
\begin{equation}
\mathcal{H}_{int} = g_n[\Psi(x)\Psi^\dagger(x)]^n,
\label{lambdaNi}
\end{equation}
with $n$ a real number and $g_n$ the associated coupling constant. 
Since $\Psi$ and $\Psi^\dagger$ have the same exponent, we can use 
eq.~(\ref{ikg}) and the same arguments above to prove the equivalence between
the Klein-Gordon and the Lorentz covariant Schr\"odinger theories.

On the other hand, the requirement for Lorentz covariance forbids interactions 
given by
\begin{equation}
\mathcal{H}_{int} = g_{n\tilde{n}}[\Psi(x)]^n[\Psi^\dagger(x)]^{\tilde{n}},
\hspace{.25cm}\mbox{with}\hspace{.25cm} n\neq \tilde{n}.
\label{lambdaNMi}
\end{equation}
Indeed, for different values of $n$ and $\tilde{n}$ we have after a proper Lorentz 
transformation
\begin{equation}
[\Psi(x)]^n[\Psi^\dagger(x)]^{\tilde{n}} \longrightarrow 
e^{i(n-\tilde{n})\theta(x)/\hbar}[\Psi(x)]^n[\Psi^\dagger(x)]^{\tilde{n}}, 
\end{equation}
which clearly shows the lack of Lorentz covariance if $n\neq \tilde{n}$. 

When we have two (or more) types of particles with different masses, similar 
arguments forbid interaction terms proportional to 
$\Psi_1^{n_1}(\Psi_1^\dagger)^{\tilde{n}_1}\Psi_2^{n_2}(\Psi_2^\dagger)^{\tilde{n}_2}$, 
for $n_j\neq \tilde{n}_j$, $j=1,2$. Here $\Psi_j$ 
corresponds to the field associated to the particle of mass $m_j$. For Klein-Gordon fields, 
interactions given by terms similar to those of eq.~(\ref{lambdaNMi}) are possible
since they remain invariant under a proper Lorentz transformation.

\subsection{Scalar electrodynamics}
\label{sedsection}

The proof of the equivalence between the Klein-Gordon and the Lorentz covariant 
Schr\"o\-din\-ger quantum scalar electrodynamics is more subtle. 
Before presenting this proof, we
need first to develop the classical scalar electrodynamics
of the Lorentz covariant Schr\"odinger fields. After obtaining the classical
Hamiltonian density describing all fields and the interactions among them, we are
ready to proceed to its canonical quantization and give the  
proof of the equivalence of both theories.

Applying the minimal coupling prescription, eq.~(\ref{minimalcoupling}), 
to the free Lorentz covariant 
Schr\"odinger Lagrangian density, eq.~(\ref{symL2}), we get
\begin{eqnarray}
\mathcal{L} &=& (D_\mu\Psi)(D^\mu\Psi)^*
+ i\mu \left[ \Psi^* (D_0\Psi) - \Psi (D_0\Psi)^*\right] \nonumber \\
&=&\mathcal{L}_{LS} + \mathcal{L}_{int}. \label{L0LI}
\end{eqnarray}
Here the free field and interaction Lagrangian
densities are
\begin{eqnarray}
\mathcal{L}_{LS} &=& \partial_\mu\Psi\partial^\mu\Psi^*
+ i\mu \Psi^*\overleftrightarrow{\partial}_{\hspace{-.15cm}0}\Psi,
\label{Lfree}\\
\mathcal{L}_{int} &=& 
-\frac{iq}{\hbar}(\Psi^*\overleftrightarrow{\partial}_{\hspace{-.15cm}\mu}\Psi)A^\mu
+ \frac{q^2}{\hbar^2}\Psi\Psi^*A_\mu A^\mu 
-\frac{2q\mu}{\hbar}\Psi\Psi^*A^0. 
\label{Lint}
\end{eqnarray}
Equation (\ref{L0LI}) is invariant under a gauge transformation, namely,
if 
\begin{eqnarray}
\Psi(x) &\longrightarrow& e^{\frac{iq}{\hbar}\chi(x)}\Psi(x), \\
A_\mu &\longrightarrow& A_\mu -\partial_\mu\chi(x),
\end{eqnarray}
we have that
\begin{equation}
\mathcal{L} \longrightarrow \mathcal{L}.
\end{equation}

In eq.~(\ref{Lint}), the third term at the right hand side is characteristic of 
the Lorentz covariant Schr\"odinger Lagrangian, while the first and second ones 
are formally equivalent to what one would get by applying the 
minimal coupling prescription to the Klein-Gordon Lagrangian.

If we use eq.~(\ref{transformationKG2}),
\begin{equation}
\Psi(x) = e^{i\mu x^0}\Phi(x),
\label{transformationKG2b}
\end{equation}
which connects the Lorentz covariant Schr\"odinger and Klein-Gordon equations, we get
after inserting it in eq.~(\ref{L0LI}),
\begin{eqnarray}
\mathcal{L}_{LS} &\longrightarrow& \mathcal{L}_{KG}, \\
\mathcal{L}_{int} &\longrightarrow& 
-\frac{iq}{\hbar}(\Phi^*\overleftrightarrow{\partial}_{\hspace{-.15cm}\mu}\Phi)A^\mu
+ \frac{q^2}{\hbar^2}\Phi\Phi^*A_\mu A^\mu.  \label{intKG} 
\end{eqnarray}
Here $\mathcal{L}_{KG}$ is the Klein-Gordon free field Lagrangian density and
the right hand side of eq.~(\ref{intKG}) is exactly the interaction term one gets
by applying the minimal coupling prescription to $\mathcal{L}_{KG}$.

We should also mention
that similarly to $\mathcal{L}_{LS}$, the interaction Lagrangian density 
$\mathcal{L}_{int}$, eq.~(\ref{Lint}), is invariant under a proper Lorentz transformation. 
This is proved via a direct calculation using  eq.~(\ref{transformationL}), the transformation 
rule for $\Psi(x)$ under a proper Lorentz transformation, and the respective covariant 
and contravariant transformation rules for the vectors $\partial_\mu$ and $A^\mu$.

The complete scalar electrodynamics Lagrangian density also has 
the electromagnetic 
(EM) free-field term $\mathcal{L}_{EM}$, 
which is a function of the four-potential $A^\mu$ (the four components of $A^\mu$ are 
considered independent variables) 
\cite{man86,gre95}. Therefore, the total Lagrangian density
for the scalar electrodynamics (SED) becomes
\begin{equation}
\mathcal{L}_{SED} = \mathcal{L}_{LS} + \mathcal{L}_{EM} + \mathcal{L}_{int}.
\label{sed}
\end{equation}

In order to carry out the canonical quantization of the scalar electrodynamics, we 
need the Hamiltonian density associated to $\mathcal{L}_{SED}$. 
Since the interaction term $\mathcal{L}_{int}$ does not contain time derivatives of
the fields $A^\mu$, the canonically conjugate fields related to them are the same 
as those of the free-field case. As such, after the Legendre transformation leading 
from the Lagrangian to the Hamiltonian density we have 
$\mathcal{L}_{EM} \rightarrow \mathcal{H}_{EM}$, where $\mathcal{H}_{EM}$ is the 
standard free-field
Hamiltonian density for the electromagnetic field \cite{man86,gre95}. 
However, $\mathcal{L}_{int}$ 
contains terms involving time derivatives of the fields $\Psi$ and $\Psi^*$.
This means that their canonically conjugate fields are different from the free-field case,
which ultimately implies that $\mathcal{H}_{int}$ is no longer 
simply $-\mathcal{L}_{int}$. 

The new conjugate field to $\Psi$ is
\begin{eqnarray}
\Pi_{\Psi} &=& \frac{\partial \mathcal{L}_{SED}}{\partial(\partial_t\Psi)} 
=\frac{\partial \mathcal{L}_{LS}}{c\partial(\partial_0\Psi)}
+ \frac{\partial \mathcal{L}_{int}}{c\partial(\partial_0\Psi)} 
=\frac{1}{c}\left( \partial_0\Psi^* +i\mu\Psi^* -\frac{iq}{\hbar}\Psi^*A^0\right).
\label{pinovo}
\end{eqnarray}
In a similar way, the conjugate field to $\Psi^*$ is
\begin{equation}
\Pi_{\Psi^*} =
\frac{1}{c}\left( \partial_0\Psi -i\mu\Psi +\frac{iq}{\hbar}\Psi A^0\right)
=\Pi^*_{\Psi}.
\label{pinovo*}
\end{equation}

Using eqs.~(\ref{pinovo}) and (\ref{pinovo*}), the Hamiltonian density 
\begin{equation}
\mathcal{H} = \Pi_{\Psi}\partial_t\Psi + \Pi_{\Psi^*}\partial_t\Psi^* 
- \mathcal{L}_{LS} - \mathcal{L}_{int} + \mathcal{H}_{EM} 
\end{equation}
becomes
\begin{equation}
\mathcal{H} = \mathcal{H}_{LS} + \mathcal{H}_{EM} + \mathcal{H}_{int}, 
\end{equation}
where $\mathcal{H}_{LS}$ and $\mathcal{H}_{EM}$ are, respectively, 
the free-field Hamiltonian densities
for the Lorentz covariant Schr\"odinger and electromagnetic fields, while
\begin{equation}
\mathcal{H}_{int} = \frac{iqc}{\hbar}(\Pi_{\Psi^*}\Psi^*-\Pi_{\Psi}\Psi )A^0
+\frac{iq}{\hbar}(\Psi^* \overleftrightarrow{\partial}_{\hspace{-.15cm}j}
\Psi)A^j 
-\frac{q^2}{\hbar^2}\Psi\Psi^*A_\mu A^\mu  + \frac{q^2}{\hbar^2}\Psi\Psi^*A_0 A^0.
\label{HintSED}
\end{equation}
Equation (\ref{HintSED}) is formally the same as the one we would get if we had worked 
with the Klein-Gordon Lagrangian \cite{gre95}. Here, however, the expressions
for $\Pi_{\Psi}$ and $\Pi_{\Psi^*}$ are given by eqs.~(\ref{pinovo}) and (\ref{pinovo*})
while in the Klein-Gordon case we do not have the terms $i\mu\Psi^*$ and
$-i\mu\Psi$ in eqs.~(\ref{pinovo}) and (\ref{pinovo*}), respectively.

In order to quantize eq.~(\ref{HintSED}), the independent field variables $\Psi$, $\Psi^*$,
$\Pi_{\Psi}$, $\Pi_{\Psi^*}$, and $A^\mu$ are promoted to operators, satisfying the usual
equal-time commutation relations. The commutation relations for the 
scalar fields are given in eqs. (\ref{com1}) and (\ref{com2}) and the commutation relations
for the four-vector potential can be found in refs. \cite{man86,gre95}. Also, to avoid
ordering ambiguities among the operators as well as to eliminate non-physical vacuum contributions, we normal order all products of operators.

Going to the interaction picture we have
\begin{equation}
 \mathcal{H}^I = \mathcal{H}_{LS}^I + \mathcal{H}_{EM}^I + \mathcal{H}_{int}^I,
\end{equation}
where the operator $\mathcal{O}$ in the Heisenberg picture is connected to its
representation in the interaction picture by the following unitary transformation,
\begin{equation}
\mathcal{O}^I = U(t)\mathcal{O}U^\dagger(t).
\label{unitaryHPIP}
\end{equation}
The unitary operator above is \cite{man86,gre95}
\begin{equation}
 U(t) = e^{iH_0^St/\hbar}e^{-iHt/\hbar},
 \label{unitaryHPIP2}
\end{equation}
with 
\begin{equation}
H_0^S = H_{LS}^S + H_{EM}^S
\end{equation}
being the total free-field Hamiltonian in the Schr\"odinger picture. Note that
$H_j^S=\int d^3x$ $\mathcal{H}_{j}^S(t,\mathbf{x})$, $j=LS$ or $EM$. The quantity
$H=\int d^3x\mathcal{H}(t,\mathbf{x})$ is the complete Hamiltonian, 
having the same form in the Schr\"odinger and Heisenberg
pictures. 

In order to apply the Dyson series we have to write the interaction part of the Hamiltonian
density in terms of operators in the interaction picture. Using eqs.~(\ref{HintSED}),  
the unitarity of $U(t)$, and that $U(t)$ does not depend on the spatial variables, we obtain
with the help of eq.~(\ref{unitaryHPIP}),
\begin{eqnarray}
\mathcal{H}_{int}^I &=& \frac{iqc}{\hbar}[\Pi^I_{\Psi^\dagger}(\Psi^I)^\dagger
-\Pi^I_{\Psi}\Psi^I ](A^I)^0 
+\frac{iq}{\hbar}[(\Psi^I)^\dagger \overleftrightarrow{\partial}_{\hspace{-.15cm}j}
\Psi^I](A^I)^j \nonumber \\
&&-\frac{q^2}{\hbar^2}\Psi^I(\Psi^I)^\dagger A^I_\mu (A^I)^\mu  
+ \frac{q^2}{\hbar^2}\Psi^I(\Psi^I)^\dagger A^I_0 (A^I)^0.
\label{HintSED2}
\end{eqnarray}

We now eliminate the conjugate fields $\Pi^I_{\Psi}$ and  
$\Pi^I_{\Psi^\dagger}$ in favor of the time derivatives of the fields
$\partial_0\Psi^I$ and $\partial_0(\Psi^I)^\dagger$. To that end, we note
first that
\begin{eqnarray}
\partial_0\Psi^I \!&\!=\!&\! \partial_0 ( U\Psi U^\dagger) 
=(\partial_0 U)\Psi U^\dagger + U(\partial_0 \Psi) U^\dagger 
+ U\Psi (\partial_0 U^\dagger). \label{partial0PsiI}
\end{eqnarray}
Using the definitions of $U(t)$ and $U^\dagger(t)$, a direct calculation leads to
\begin{eqnarray}
\partial_0 U &=& -\frac{i}{\hbar c}UH_{int}, \label{delU}\\
\partial_0 U^\dagger &=& \frac{i}{\hbar c}H_{int}U^\dagger \label{delUdagger}.
\end{eqnarray}
Moreover, $\partial_0 \Psi$ is obtained from the time evolution of $\Psi$ in the 
Heisenberg picture,
\begin{equation}
\partial_0 \Psi = \frac{1}{i\hbar c}[\Psi,H]. \label{eqHP}
\end{equation}
Inserting eqs.~(\ref{delU}), (\ref{delUdagger}), and (\ref{eqHP}) into (\ref{partial0PsiI})
we get
\begin{equation}
\partial_0\Psi^I = \frac{i}{\hbar c}U[H_0,\Psi]U^\dagger, \label{partial0PsiI2}
\end{equation}
where
\begin{equation}
H_0 = H_{LS} + H_{EM}.
\end{equation}
Since $H_{EM}$ has no dependence on the matter fields, it commutes with $\Psi$ and
eq.~(\ref{partial0PsiI2}) is reduced to
\begin{equation}
\partial_0\Psi^I = \frac{i}{\hbar c}U[H_{LS},\Psi]U^\dagger. \label{partial0PsiI3}
\end{equation}

The commutator above, with the aid of eqs.~(\ref{mathcalHLS}), (\ref{com1}), 
and (\ref{com2}), is given by 
\begin{eqnarray}
[H_{LS},\Psi(t,\mathbf{x})] &=& \int d^3x'[\mathcal{H}_{LS}(t,\mathbf{x'}),
\Psi(t,\mathbf{x})] \nonumber \\
&=&  \int d^3x'c^2[\Pi_\Psi(t,\mathbf{x'}),\Psi(t,\mathbf{x})]
\Pi_{\Psi^\dagger}(t,\mathbf{x'})  \nonumber \\
&& + \int d^3x'i\mu c\Psi(t,\mathbf{x'})[\Pi_\Psi(t,\mathbf{x'}) ,\Psi(t,\mathbf{x})]
\nonumber \\
&=& -i\hbar c^2 \Pi_{\Psi^\dagger}(t,\mathbf{x}) + \mu c\hbar  \Psi(t,\mathbf{x}).
\label{comHLSPsi}
\end{eqnarray}

Inserting eq.~(\ref{comHLSPsi}) into (\ref{partial0PsiI3}) we finally get
\begin{equation}
\partial_0\Psi^I = c \Pi^I_{\Psi^\dagger} + i\mu \Psi^I.
\label{partial0PsiI4}
\end{equation}
An analogous calculation leads to
\begin{equation}
\partial_0(\Psi^I)^\dagger = c \Pi^I_{\Psi} - i\mu (\Psi^I)^\dagger.
\label{partial0PsiI5}
\end{equation}

Equations (\ref{partial0PsiI4}) and (\ref{partial0PsiI5}) are what we need to 
eliminate the conjugate fields in eq.~(\ref{HintSED2}). Remembering that we are
adopting the normal ordering prescription (we can freely write, for example, 
$(\Psi^I)^\dagger\Psi^I$ instead of $\Psi^I(\Psi^I)^\dagger$),
we obtain
\begin{eqnarray}
\mathcal{H}_{int}^I &=& -\mathcal{L}_{int}^I 
+ \frac{q^2}{\hbar^2}(\Psi^I)^\dagger\Psi^I A^I_0 (A^I)^0,
\label{HintSED3}
\end{eqnarray}
where 
\begin{eqnarray}
\mathcal{L}_{int}^I & = & 
-\frac{iq}{\hbar}[(\Psi^I)^\dagger \overleftrightarrow{\partial}_{\hspace{-.15cm}\mu}
\Psi^I](A^I)^\mu
+\frac{q^2}{\hbar^2}(\Psi^I)^\dagger\Psi^I A^I_\mu (A^I)^\mu 
-\frac{2\mu q}{\hbar}(\Psi^I)^\dagger\Psi^I (A^I)^0. \label{LintSED}
\end{eqnarray}
Note that $\mathcal{L}_{int}^I$ is invariant under a proper Lorentz transformation while 
the so-called ``normal-dependent term'' $(q^2/\hbar^2)$ $(\Psi^I)^\dagger\Psi^IA^I_0 (A^I)^0$ is not \cite{gre95}. 

Equation (\ref{HintSED3}) differs from the one we would obtain if we worked with
the Klein-Gordon fields by the last term in eq.~(\ref{LintSED}),
\begin{equation}
\frac{2\mu q}{\hbar}(\Psi^I)^\dagger\Psi^I (A^I)^0.
\label{LSKGdiff}
\end{equation}
However, the extra term (\ref{LSKGdiff}), together with the rest of $\mathcal{H}^I_{int}$,
will not lead to different physical processes that are not contained in the 
Klein-Gordon theory. 

To see this we rewrite eq.~(\ref{HintSED3}) as
\begin{eqnarray}
\mathcal{H}_{int}^I &=&  
\frac{q}{\hbar}\left\{[i(\Psi^I)^\dagger \overleftrightarrow{\partial}_{\hspace{-.15cm}\mu}
\Psi^I](A^I)^\mu+2\mu (\Psi^I)^\dagger\Psi^I (A^I)^0\right\} \nonumber \\
&+& 
\frac{q^2}{\hbar^2}[(\Psi^I)^\dagger\Psi^I A^I_0 (A^I)^0 - (\Psi^I)^\dagger\Psi^I A^I_\mu (A^I)^\mu].
\label{HintSED4}
\end{eqnarray}
In eq.~(\ref{HintSED4}) we have grouped terms proportional to the char\-ge $q$ 
and those proportional to $q^2$. 

If we now insert eq.~(\ref{transformationKG2b}) into (\ref{HintSED4}) we obtain
\begin{eqnarray}
\mathcal{H}_{int}^I \!\!&\!=\!&\!\!  
\frac{q}{\hbar}\left\{[i(\Phi^I)^\dagger \overleftrightarrow{\partial}_{\hspace{-.15cm}\mu}
\Phi^I](A^I)^\mu\right\} 
+\frac{q^2}{\hbar^2}[(\Phi^I)^\dagger\Phi^I\! A^I_0 (A^I)^0 \!-\! (\Phi^I)^\dagger\Phi^I\!
A^I_\mu (A^I)^\mu].
\label{HintSED5}
\end{eqnarray}
Note that terms proportional to $q$ and $q^2$ in eq.~(\ref{HintSED4})
are respectively mapped to the ones proportional to $q$ and $q^2$ in eq.~(\ref{HintSED5}).

Furthermore, eq.~(\ref{HintSED5}) is exactly the interaction Hamiltonian density 
we obtain when dealing 
with the Klein-Gor\-don theory and it is not difficult to see that the free-field Hamiltonians
are all transformed to the ones of the Klein-Gordon theory after eq.~(\ref{transformationKG2b}). 
Therefore, the S-matrix of the Klein-Gordon and Lorentz
covariant Schr\"odinger theories are the same. This implies, using the same arguments
given in section \ref{lambda4b}, that in the framework of scalar electrodynamics 
both theories lead to equivalent predictions when it comes to 
scattering and decay processes.

\section{Transformation law for $\mathbf{\Psi(x)}$ under finite proper Lorentz transformations }
\label{apB}

After an infinitesimal counterclockwise spatial rotation of a reference frame 
about its $x^3$-axis, the original coordinates $x^\mu$ 
describing a four-vector are related
to those in the new frame ($x^{\mu'}$) as follows,
\begin{eqnarray}
x^0 &=& x^{0'}, \\
x^1 &=& x^{1'}-\epsilon x^{2'}, \label{apb2}\\
x^2 &=& x^{2'}+\epsilon x^{1'}, \label{apb3}\\
x^3 &=& x^{3'},
\end{eqnarray}
where $\epsilon$ is an infinitesimal angle of rotation. 

To simplify the notation 
in the following calculations, it is convenient to use $ct,x,y,z$ to label the
coordinates of the above contravariant four-vector. Also, the inertial frame
before any rotation is implemented is called $S_0$ and the inertial frame after
$n$ infinitesimal rotations is denoted by $S_n$. The coordinates of
a four-vector in $S_n$ is given by $ct_n,x_n,y_n,z_n$. Therefore, according to
eqs.~(\ref{apb2}) and (\ref{apb3}), after $n$ spatial rotations about the $z$-axis we have
\begin{eqnarray}
x_{n-1} &=& x_{n}-\epsilon y_{n}, \label{apbx}\\
y_{n-1} &=& \epsilon x_{n} + y_{n}. \label{apby}
\end{eqnarray}

After $N$ infinitesimal rotations, we end up at the frame $S_N$. Solving the system
of recursive relations (\ref{apbx}) and (\ref{apby}) with the ``final conditions'' 
$x_N$ and $y_N$ we get
\begin{eqnarray}
x_n &=& \frac{1}{2}[(1-i\epsilon)^{N-n}+(1+i\epsilon)^{N-n}]x_N 
-\frac{i}{2}[(1-i\epsilon)^{N-n}-(1+i\epsilon)^{N-n}]y_N, \label{xn}\\
y_n &=& \frac{i}{2}[(1-i\epsilon)^{N-n}-(1+i\epsilon)^{N-n}]x_N 
+\frac{1}{2}[(1-i\epsilon)^{N-n}+(1+i\epsilon)^{N-n}]y_N.\label{yn}
\end{eqnarray}

According to eqs.~(\ref{psiinf}) and (\ref{bmu}), after an infinitesimal rotation
about the $z$-axis, the wave functions in $S_{n-1}$ and $S_n$ are connected to each
other according to the following transformation law,
\begin{equation}
\Psi_{n-1} = e^{-i\epsilon(\kappa_y x_n-\kappa_x y_n)}\Psi_n. \label{psin}
\end{equation}
To arrive at eq.~(\ref{psin}) we used that 
$\epsilon^{21}=-\epsilon^{12}=\epsilon$, with all other $\epsilon^{\mu\nu}$ being zero,
and renamed $\kappa^1$ and $\kappa^2$ to $\kappa_x$ and $\kappa_y$. 

Using repeatedly eq.~(\ref{psin}), we obtain after $N$ infinitesimal rotations $\epsilon$
about the $z$-axis that
\begin{equation}
\Psi_{N} = e^{i\epsilon\left(\kappa_y \sum_{n=1}^N x_n
-\kappa_x \sum_{n=1}^N y_n\right)}\Psi_0. \label{psiN}
\end{equation}
If we now employ eqs.~(\ref{xn}) and (\ref{yn}), the sums in eq. (\ref{psiN}) become
\begin{eqnarray}
\sum_{n=1}^N x_n & = & -\frac{i}{2\epsilon}[(1+i\epsilon)^N-(1-i\epsilon)^N]x_N 
- \frac{1}{2\epsilon}[2\!-\!(1-i\epsilon)^N \!-\! (1+i\epsilon)^N]y_N,
\label{b11}\\
\sum_{n=1}^N y_n & = & \frac{1}{2\epsilon}[2-(1-i\epsilon)^N - (1+i\epsilon)^N]x_N 
- \frac{i}{2\epsilon}[(1+i\epsilon)^N - (1-i\epsilon)^N  ]y_N.
\label{b12}
\end{eqnarray}
Inserting eqs.~(\ref{b11}) and (\ref{b12}) into (\ref{psiN}) we obtain
\begin{eqnarray}
\Psi_N &=& \exp\left\{-\frac{i\kappa_x x_N}{2} f(\epsilon) + \frac{\kappa_y x_N}{2} g(\epsilon)
-\frac{\kappa_x y_N}{2} g(\epsilon) - \frac{i\kappa_y y_N}{2} f(\epsilon)
\right\}\Psi_0, \label{psiNpsi0}
\end{eqnarray}
where
\begin{eqnarray}
f(\epsilon) &=& 2-(1-i\epsilon)^N - (1+i\epsilon)^N, \label{fe}\\
g(\epsilon) &=& (1+i\epsilon)^N - (1-i\epsilon)^N. \label{ge}
\end{eqnarray}

A finite rotation $\phi$ can be split into $N$ infinitesimal ones such that
\begin{equation}
\epsilon = \phi/N.\label{epsphiN} 
\end{equation}
Inserting eq.~(\ref{epsphiN}) into eqs.~(\ref{fe}) and (\ref{ge}) and taking
the limit for large $N$ we get
\begin{eqnarray}
\lim_{N\rightarrow \infty} f(\phi/N) &=& 2(1-\cos\phi), \label{Nfe} \\
\lim_{N\rightarrow \infty} f(\phi/N) &=& 2i\sin\phi. \label{Nge}
\end{eqnarray}

If we identify $\Psi_0$ as the wave function in the rest frame $S$ before the finite 
rotation $\phi$ is implemented, $\lim_{N\rightarrow \infty}\Psi_N$ as the 
wave function at the rest frame $S'$ after the rotation, 
and if we go back to the four-vector notation, 
eqs.~(\ref{Nfe}) and (\ref{Nge})
when inserted into (\ref{psiNpsi0}) give
\begin{equation}
\Psi(x) = e^{i[(\kappa^1x^{1'}+\kappa^2x^{2'})(1-\cos\phi)
+(\kappa^1x^{2'}-\kappa^2x^{1'})\sin\phi]}\Psi'(x'). \label{b19}  
\end{equation}
Equation (\ref{b19}) is nothing but the transformation law given by eq.~(\ref{rotationx3}) 
for a finite rotation $\phi$ about the $x^3$-axis. 
By similar calculations we can get the transformation laws for $\Psi(x)$ when we rotate
about the $x^2$ and $x^1$ axes.

Let us now show how the transformation law for finite boosts can be derived for $\Psi(x)$.
To that aim we write a finite boost along the $x^1$-axis as a rotation 
in a hyperbolic space. The coordinate transformation is thus
\begin{eqnarray}
x^0 &=& x^{0'}\cosh\xi  + x^{1'}\sinh\xi, \label{b20}\\
x^1 &=& x^{0'}\sinh\xi  + x^{1'}\cosh\xi, \label{b21}\\
x^2 &=& x^{2'}, \\
x^3 &=& x^{3'},
\end{eqnarray}
where the ``hyperbolic angle'' $\xi$ is called the ``rapidity'' and is given by
\begin{equation}
\tanh\xi = \beta = v/c.
\end{equation}
Here $v$ is the speed of frame $S'$ with respect to $S$, directed along the $x^1$-axis.
Note that $\cosh\xi = \gamma$ and $\sinh\xi = \beta \gamma$. 

Expressed as given by eqs.~(\ref{b20}) and (\ref{b21}), two successive boosts along the 
$x^1$-axis with rapidity $\xi$ and $\xi'$ gives $x^0 = x^{0''}\cosh(\xi+\xi')  
+ x^{1''}\sinh(\xi+\xi')$ and $x^1 = x^{0''}\sinh(\xi+\xi')  + x^{1''}\cosh(\xi+\xi')$,
where $x^{\mu''}$ are the coordinates of the four-vector $x^\mu$ in the frame $S''$. 
This is formally equivalent to spatial rotations if
the hyperbolic sines and cosines are changed to the usual trigonometric ones. 
Therefore, the calculations above leading to the transformation law for $\Psi(x)$ after 
a spatial rotation can be readily adapted to a boost if we note that for an 
infinitesimal boost ($\xi\ll 1$) we have
\begin{eqnarray}
x^0 &=& x^{0'}  + \xi x^{1'}, \label{b25}\\
x^1 &=& x^{1'} +  \xi x^{0'}. \label{b26}
\end{eqnarray}

The analog to eqs.~(\ref{apbx}) and (\ref{apby}) are
\begin{eqnarray}
ct_{n-1} &=& ct_{n} + \xi x_{n}, \label{apbx2}\\
x_{n-1} &=& \xi ct_{n} + x_{n}, \label{apby2}
\end{eqnarray}
whose solution is
\begin{eqnarray}
ct_n &=& \frac{1}{2}[(1+\xi)^{N-n}+(1-\xi)^{N-n}]ct_N 
+\frac{1}{2}[(1+\xi)^{N-n}-(1-\xi)^{N-n}]x_N, \label{xn2}\\
x_n &=& \frac{1}{2}[(1+\xi)^{N-n}-(1-\xi)^{N-n}]ct_N 
+\frac{1}{2}[(1+\xi)^{N-n}+(1-\xi)^{N-n}]x_N.\label{yn2}
\end{eqnarray}

Repeating all the steps of the previous calculation we get
\begin{equation}
\Psi_{N} = e^{i\xi\left(\kappa_x \sum_{n=1}^N ct_n
-\kappa_{ct} \sum_{n=1}^N x_n\right)}\Psi_0 \label{psiN2}
\end{equation}
and
\begin{eqnarray}
\sum_{n=1}^N ct_n & = & \frac{1}{2\xi}[(1+\xi)^N-(1-\xi)^N]ct_N 
- \frac{1}{2\xi}[2\!-\!(1+\xi)^N \!-\! (1+\xi)^N]x_N,
\label{b112}\\
\sum_{n=1}^N x_n & = & -\frac{1}{2\xi}[2-(1+\xi)^N - (1+\xi)^N]ct_N 
+\frac{1}{2\xi}[(1+\xi)^N - (1-\xi)^N  ]x_N.
\label{b122}
\end{eqnarray}

Inserting eqs.~(\ref{b112}) and (\ref{b122}) into (\ref{psiN2}) we arrive at
\begin{eqnarray}
\Psi_N &=& \exp\left\{ \frac{i\kappa_{ct} ct_N}{2} f(\xi) + \frac{i\kappa_{x} ct_N}{2} g(\xi) 
-\frac{i\kappa_{ct} x_N}{2} g(\xi) - \frac{i\kappa_x x_N}{2} f(\xi)
\right\}\Psi_0, \label{psiNpsi02}
\end{eqnarray}
where
\begin{eqnarray}
f(\xi) &=& 2-(1+\xi)^N - (1-\xi)^N, \label{fe2}\\
g(\xi) &=& (1+\xi)^N - (1-\xi)^N. \label{ge2}
\end{eqnarray}

To go from successive infinitesimal boosts to a finite one we set 
\begin{equation}
\xi = \Xi/N 
\end{equation}
and take the appropriate limits in eqs.~(\ref{fe2}) and (\ref{ge2}),
\begin{eqnarray}
\lim_{N\rightarrow \infty} f(\Xi/N) &=& 2(1-\cosh\Xi), \label{Nfe2} \\
\lim_{N\rightarrow \infty} f(\Xi/N) &=& 2\sinh\Xi. \label{Nge2}
\end{eqnarray}

Noting that $\cosh \Xi = \gamma$ and $\sinh \Xi = \beta \gamma$, eqs.~(\ref{Nfe2})
and (\ref{Nge2}) allow us to write (\ref{psiNpsi02}) as
\begin{equation}
\Psi(x)=e^{i[(\gamma -1)\kappa^0-\gamma \beta \kappa^{1}]x^{0'}
+i[\gamma \beta \kappa^0 -(\gamma -1)\kappa^1]x^{1'}}\Psi'(x'). \label{b192}
\end{equation}

To arrive at eq.~(\ref{b192}) we have reversed to the usual four-vector notation,
namely, $ct_N=x^{0'}, x_N = x^{1'}, \kappa_{ct}=\kappa^0,\kappa_{x}=\kappa^1$, and 
identified $\Psi_0$ with $\Psi(x)$ and $\Psi_N$ with $\Psi'(x')$. Equation
(\ref{b192}) is the transformation law for $\Psi(x)$ when it is subjected to 
a finite boost along the $x^1$-axis, namely, eq.~(\ref{boostx1}) given in the main text.

\section*{Data availability}

This is a theoretical work and 
no data were used in writing it.

\section*{Conflicts of interest}

The author declares that he has no conflicts of interest.

\acknowledgments
The author thanks the Brazilian agency CNPq
(National Council for Scientific and Technological Development) for partially
funding this research.


\begin{thebibliography}{200}

\bibitem{bal98} L. E. Ballentine, \textit{Quantum Mechanics: A Modern Development}, World Scientific, Singapore (1998).

\bibitem{gre00} W. Greiner, \textit{Relativistic Quantum Mechanics: Wave Equations}, Springer-Verlag, Berlin (2000).

\bibitem{gri95} D. J. Griffiths, \textit{Introduction to Quantum Mechanics}, Prentice-Hall, Upper Saddle River (1995).

\bibitem{man86} F. Mandl and G. Shaw, \textit{Quantum Field Theory}, John Wiley \& Sons, Chichester (1986).

\bibitem{gre95} W. Greiner and J. Reinhardt, \textit{Field Quantization}, Springer-Verlag, Berlin (1996).

\bibitem{sch26} E. Schr\"odinger, \emph{An Undulatory Theory of The Mechanics of
Atoms and Molecules}, \emph{Phys. Rev.} \textbf{28} (1926) 1049. 

\bibitem{fol50} L. L. Foldy and S. A. Wouthuysen, 
\emph{On the Dirac Theory of Spin 1/2 Particles and Its Non-Relativistic Limit}, \emph{Phys. Rev.} \textbf{78} (1950) 29. 

\bibitem{tuc99} M. A. H. Tucker and A. F. G. Wyatt, 
\emph{Direct Evidence for R-Rotons Having Antiparallel Momentum and Velocity}, \emph{Science} \textbf{283} 
(1999) 1150.

\bibitem{pol01} B. Julsgaard, A. Kozhekin and E. S. Polzik, 
\emph{Experimental long-lived entanglement of two macroscopic objects}, 
\emph{Nature (London)} \textbf{413} (2001) 400.

\bibitem{cav12} M. Tsang and C. M. Caves, \emph{Evading Quantum Mechanics: Engineering a Classical Subsystem within a Quantum Environment}, \emph{Phys. Rev. X} \textbf{2} (2012) 031016.

\bibitem{con15} M. Conforti, S. Trillo, A. Mussot and A. Kudlinski, \emph{Parametric excitation of multiple resonant radiations from localized wavepackets}, 
\emph{Sci. Rep.} \textbf{5} (2015) 9433.

\bibitem{kha17} M. A. Khamehchi, K. Hossain, M. E. Mossman, Y. Zhang, Th. Busch,
M. McNeil Forbes and P. Engels, \emph{Negative-Mass Hydrodynamics in a Spin-Orbit–Coupled Bose-Einstein Condensate}, \emph{Phys. Rev. Lett.} \textbf{118} (2017) 155301.

\bibitem{stu42} E. C. G. Stueckelberg, \emph{La m\'ecanique du point mat\'eriel en
th\'eorie de la relativit\'e et en th\'eorie des quanta}, \emph{Helv. Phys. Acta} \textbf{15} (1942) 23.

\bibitem{nam50} Y. Nambu, \emph{The Use of the Proper Time in Quantum 
Electrodynamics I}, \emph{Prog. Theor. Phys.} \textbf{5} (1950) 82.

\bibitem{hor73} L. P. Horwitz and C. Piron, \emph{Relativistic dynamics}, \emph{Helv. Phys. Acta} \textbf{46} (1973) 316. 

\bibitem{fan78} J. R. Fanchi and R. E. Collins, 
\emph{Quantum Mechanics of Relativistic Spinless Particles}, \emph{Found. Phys.} \textbf{8} (1978) 851.

\bibitem{stu41} E. C. G. Stueckelberg, \emph{Remarque \`a propos de la cr\'eation de paires de particules en th\'eorie de relativit\'e}, \emph{Helv. Phys. Acta} \textbf{14} (1941) 588.

\bibitem{hea87a} O. Heaviside, \emph{Electromagnetic Induction and its Propagation}, \emph{Electrician} \textbf{19} (1887) 79.

\bibitem{hea87b} O. Heaviside, \textit{Electromagnetic Theory, Vol. 1}, 
The Electrician Printing \& Publishing Company, London (1893).

\bibitem{pan97} P. Sancho, \emph{The quantum telegraph equation}, \emph{Il Nuovo Cimento} \textbf{112B} (1997) 1437.

\bibitem{wil74} C. M. Will, \emph{Gravitational red-shift measurements as tests of nonmetric theories of gravity}, \emph{Phys. Rev. D} \textbf{10} (1974) 2330.

\bibitem{tou76} P. Tourrenc and J. L. Grossiord, 
\emph{Modification du spectre de l'hydrog\`ene atomique sous l'influence
de champs de gravitation}, 
\emph{Il Nuovo Cimento} \textbf{32B} (1976) 163.

\bibitem{par80} L. Parker, \emph{One-Electron Atom in Curved Space-Time}, 
\emph{Phys. Rev. Lett.} \textbf{44} (1980) 1559.

\bibitem{fis81} E. Fischbach, B. S. Freeman and W.-K Cheng, 
\emph{General-relativistic effects in hydrogenic systems}, 
\emph{Phys. Rev. D} \textbf{23} (1981) 2157.

\bibitem{asp20} T. Westphal, H. Hepach, J. Pfaff and M. Aspelmeyer, 
\emph{Measurement of Gravitational Coupling between Millimeter-Sized Masses}, 
[arXiv:2009.09546].

\bibitem{din04} M. Dine and A. Kusenko, 
\emph{Origin of the matter-antimatter asymmetry}, 
\emph{Rev. Mod. Phys.} \textbf{76} (2004) 1.

\bibitem{ber97} O. Bertolami, D. Colladay, V. A. Kosteleck\'y and
R. Potting, \emph{CPT violation and baryogenesis}, 
\emph{Phys. Lett. B} \textbf{395} (1997) 178.

\bibitem{col98} D. Colladay and V. A. Kosteleck\'y,
\emph{Lorentz-violating extension of the standard model},
\emph{Phys. Rev. D} \textbf{58} (1998) 116002.

\bibitem{car06} S. M. Carroll and J. Shu, \emph{Models of baryogenesis via spontaneous Lorentz violation}, \emph{Phys. Rev. D} \textbf{73} 
(2006) 103515.

\bibitem{ces15} M. de Cesare, N. E. Mavromatos and S. Sarkar, \emph{On the possibility of tree-level leptogenesis from Kalb–Ramond torsion background}, 
\emph{Eur. Phys. J. C} \textbf{75} (2015) 514.

\bibitem{sak17} J. Sakstein and A. R. Solomon, \emph{Baryogenesis in Lorentz-violating gravity theories}, \emph{Phys. Lett. B} \textbf{773} (2017) 186. 

\bibitem{edw18} B. R. Edwards and V. A. Kosteleck\'y,
\emph{Riemann-Finsler geometry and Lorentz-violating scalar fields},
\emph{Phys. Lett. B} \textbf{786} (2018) 319.

\bibitem{and04} D. L. Anderson, M. Sher and I. Turan, 
\emph{Lorentz and CPT violation in the Higgs sector},
\emph{Phys. Rev. D} \textbf{70} (2004) 016001.


\bibitem{saf93}  V. I. Safonov, \emph{Conduction Electron in the Anisotropic Medium}, \emph{Int. J. Mod. Phys. B} \textbf{7} (1993) 3899.

\bibitem{zha19} A. Zhao, J. Zhang, Q. Gu and R. A. Klemm, 
\emph{A relativistic electron in an anisotropic conduction band}, [arXiv:1905.03127].

\bibitem{saf18} M. S. Safronova, D. Budker, D. DeMille, D. F. J. Kimball, 
A. Derevianko and C. W. Clark, \emph{Search for new physics with atoms and molecules}, \emph{Rev. Mod. Phys.} \textbf{90} (2018) 025008.

\bibitem{bon57} H. Bondi, \emph{Negative Mass in General Relativity}, 
\emph{Rev. Mod. Phys.} \textbf{29} (1957) 423.

\bibitem{kow96} M. Kowitt, \emph{Gravitational Repulsion and Dirac Antimatter}, \emph{Int. J. Theor. Phys.} \textbf{35} (1996) 605. 

\bibitem{vil11} M. Villata, \emph{CPT symmetry and antimatter gravity in general
relativity}, \emph{EPL} \textbf{94} (2011) 20001.

\bibitem{far18} J. S. Farnes, \emph{A unifying theory of dark energy and dark matter: Negative masses and matter creation within a modified $\Lambda CDM$ framework}, \emph{A\&A} \textbf{620} (2018) A92.


\end{thebibliography}
\end{document}